     \documentclass[10pt]{article}
     \usepackage{amstex,amssymb}
     \usepackage{epsfig}
     \usepackage{subfigure}
     \usepackage{afterpage}

     \newcommand{\e}{\ensuremath{\,\mathrm{e}}}
     \newcommand{\ie}{i.e.}    
     \newcommand{\eg}{e.g.} 
     \newcommand{\ea}{et al.}

\begin{document}

\bibliographystyle{PhysRepAlpha}  



{\mbox{ }}\hfill IASSNS-HEP-96/25

{\mbox{ }}\hfill CERN-TH/96-97

          {\mbox{ }}{\vspace{2cm}}\\
          \noindent 
          {\Large\bf Supergravity Domain Walls}\\ 
          {\mbox{ }}{\vspace{0.7cm}}\\
          {\bf Mirjam Cveti{\v{c}}}\\
          Institute for Advanced Study, 
          School of Natural Sciences\\
          Olden Lane, Princeton, NJ 80540. U.S.A.\\
          Phone: (+1) 609 734 8176\\
          FAX:   (+1) 609 924 8399\\
          e-mail: {\verb+cvetic@ias.edu+}\\[1em] 
          and\\[1em] 
          Department of Physics and Astronomy,
          University of Pennsylvania,\\
          Philadelphia, PA 19104-6396, 
          U.S.A.\\
          Phone: (+1) 215 898 8153\\
          FAX:   (+1) 215 898 8512\\
          e-mail: {\verb+cvetic@cvetic.hep.upenn.edu+} 
          {\mbox{ }}{\vspace{0.9cm}}\\
          {\bf Harald H. Soleng}\\ 
          CERN, Theory Division,\\
          CH-1211 Geneva 23,
          Switzerland\\
          Phone: (+41) 22--767 2139\\
          FAX:   (+41) 22--768 3914\\
          e-mail: {\verb+soleng@surya11.cern.ch+} 

          {\mbox{ }}{\vspace{2cm}}\\
          April 16, 1996


\tableofcontents

\title{Supergravity domain walls} 

\author{Mirjam Cveti{\v{c}}\footnotemark\\
        Institute for Advanced Study,
        School of Natural Sciences,\\
        Olden Lane, Princeton, NJ 80540. U.S.A.\\
       \and
        Harald H. Soleng\\ 
        Theory Division, CERN, CH-1211 Geneva 23, Switzerland}

\date{April 16, 1996}

\maketitle

\renewcommand{\thefootnote}{\fnsymbol{footnote}}
\addtocounter{footnote}{1}
\footnotetext{On 
        sabbatical leave from  Department of Physics and Astronomy,
        University of Pennsylvania,
        Philadelphia, PA 19104-6396, U.S.A.}
\renewcommand{\thefootnote}{\arabic{footnote}}
\addtocounter{footnote}{-\value{footnote}}

\begin{abstract}
We review the status of domain walls in $N=1$ supergravity
 theories for both the \emph{vacuum}
domain walls as well as \emph{dilatonic} domain walls. 
We concentrate on a systematic analysis of  
the nature of the space--time  in such domain 
wall backgrounds and the special r\^ole that supersymmetry 
is playing in determining the nature of such 
configurations. 
Isotropic vacuum 
domain
walls that can exist between isolated minima of a $N=1$ supergravity
matter potential fall into three classes:
(i)~extreme walls which are  static planar walls between supersymmetric
minima,
 (ii)~non-extreme walls which are expanding
bubbles with two centres and  (iii)~ultra-extreme walls which are
 bubbles of false vacuum decay.  
Dilatonic walls arise  in 
$N=1$ supergravity with the  linear supermultiplet, an additional scalar field-the dilaton- has no perturbative self-interaction, which however, couples to
the  matter potential responsible for the formation of
the wall. The dilaton drastically 
changes the global space--time properties of
the wall. For the extreme ones the
space--time structure
 depends on the strength of the
dilaton coupling, while  for non-\
and ultra-extreme solutions one always
encounters naked singularities (in the absence of 
non-perturbative corrections to the dilaton potential).
Non-perturbative  effects may modify the 
dilaton coupling so that it has 
a discrete non-compact symmetry ($S$-duality). In this case 
the non- and ultra-extreme solutions  can reduce to the singularity-free 
vacuum domain wall solutions.
We also summarize domain wall configurations within  effective 
theory of $N=1$ superstring vacua, 
with or 
with\-out in\-clu\-sion of non-per\-tur\-ba\-tive string effects,
and also provide a comparison with other topological
defects of perturbative string vacua. 
\end{abstract}


           \section{Introduction}
\label{Sect:Introduction}

Topological defects can occur in a physical system when 
the vacuum manifold of the system possesses a non-trivial topology.  
Domain walls correspond to a  specific type  of topological 
defects  that can occur  when the vacuum manifold 
consists of (energetically degenerate) disconnected components.  
Among  the topological defects, domain walls  are the 
most extended ones, and thus may have the most 
``disruptive'' implications for
the nature of the  non-trivial ground  state of the physical system.

In fundamental
theories of elementary particles  the spontaneous (gauge)
symmetry breaking via the 
 Higgs mechanism 
\cite{Goldstone:INC61,%
EB:PRL64,Higgs:PL64,Higgs:PRL64,GHK:PRL64,Higgs:PR66}
plays a central r{\^{o}}le.  Such theories in general possess
a degenerate vacuum manifold with a non-trivial topology, 
as specified by the potential of the scalar (Higgs) field(s). 
Therefore it is  plausible that topological defects  
could be important in basic theory, in particular in its 
application to cosmology. 

In particular, domain wall solutions  exist in 
theories where the scalar field potential has 
isolated minima.   The walls are surfaces interpolating 
between separate minima of the scalar potential  
with different vacuum expectation values of the  scalar 
field. In this case the scalar field 
changes with spatial position and 
settles in one minimum 
at one spatial 
infinity 
while in the other direction 
it settles in another disconnected minimum.  The 
interpolating
region of rapid change of the scalar field 
corresponds to  the domain wall. 
In the thin wall approximation  the  variation 
of the scalar field energy density is localized 
at the domain wall surface, and  is replaced by the 
delta function. In the case when all the matter fields are 
constant on each side of the wall, \ie\ they are 
settled at the minimum of the potential, the domain walls are 
referred to as vacuum domain walls.

In the early universe such a domain structure can form by the Kibble
\cite{Kibble:LesHouches80,Vilenkin:PRep85} 
mechanism whereby 
different regions of  a hot
universe cool into
different isolated minima of the matter potential.
Domain walls \cite{Vilenkin:PRep85}
can also form as the boundary of a (true)\ vacuum bubble
created by the quantum tunneling process of false
vacuum decay \cite{Coleman:Erice77}.
Additionally,
the universe could be born through
a quantum tunneling process from nothing
\cite{Vilenkin:PLB82,Vilenkin:PRD83,HH:PRD83,%
Vilenkin:PRD84,Linde:LNC84,GH:PRD90}
into different domains with walls in between.

\subsection{Classes of domain walls}
\label{Subsect:Classes}

Because of its extended nature,  the space--time 
around the domain wall is drastically affected. 
Tension reduces the 
gravitational mass, and in the case of a domain wall
where
tension 
is equal to the energy-density and  
where there are two spatial directions contributing with
tension and only one time-direction contributing with 
an energy-density, the total gravitational mass is negative.
The nature of space--time in the presence of 
domain walls is a central topic of this review. 
Let us first briefly summarize some earlier developments in this direction.

The first solution 
of Einstein's field equations for the gravitational field
produced by a thin planar domain wall 
was found by Vilenkin \cite{Vilenkin:PRD81}
in the linear approximation for the gravitational field. 
In this solution, the vacuum energy or the cosmological constant
vanishes on both sides
of the wall.
Since the linear approximation  for the gravitational 
field breaks down at large distances,
this solution could not say anything about the global 
structure of the gravitational field,
but as one should expect, test-particles near the
wall were found to be gravitationally repelled by the wall. 
However, the physical meaning of this  
approximate solution remained obscure 
since it does not correspond 
to \emph{any} exact static solution of
Einstein's field equations: 
in fact,
no such solution exists \cite{DK:GRG89}. 
Instead, 
the corresponding 
exact (thin wall)\ 
solution 
\cite{Vilenkin:PLB83}
has a \emph{time-dependent} metric;
it is the (2+1)-dimensional de~Sitter space on the wall's world
volume and Minkowski space--time elsewhere.
A coordinate transformation revealed that 
Vilenkin's 
\cite{Vilenkin:PLB83}
exact
thin ``planar wall'' 
solution
is a segment of an accelerated sphere \cite{IS:PRD84,Ipser:PRD84} 
which comes in from infinity, turns around,  and heads back out to 
infinity. 

The gravitational effects of spherically symmetric 
thin vacuum bubbles 
had earlier been 
studied in connection with 
vacuum decay \cite{CDeL:PRD80}.  
Using Israel's method \cite{Israel:INC66}
of singular layers
Berezin, Kuzmin and Tkachev 
\cite{BKT:PLB83a,BKT:Moscow84} studied a spherically symmetric 
domain wall separating regions of true or false vacua
with arbitrary non-negative energy densities
(and also allowing for a non-vanishing Schwarzschild mass parameter). 
The vacuum decay bubbles 
were distinguished from 
those originating as results of phase transitions
by their different surface energy density.

Domain walls between Minkowski, de~Sitter, 
Schwarz\-schild, and Schwarz\-schild--de~Sitter spaces 
are discussed in 
Refs.~\cite{IS:PRD84,%
LM:PRD86,%
Sato:PTP86,%
BGG:PRD87,%
AKMS:PRD87,BKT:PRD87,%
GV:CQG89}.

Another class of domain walls arise in 
theories where  certain scalar fields (dilatons)\  do not 
possess isolated minima of the matter potential, 
but can couple to the matter potential responsible 
for the formation of the wall. In this case the dilaton can vary with the spatial distance from the wall, thus forming a new type of walls:
dilatonic walls. Particular   
examples are dilatonic walls in the Brans--Dicke theory and 
certain effective  theories describing perturbative string 
vacua. 
As specific examples, a static planar domain wall 
in general relativity coupled to a
conformally coupled massless scalar 
field  \cite{GS:PLA92} 
and 
a solution for general Brans--Dicke coupling $\omega$
\cite{La:PLB93} have been found.

Most of the analysis of the domain wall solutions 
has been done within the thin wall approximation.  
However, also \emph{thick} domain 
wall solutions  have 
been addressed \cite{Widrow:PRD89a,Widrow:PRD89b,%
Goetz:JMP90,Mukherjee:CQG93,AL:PRD94,Arodz:PRD95}.
The interest in such solutions was boosted by the
suggestion that late-time phase transitions could produce
soft topological defects, such as very light domain walls \cite{HSF:CNPP89}
and even more when it was suggested that the Great Attractor could
be such a domain wall \cite{ST:APJL89}. 

\subsection{Walls in $N=1$ supergravity}
\label{Subsect:Sgwalls}

Spontaneously broken $N=1$ supergravity theory  
coupled to Yang--Mills fields provides one of a very few
viable theories, which can  explain   low energy phenomena.  
Such theories are therefore a subject of intense 
research of their phenomenological \cite{Langacker:PROC95} 
as well as certain cosmological implications. The  problem 
of non-renormalizability of supergravity theory has been 
removed by  realizing that supergravity 
is an effective low-energy theory of
superstring theory, which is  
is believed to be a \emph{finite} theory of gravity and gauge interactions. 

It is therefore important  to address different  aspects of 
$N=1$ supergravity theory, and in particular 
those that arise as an effective theory of four 
dimensional superstring vacua. 
In this paper we shall review the status of 
domain wall configurations in  a specific theory of  
elementary particles, \ie\ in a theory  of 
gravity and gauge interactions, that 
possesses (spontaneously broken) $N=1$ supersymmetry.

The first  type of $N=1$ supergravity walls  are vacuum domain walls between vacua
with non-positive cosmological constants  (Minkowski 
and anti--de~Sitter vacua). 
They can be 
classified \cite{CGS:PRL93,CGS:PRD93} according to 
the values of their energy densities $\sigma$.
In particular, for a special value of the domain 
wall energy density there exist  \cite{Linet:IJTP85} static, 
reflection symmetric, planar domain wall with anti-de~Sitter
space--time on both sides. They turn out to correspond to 
a special case of  
a \emph{static}
planar \emph{extreme} domain 
walls with a \emph{supersymmetric 
embedding} in $N=1$ supergravity theory 
\cite{CGR:NPB92,CG:PLB92,CDGS:PRL93} interpolating 
between supersymmetric minima of the scalar potential. 
They have (in the thin wall approximation) 
a fixed energy density $\sigma=\sigma_{\text{ext}}$, 
which is specified by  the values of the 
cosmological constant on each side of the wall. 
The  non-extreme (isotropic) walls  have $\sigma>\sigma_{\text{ext}}$ 
and correspond to accelerating two-centered bubbles,
while ultra-extreme  wall solutions  
have $\sigma<\sigma_{\text{ext}}$ and correspond to 
the false vacuum decay bubbles \cite{CGS:PRL93,CGS:PRD93}. 
Non-\ and ultra-extreme walls do not 
have supersymmetric embeddings,  with 
the extreme solution  (with supersymmetric embedding) 
providing a dividing line between the two classes of them.

Another class, dilatonic  domain walls, 
arise within   $N=1$ 
supergravity coupled to the linear supermultiplet.  For  special couplings of the dilaton field describe  
certain effective  theories of  perturbative string 
vacua. 
A specific example, a static planar domain wall 
in general relativity coupled to a
conformally coupled massless scalar 
field  \cite{GS:PLA92}, 
 corresponds to a special case of  a supersymmetric 
dilatonic domain walls \cite{Cvetic:PRL93} of $N=1$ 
supergravity coupled to the linear supermultiplet  
with a particular value for the dilaton coupling.  
Classification of dilatonic domain walls 
according to the value of its energy density 
has been given in Ref.~\cite{CS:PRD95}.

 Here 
we shall review the status of domain 
wall configurations within a general class of 
$N=1$ supergravity  theories,  and within   
effective theories of superstring vacua as a
special example.
The emphasis will be on a systematic analysis of  
the nature of the space--time  in such domain 
wall backgrounds and the special r\^ole that supersymmetry 
is playing in determining the nature of such 
configurations.  We will summarize the analysis for both the \emph{vacuum}
domain walls as well as \emph{dilatonic} domain walls. 
In addition, the analysis will 
incorporate the  non-supersymmetric generalizations 
of these domain wall backgrounds, and will include 
the majority of the results for the 
domain wall examples discussed in the literature. 

The review is organized in the following way:
In
Section~\ref{Sect:Supergravity}
we present the structure of
the Lagrangian of $N=1$ supergravity coupled
to
Yang--Mills and matter fields focusing on the
bosonic  
part of the action responsible for formation of defects.
Section~\ref{Sect:Topolog} gives an overview of the
physics of topological defects as a general 
background to the 
subject of this review.
Then in Section~\ref{Sect:IsoWalls}
the space--time symmetry assumptions,
the metric ansatz, 
and
the thin wall formalism and the general relativistic field equations 
are presented.
In Section~\ref{Sect:SUSY_Embedding} we
show how these domain walls can be embedded
in $N=1$ supergravity theory. 
Section~\ref{Sect:VacuumWalls} is devoted to
a study of vacuum domain walls and their induced space--times. 
These results 
are generalized to the dilatonic case in Section~\ref{Sect:DilatonicWalls}.
In Section~\ref{Sect:ConnectionOTD}
we discuss the connection of the supergravity domain walls 
to other topological defects of four-dimensional
supergravity theories. Here we also discuss implications for the domain walls for effective supergravity vacua from superstring theory.
Conclusions are given in Section~\ref{Sect:Conclusions}.

           \addtocounter{equation}{-\value{equation}}
%


           \section{Supergravity theory}
\label{Sect:Supergravity}

In this section  we shall review  the structure of the effective 
Lagrangian of $N=1$  supergravity coupled to the 
Yang--Mills and matter fields. 
$N=1$ refers to the fact that the Poincar{\'e} 
algebra is extended with a single 
spinor generator.
We shall primarily concentrate on the bosonic part of
the Lagrangian, since this is the one  
responsible for the formation of the 
topological defects.  In addition,
since the matter fields which make up the 
defects  are  assumed to be neutral under the gauge symmetry, 
we will be most specific about the part of the Lagrangian that
involves the  gravitational and gauge neutral
matter super-multiplets. 

We also spell out the supersymmetry transformations for the
gravitational and matter super-fields.
 Again, we concentrate on supersymmetry
transformations of the  fermionic fields, since  those are the 
non-trivial supersymmetry transformations that are
preserved  by the supersymmetric
(classical)\ bosonic  domain wall backgrounds.

There is a number of excellent reviews addressing  the
supergravity theories in general 
and $N=1$ supergravity theory in particular,
and in much more  details  than   covered in this chapter. 
We  refer the interested reader 
to Refs.~\cite{FF:PRep77,Nieuwenhuizen:PRep81,Nilles:PRep84,%
GGRS:Benjamin83,West:WorldScientific90,WB:Princeton92}. 

Throughout this paper
we use units such that $\kappa\equiv 8\pi G = c=1$.
Our sign convention for the metric, the Riemann tensor,
and the Einstein tensor,
is
of the type 
($-\ +\ +$)\ 
as classified by Misner, Thorne and Wheeler \cite{MTW:Freeman73}.
Also, we use
the conventions: $\gamma^{\mu}={e^\mu}_a\gamma^a$ where
$\gamma^a$ are the flat  space--time Dirac matrices satisfying
$\{\gamma^a,\gamma^b\}=2\eta^{ab} I_4$,
$\gamma^5=\gamma^0\gamma^1\gamma^2\gamma^3$; 
${e^a}_\mu {e^\mu}_b
= {\delta^a}_b$; $a\in \{0, 1, 2, 3\}$; 
$\mu\in\{t, x, y, z\}$.

\subsection{Field content of $N=1$ supergravity}
\label{FCN1SG}
$N=1$ supergravity theory preserves local (space--time)\ dependent 
supersymmetric transformations.
The field content  of $N=1$ supergravity theory coupled to the 
Yang--Mills fields and matter 
fields consists of the  gravitational,  the gauge field, 
and  matter supermultiplets.
The  physical field components of the superfields correspond to  
dynamical  fields, which  remain in the
Lagrangian after elimination of the auxiliary fields 
through the equations 
of
motion.
The  physical particle spectrum of the three types 
of  superfields is the following: 

\begin{itemize}

\item  The \emph{gravitational supermultiplet} ${\bf \Psi}_{\mu\nu}$ contains 
the  spin-$2$ component---%
the graviton field $g_{\mu\nu}$, and  
spin-$\frac{3}{2}$ component---the gravitino $\psi^\alpha_\mu$,

\item  The \emph{vector (Yang--Mills) superfields} 
${\mathcal{W}}_\alpha^{(a)}$,  whose components in the 
Wess--Zumino gauge are spin-1 
gauge field $A_\mu^{(a)}$ and spin-$\frac{1}{2}$  
gaugino field $\lambda^{(a)}$, 

\item  The \emph{chiral superfields} ${\mathcal{T}}_j$  
whose spin-$0$ 
component
is 
a complex scalar field  $T_j$ and spin-$\frac{1}{2}$ 
component is  its supersymmetric partners $\chi_j$, 

\item The \emph{linear supermultiplet} ${\mathcal{L}}$, 
which are gauge neutral fields,
and   contain a spin-0 scalar field $\phi$,   
an anti-symmetric field $b_{\mu\nu}$, which 
is related to a pseudo-scalar field 
through a duality transformation, and the supersymmetric partner
spin-$\frac{1}{2}$ field $\eta$.

\end{itemize}

The linear supermultiplet  
$\mathcal{L}$ can be rewritten (on-shell) in terms
of 
a chiral supermultiplet $\mathcal{S}$, 
by performing  a duality transformation 
\cite{FV:PLB87,BGGM:PLB87,OR:PLB91,ABGG:NPB93}\@.%
\footnote{We confine ourselves to one linear supermultiplet, only. 
For a discussion of more than one  linear supermultiplet see, \eg\ 
Ref.~\cite{BGG:PLB91}.} 
The chiral multiplet $\mathcal{S}$  has the property
that, due to the left-over Peccei--Quinn-type  symmetry,
it cannot appear in the superpotential, 
and thus, its bosonic component cannot have a  potential.
In the following we shall follow the 
description in terms of chiral superfields, only.

In superstring theory,\footnote{For a review of the  structure 
$N=1$ effective Lagrangian from  superstring theory, see  \eg\ 
Refs.~\cite{KL:NPB95,Quevedo:96}.}
some of the  chiral supermultiplets  do not have any  
self-interaction, \ie\ their superpotential is zero. 
They are referred  to as moduli, since  the 
vacuum expectation values of their scalar components 
parameterize the compactification space of string theory. 
In  superstring theory the dilaton superfield  is the linear supermultiplet 
whose   
vacuum expectation value of its scalar component 
parameterizes the strength of the gauge coupling in  string theory.

\subsection{Bosonic part of the Lagrangian}
\label{BPL}

In the following we shall write down the  
bosonic part of the  $N=1$ supergravity
Lagrangian  which involves the   graviton,  
the Yang--Mills vector field and 
the scalar components of the chiral supermultiplets\@.%
\footnote{For the full  $N=1$ supergravity Lagrangian with  
bosonic as well as femionic fields  see for example  
Ref.~\cite{WB:Princeton92}, chapters  XXI through XXV and appendix G.} 

The $N=1$ supergravity Lagrangian  is of a constrained form,
specified by the  gauge function $f_{ab}$, the
K{\"a}hler potential $K$ and 
the super-potential 
$W$, which are specific functions of 
matter fields, \ie\  scalar components 
$T_j$
of matter chiral-superfields $\mathcal{T}_j$. Some of the 
chiral super-multiplets may consist of moduli. 
We also allow for the existence of a linear 
super-multiplet, which in the
K{\"a}hler superspace formalism can be rewritten as 
a chiral superfield $\mathcal{S}$
(dilaton). 	It has no superpotential and its 
K{\"a}hler potential  is of a special kind,
which decouples
from the one of $\mathcal{T}_j$.%

The three functions specifying the 
$N=1$ supergravity Lagrangian are  specified as: 
\begin{itemize}
\item 
The gauge function $f_{ab}$ is a holomorphic function of  the 
chiral  superfields  $\mathcal{T}_j$ and $\mathcal{S}$. In particular, 
in
the bosonic Lagrangian it determines the 
 gauge couplings to
the field strengths of the  Yang--Mills vector fields 
$A_{\mu}^{(a)}$. At the  tree level of the Lagrangian, the gauge 
function is specified by the (gauge neutral)\ fields $\mathcal{S}$, 
only, \ie\ $f_{ab}=\delta_{ab}\mathcal{S}$, however, 
at the loop-level the chiral superfields $\mathcal{T}_j$ 
can also contribute and thus in general:
\begin{equation}
f_{ab}=f_{ab}(\mathcal{S},\mathcal{T}_j).
\label{gaugefunc}
\end{equation}

\item 
The  superpotential $W$
is a holomorphic function of the 
chiral matter superfields, 
$\mathcal{T}_j$. In particular, in the 
bosonic part of the Lagrangian $W$  
specifies 
the potential for the matter fields. The superfield  
$\mathcal{S}$ has \emph{no} (perturbative)\  
superpotential%
\footnote{However, 
non-perturbative effects, \eg\ gaugino condensation  in certain 
superstring models, can induce 
non-perturbative superpotential for $S$. 
For the status of such effects see \eg\  Ref.~\cite{Quevedo:95}.} 
($W_{\text{dil}}(\mathcal{S})=0$)\ 
thus: 
\begin{equation} 
W=W_{\text{matt}}(\mathcal{T}_j).
\label{superpot}
\end{equation} 
\item
The   K{\"a}hler potential $K$
is a real function of 
chiral superfields, 
$\mathcal{T}_j$, and of $\mathcal{S}$. 
In particular,  in the bosonic part of 
the Lagrangian $K$ specifies the 
(K\"ahler) metric   of the kinetic energy terms  for the scalar 
components of the matter superfields.
At the tree level of the Lagrangian, 
$\mathcal{T}_j$, and $\mathcal{S}$\  do not couple to each other
in  the  K\"ahler potential,\footnote{At the loop level, there could 
be, however, corrections, which induce mixing interactions  either in the  
K\"ahler potential or in the gauge function. 
For such effects within the effective $N=1$ supergravity 
from superstring theory, see \eg\  
Refs.~\cite{FV:PLB87,BGGM:PLB87,OR:PLB91,ABGG:NPB93}. 
We chose to insert the loop effects into the 
gauge function $f_{ab}$ (\ref{gaugefunc}). }
and thus the K\"ahler potential is of the form:
\begin{equation}
\label{Eq:Kahlerpot}
K=K_{\text{dil}}(\mathcal{S}, \mathcal{S}^*)+ 
K_{\text{matt}}(\mathcal{T}_j,\mathcal{T}_j^*).
\tag{\ref{Eq:Kahlerpot}a}
\label{Eq:Kahlerpota}
\end{equation}
where 
\begin{equation}
K_{\text{dil}}(\mathcal{S},\mathcal{S}^*)=
-\alpha\ln(\mathcal{S}+\mathcal{S}^*).
\tag{\ref{Eq:Kahlerpot}b}
\label{Eq:kald}
\end{equation}
\addtocounter{equation}{1}
In $N=1$ supergravity $\alpha\ge 0$ is a free parameter, 
while in string theory
$\alpha=1$.
\end{itemize}
Note that $N=1$ supergravity Lagrangian possesses 
the  K\"ahler invariance associated with the transformation:
\[
K_{\text{matt}}(\mathcal{T}_j,\mathcal{T}_j^*)\rightarrow 
K_{\text{matt}}(\mathcal{T}_j,\mathcal{T}_j^*)+
W_{\text{matt}}(\mathcal{T}_j)+W_{\text{matt}}(\mathcal{T}_j^*)^*
\]

The bosonic part of the Lagrangian is  
fully determined by the three functions 
$f_{ab}$ (\ref{gaugefunc}), 
$W$ (\ref{superpot}) and 
$K$ (\ref{Eq:Kahlerpot}).  When confined to  the  lowest  
derivative terms, it assumes the  following form:
\begin{align}
\mathcal{L} &=  -\frac{1}{2}R 
-\frac{1}{4} \Re (f_{ab})F_{\mu\nu}^{(a)}F^{(b)\mu\nu}  
+\frac{1}{8}  
\Im (f_{ab})\epsilon^{\mu\nu\rho\sigma}F_{\mu\nu}^{(a)} F_{\rho\sigma}^{(b)}
\nonumber
\\  
& \quad \quad {} - K_{T_iT_j^*} \mathcal{D}_{\mu}T_i\,
\mathcal{D}^{\mu}T_j^*-K_{SS^*}\partial_{\mu}S\,
\partial^{\mu}S^*
-V
\label{Eq:SUSYLagrangian}
\end{align}
where
the potential $V$ is
\begin{align}
V& = {\e}^{K}\left[K^{T_i\,T_j^*} D_{T_i}WD_{T^*_j}W^*
-\left(3-K_{S}K_{S^*}K^{S\,S^*}\right)|W|^2\right]
\nonumber
\\
&\quad \quad {} + 
\frac{1}{2}D_{(a)}D^{(a)}, 
\label{susypot}
\end{align}
where $D^{(a)}$ (``D-term'') is the Killing potential related to the 
holomorphic Killing vectors $X^{(a)}$ of the K\"ahler manifold
 in  the following way:
\begin{align*}
K_{T_iT_j^*}X^{j(a)*}&=i{{\partial D^{(a)}}\over {\partial T_i}},
\\
K_{T_iT_j^*}X^{i(a)}&=-i{{\partial D^{(a)}}\over {\partial T_j^*}}.
\end{align*}
The holomorphic Killing vectors  generate 
the  representation of the gauge group, \ie\ 
$[X^{(a)}, X^{(b)}]=-{f_{c}}^{ab} X^{(c)}$ etc.\  
where ${f_c}^{ab}$ are the structure constants of the gauge group. 
The D-term  contribution to the scalar potential is due to the 
scalar  components of the  
chiral supermultiplets  $\mathcal{T}_j$,  which 
transform as gauge non-singlets under the Yang--Mills gauge group.

Above, in Eq.~(\ref{Eq:SUSYLagrangian}), 
the gauge covariant derivative $\mathcal{D}_\mu$ is defined as 
$\mathcal{D}_\mu T_i 
\equiv\partial_\mu -A_\mu^{(a)}X^{i(a)}$, and the gauge field  strength 
as $F_{\mu\nu}^{(a)}\equiv\partial_\mu 
A^{(a)}_\nu-\partial_\nu 
A^{(a)}_\nu -{f^{a}}_{bc}A_\mu^{(b)}A_\nu^{(c)}$. 
The scalar components of the chiral superfields $\mathcal{T}_i$ 
and $\mathcal{S}$, \ie\  $T_i$ and $S$, respectively,  specify 
the following quantities in the bosonic Lagrangian (\ref{Eq:SUSYLagrangian}):
$K_{T_i}=\partial_{T_i} K$
and
$K_{T_iT_j^*}=\partial_{T_i}\partial_{T^*_j} K$ is the
positive definite K{\"a}hler
metric, and  $D_{T_i}W=\partial_{T_i}W+K_{T_i}W$. As usual,  
summation over repeated 
indices is implied. 

\subsection{Bosonic Lagrangian and  topological defects}
\label{Sect:Lagtopdef}
The above bosonic Lagrangian (\ref{Eq:SUSYLagrangian}) is a starting point 
for addressing the topological defect in $N=1$  Yang--Mills supergravity 
theory, in particular, (charged) domain walls,  (charged) strings, 
monopoles and (charged) black holes. In general, the existence of 
$S$ (gauge singlet without  a perturbative potential) 
allows for existence of a new types of topological defects, where in the 
presence of the curved space--time and the non-trivial gauge fields 
$A_\mu^{(a)}$  as well as  
the scalar fields  $T_i$, $S$ vary with the space--time coordinates.

In the following chapters we shall primarily concentrate on the  
gauge-neutral domain wall configurations. 
Namely, we shall assume  the  
potential (\ref{susypot}) has a  non-trivial structure, \eg\  
isolated
minima,  and that 
the   matter field(s)  $T_i$---which are  
gauge singlets, \ie\  with  flat D-terms---are
responsible for the formation of domain  walls.
Supersymmetric minima of the potential correspond to those 
with  $\left. D_{T_i}W\right|_{T_{i}=T_0}=0$. 

A particular case which will be addressed in 
detail is a $N=1$ supergravity theory with one matter
chiral superfield $\mathcal{T}$ and a 
linear super\-mul\-ti\-plet, expressed in  terms of a 
chiral super\-mul\-ti\-plet $\mathcal{S}$ with the
K{\"a}hler potential  (\ref{Eq:Kahlerpot})\ 
\cite{FV:PLB87,BGGM:PLB87,OR:PLB91,ABGG:NPB93}.
With the choice of the scalar component of $\mathcal{S}$ written as
$S= \e^{-2\phi/\sqrt\alpha}+i\mathcal{A}$, 
where $\mathcal{A}$ is the axion, 
the  potential
in Eq.~(\ref{susypot})\ is of the form:
\begin{equation}
V= \e^{2\sqrt\alpha\phi}\e^{K_M}\left[|D_TW|^2K^{T\,T^*}-(3-\alpha)|W|^2\right].
\label{susypota}
\end{equation}
The above potential  shall be a starting point for addressing 
domain wall configurations  $N=1$ supergravity theory.

We shall also compare  the space--time of such  domain walls to that 
of certain charged black holes, which are specified by the space--time 
metric, the gauge fields $A_\mu^{(a)}$ and the field $S$, all 
of them respecting the spherical symmetry.

\subsection{Supersymmetry
transformations} \label{ST}

The  supersymmetry transformations involve the  physical components%
\footnote{%
The  auxiliary  field components  are  again eliminated by their equations of 
motion.} 
of
the   gravitational, gauge and 
chiral superfields.
We shall display the 
relevant supersymmetry transformations  of the fermionic fields, since 
those are the ones that  specify  the Killing spinor equations,  which  in turn 
determine the supersymmetric topological defects.
Namely, setting 
such supersymmetry transformations to  zero in the presence of nontrivial 
bosonic field background defines the non-trivial bosonic  field 
configuration 
 with the minimal energy in its class.  The fermionic field supersymmetry 
transformations  with only bosonic fields turned on%
\footnote{For the full set of supersymmetry transformations for the 
fermionic as well as bosonic fields 
see \eg\  Ref.~\cite[chapter 23]{WB:Princeton92}.}
  is of the following 
form: 
\begin{align}
\delta\psi_\mu=&\left[\ 2\nabla_{\rho}+i {\e}^{K \over 2}
\left( \Re (W) +\gamma^5 \Im (W)\right)\gamma_{\rho}\right. \nonumber\\ 
& \left. {} - \gamma^5 \Im (K_{T_j}\mathcal{D}_{\rho}T_j) -
 \gamma^{5}\Im (K_{S}\mathcal{D}_{\rho}S)\right]\epsilon,
\label{Eq:deltapsi}\\
\delta\lambda^{(a)}=&\left[{ F}_{\mu\nu}^{(a)}\gamma^{\mu\nu}-i 
 D^{(a)}\right]\epsilon,\\ 
\delta\chi_j =&  -\sqrt2\left[ {\e}^{K\over2}K^{T_i  T_j^*}
\left(\Re( D_{T_j}W)+\gamma^5\Im(  D_{T_j}W)\right) \right. \nonumber \\
&   \left. {} 
+ i \left(\Re (\mathcal{D}_{\mu}T_i )
+ \gamma^5(\mathcal{D}_{\mu} T_i) \right)
 \gamma^{\mu} \right]\epsilon
\label{Eq:deltachi}\\
\delta\eta =&  -\sqrt2 \left[ {\e}^{K\over2}K^{S S^*}
\left(\Re( K_{S}W)+\gamma^5\Im(K_{S}W)\right) \right. \nonumber \\ 
&  \left. {}
+i\left(\Re (\mathcal{D}_{\mu}S )
+ \gamma^5(\mathcal{D}_{\mu} S) \right)
 \gamma^{\mu} \right] \epsilon, 
  \label{Eq:deltaeta}
\end{align}
where $\epsilon$ is a  Majorana spinor,
and  $\nabla_{\mu}\epsilon = (\partial_{\mu}
+ {1\over2}{\omega^{ab}}_{\mu}\sigma_{ab})\epsilon$ 
and the Einstein summation convention is implied.

           \addtocounter{equation}{-\value{equation}}
%


\section{Topological defects and tunneling bubbles}
\label{Sect:Topolog}

\subsection{Overview} 

\subsubsection{Topological defects in physics}

Topological defects can be 
studied in the laboratory.
Condensed matter 
systems provide 
a wide range of such 
structures.
{}In the low-temperature regime 
one has magnetic flux lines of 
type II superconductors \cite{Huebener:SV79}, 
quantized vortex-lines in superfluids \cite{Donnelly:CUP91},
and in 
solid state physics one encounters the dislocation and
disclination lines of crystals \cite{Kleinert:WS89b}. 
When viewed under the polarizing microscope, liquid crystals 
exhibit a
variety 
of 
optical textures, each
characteristic of defects
peculiar to the state of molecular order prevailing 
in the substance
\cite{CR:AdvP86}.
But probably the most accessible example is
the domain structure of ferromagnetic materials.
For temperatures $T$ above the
Curie temperature
$T_{\text{C}}$, all 
the dipoles are randomly oriented;
the ground state is rotationally invariant.
For $T<T_{\text{C}}$, we have spontaneous magnetization and the dipoles
are aligned in some arbitrary direction.
The magnetic energy 
is minimized when 
the ferromagnet
splits into domains with different magnetizations.
Domain walls appear 
at the domain boundaries, and must therefore be 
present in the equilibrium state.  
There are also other  examples which have defects formed in non-equilibrium states.

Topological defects are 
related to some form of symmetry breaking which gives rise
to a non-trivial set of degenerate ground states. 
Spontaneous (gauge)
symmetry breaking via the 
 Higgs mechanism 
\cite{Goldstone:INC61,%
EB:PRL64,Higgs:PL64,Higgs:PRL64,GHK:PRL64,Higgs:PR66}
has come to play a central r{\^{o}}le in
modern
elementary particle theory. 
Since such theories 
generally give rise to a degenerate vacuum manifold 
with a non-trivial topology, 
it is plausible that topological defects 
could be important also
in particle physics, in particular in its application to cosmology.

Within field theory, Skyrme \cite{Skyrme:PRSA61} found the first three-dimensional
defect solution, \ie\ the skyrmion,   and proposed
that such defect states could provide a description 
of observed particle states, \ie\ mesons in nuclear physics.
On the other hand, in  high energy physics
it has by now become a prevailing view
that rather than explaining familiar particle
excitations, the defect states---%
as by-products of spontaneous 
symmetry breaking \cite{Nambu:Kyoto65a}---provide 
additional non-perturbative states in the theory, 
which along with the perturbative spectrum  
of  (particle) excitations, complete  the full 
spectrum of the theory. Such defect states
provide diverse and 
supplementary sectors in fundamental theory with interesting dynamics.
Non-perturbative defect states  turn out to be 
particularly important in string theory (for recent reviews see
 \cite{CHS:Trieste91,DKL:PRep95}.).

In addition, it has been recognized 
recently that supersymmetric topological defects,   
play a crucial r\^ole \cite{HT:NPB95,Witten:NPB95} 
in establishing non-perturbative dualities  in string theory, \ie\  
establishing the equivalence of certain  
strongly coupled and   weakly coupled string vacua. 
In addition, the defects provide  important clues for  
uncovering properties of  the M-theory, \ie\ a `mystery'-theory  
of particles and extended objects, formulated in  
dimensions $d \ge 11$ and whose different 
limits  turn out to correspond to 
eleven-dimensional supergravity as well as \emph{all} the known string theories. 

\subsubsection{Cosmological implications of topological defects}
\label{Sect:CosmologicalImplications}

The possible cosmological significance of topological defects 
was anticipated by 
Nambu \cite{Nambu:Kyoto65b} in 1965.
When it was realized in  the mid seventies 
that spontaneously broken symmetries
would be restored at high 
temperatures \cite{Kirzhnits:JETPL72,KL:PL72},
and that topological defects like domain walls could be formed 
in the very early universe \cite{Weinberg:PRD74},
cosmological consequences of topological defects
had to be taken
seriously.  In addition, the relation between the topology of the vacuum manifold
and the different types of defects were pointed out by 
Coleman \cite{Coleman:Erice75} and 
by Kibble \cite{Kibble:JPA76,Kibble:LesHouches80}, thus allowing for 
existence of not only domain walls, but also strings as topological defects. In somewhat parallel developments  
 Polyakov \cite{Polyakov:JETPL74}
and `t~Hooft \cite{`tHooft:NPB74} demonstrated that  monopoles  
 can exist in a  non-Abelian gauge theory, containing 
electromagnetism within a larger compact covering group. 
These developments provided cornerstones for the study of cosmological 
implications of topological defects such as domain walls, strings,
and  monopoles 
within (grand) unified gauge theories.

Everett \cite{Everett:PRD74} 
and
Zel'dovich \ea\ \cite{ZKO:JETP75} were the first to 
quantitatively consider 
potential consequences of cosmic domain wall structures
by
discussing wave propagation across their boundaries
and their gravitational effects, respectively.
Unless the domain walls disappear at a sufficiently early stage,
such structures are incompatible with
the observed level of cosmic isotropy 
\cite{ZKO:JETP75}.
The argument is the following: 
adjoining
walls 
with opposite topological charges will move together and 
annihilate
leaving on the order of one wall per Hubble length $H_0^{-1}$. Since the
cosmic microwave background
is isotropic to an accuracy of about $10^{-5}$, the mass of
this wall $M_{\text{w}}\approx \sigma H_0^{-2}$
where $\sigma$ is the mass per area of the wall,
must be small compared with the net mass within the
Hubble length 
$M_{\text{U}}\approx 1/(G H_0)$. 
If the wall is formed by a scalar multiplet $\phi$ governed by a potential
$V(\phi)= \frac{1}{4} \lambda (\phi^2-\eta^2)^2$ where $\lambda$ is
a dimensionless coupling constant and the constant $\eta$ determines
the ground state
value of $\phi$, then the energy per area is $\sigma \approx \lambda^{1/2} \eta^3$. 
The bound on the anisotropy of the thermal cosmic background then 
translates to
\[
\delta \approx \frac{M_{\text{w}}}{M_{\text{U}}} \approx
\frac{\lambda^{1/2} \eta^3 G}{H_0} 
=
\frac{\lambda^{1/2} \eta^3 \hbar}{M_{\text{Pl}}^2 H_0} 
\lessapprox 10^{-5}
\]
where $M_{\text{Pl}}$ is the Planck mass. This constraint can be rephrased
as 
\[
 \eta \lessapprox \left(M_{\text{Pl}}^2 H_0 10^{-5} \right)^{1/3} \lambda^{-1/6}
\approx \lambda^{-1/6}\ {\text{MeV}}. 
\]
For the range of energies associated with the conventional phase 
transitions of particle physics, the coupling constant 
$\lambda$ would have to
exceedingly small:
for the GUT scale of 
$10^{15}$ GeV  
it must be
least a factor of $10^{-108}$ 
less than unity!
It is very hard to see how such a small coupling could arise naturally 
in the theory. 

In contrast to cosmic domain walls 
which had been ruled out by the an\-iso\-tro\-py argument,
cosmic strings 
could have played an important r{\^o}le 
in the early universe, \eg\ by producing local 
in\-ho\-mo\-ge\-ne\-i\-ties \cite{Kibble:JPA76}. 
This idea was later taken up by
Zel'dovich \cite{Zeldovich:MNRAS80} 
and Vilenkin  
\cite{Vilenkin:PRL81} who argued that these inhomogeneities
could be the cause of galaxy formation.
Subsequently, Silk and Vilenkin \cite{SV:PRL84}
pointed out that 
over-dense planar wakes 
are left behind relativistically moving open strings. 
This effect is a direct consequence of the conical 
geometry: matter converge behind the string from both sides 
and produce an over-dense region.
The resulting power spectrum has been computed both for
string generated
perturbations of cold dark matter \cite{AS:PRL92a} 
and hot dark matter \cite{AS:PRL92b}.
While the string-induced cold dark matter spectrum is 
in trouble for lack of power on large scales,
a combination of strings and hot dark matter
is regarded as a promising alternative \cite{HK:RPP95}
to the 
popular inflationary models.

Magnetic monopoles also provide a
serious problem since the concentration of magnetic monopoles 
left over from a phase transition in the early universe turns
out to be unacceptably large \cite{ZK:PL78,Preskill:PRL79}.    
Monopole--anti-monopole annihilation could be enhanced 
by gravitational clumping but not enough to solve 
the problem \cite{GKT:PRD81}. The only viable way  
to suppress the 
monopoles seemed to be to require that the
universe supercooled before going through 
a strongly first-order phase transition
and that the monopole mass is 
much higher than the reheating temperature 
\cite{Preskill:PRL79,GT:PRL80,EST:PRD80},
but this mechanism would require an unnatural fine tuning of
the GUT parameters \cite{ES:NPB81}.   
On the other hand,
the corresponding vacuum energy density would lead to 
an effective cosmological constant \cite{Linde:JETPL74,BR:PRL77}
which would cause a near exponential expansion of
the universe \cite{Sato:PLB81}.  
In fact, this solves the
monopole problem, which besides the horizon and flatness problems, 
was a key motivation for the cosmic inflation paradigm 
\cite{Guth:PRD81,%
Linde:PLB82,AS:PRL82} (see Ref.~\cite{Steinhardt:IJMPA95} for a
recent review of inflation which makes contact 
with the new observational data).
Because of the enormous expansion during 
the inflationary epoch, the density of primordial monopoles and
other dangerous particles are diluted
so much that only a few of them 
exist within the present Hubble horizon.

\subsection{The kink} 
\label{Sect:Kink}
\begin{figure}[hbt]
\centering{\epsfig{file=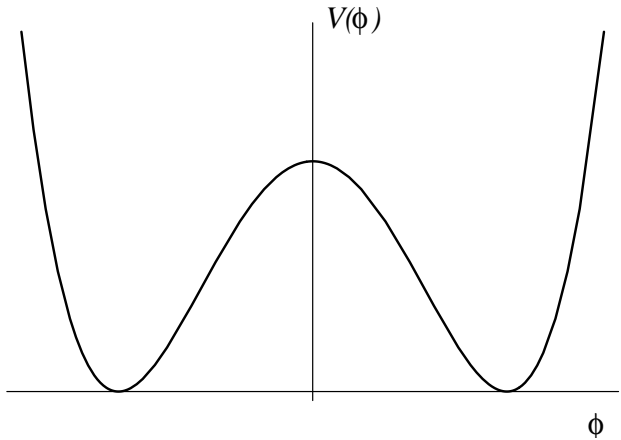,%
        height=6cm,%
        clip=}}

\caption{The double-well potential of Eq.~(\ref{Eq:DoubleWell}).} 
\label{Fig:DoubleWell}
\end{figure}

In order to prepare for  a discussion of domain walls in (super)gravity theory,
we shall first present an introduction to the classical 
finite-energy solutions of field theory: generally called
\emph{solitons}.
Let us start with a simple example of a Goldstone model \cite{Goldstone:INC61}
described by the Lagrangian
\[
{\mathcal{L}} = \frac{1}{2} \partial_\mu \phi \partial^\mu \phi - V(\phi)
\]
where $\phi$ is a scalar multiplet. Let us specialize to a real scalar field
in (1+1)-dimensional space--time and the potential
\begin{equation}
V(\phi)=\frac{\lambda}{4}\left( \phi^2-\eta^2\right)^2;  
\label{Eq:DoubleWell}
\end{equation}
$\lambda$ is a dimensionless coupling constant,
and $\eta$ defines the ground state
of the theory.  
This 
double-well 
potential is illustrated in Fig.~\ref{Fig:DoubleWell}.

\begin{figure}[hbt]
\centering{\epsfig{file=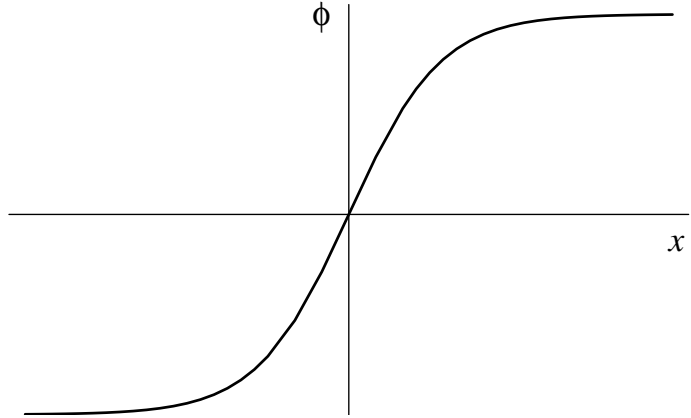,%
        height=6cm,%
        clip=}}

\caption{The kink solution $\phi_{+}$ of Eq.~(\ref{Eq:KinkSolution}).}
\label{Fig:Kink}
\end{figure}

The resulting classical field equation
\[
\Box \phi + \lambda \phi \left( \phi^2-\eta^2\right)=0
\]
has, in addition to the two ground state solutions
$\phi=\pm \eta$, the exact static solutions 
\begin{equation}
\phi_\pm = \pm \eta \tanh ( \sqrt{\lambda/2} \eta x ).
\label{Eq:KinkSolution}
\end{equation}
These are the well-known 
kink $\phi_{+}$ and anti-kink $\phi_{-}$ (for the negative sign)\ solutions. 
The scalar field
interpolates smoothly between the two ground states of the theory. 
As such it is a lower-dimensional 
analogue of a domain wall separating two Minkowski vacua.  
This behaviour is illustrated in Fig.~\ref{Fig:Kink}.

The total energy of the (an{\mbox{ti-)}}kink
is
\begin{displaymath} 
E=\int_{-\infty}^{\infty}\left( \frac{1}{2}{\phi'}^2+V(\phi) \right) dx
= 2 \sqrt{2\lambda} \eta^3/3%
\end{displaymath}%
where a prime stands for a derivative with respect to $x$.
Thus the kink is a classical finite-energy solution of a field theory 
and therefore an example of what we call a soliton: a stable
particle-like state on a non-linear system.

\subsubsection{Topological charge}
At spatial infinities we have
\[
\phi(\infty)-\phi(-\infty) = 2 n \eta
\]
where $n=0$ represents the vacuum, $n=1$ the kink and $n=-1$ the anti-kink.
This equation can be rewritten as
\[
\int_{-\infty}^{\infty} (\phi' ) dx = 2 n \eta.
\]
Let us now define a current by $j_\mu \equiv \epsilon_{\mu\nu} \partial^\nu \phi$
where $\epsilon_{\mu\nu}$ is the L\'evi-Civita tensor. The antisymmetry of
$\epsilon_{\mu\nu}$ and the commutativity of partial
derivatives automatically 
make this current
conserved. The corresponding conserved charge
is
\begin{equation}
C = \int_{-\infty}^{\infty} j_0 dx  
=\int_{-\infty}^{\infty} (\phi' ) dx = 2 n \eta, 
\label{Eq:KinkCharge}
\end{equation}
and $n$ is a conserved quantum number. 
Hence, there is no 
transition 
between kink solutions and ground states, and kinks are
stable.
The conservation of $C$ is different in nature from that
of conservation of, \eg, electrical charge.
The latter is a Noether charge conserved because of 
a symmetry in the theory, whereas
$C$ is conserved independently of the field equations. 

To see how this comes about, let 
${\mathcal{M}}_0$ be a manifold whose elements are
the points in field space which correspond to a ground state,
and let $S$ be the set of points at spatial infinity. 
A finite-energy solution must have asymptotic field
values in ${\mathcal{M}}_0$, \ie\
\[
\lim_{x\rightarrow\pm\infty} \phi(x) = \phi\in {\mathcal{M}}_0.
\]
This condition is a mapping $\Xi: S \rightarrow {\mathcal{M}}_0$. 
In the case of the (1+1)-dimensional $\lambda\phi^4$ theory,
the $\Xi_0$ of the vacuum state maps the whole of $S$ either to 
$-\eta$ or to $\eta$. 
The mapping $\Xi_{+}$ corresponding to a kink 
$\phi_{+}$ maps $-\infty$ to $-\eta$ and
$\infty$ to $\eta$, 
and
the anti-kink 
mapping $\Xi_{-}$ maps $-\infty$ to $\eta$ and
$\infty$ to $-\eta$.  
It is impossible to continuously deform these
mappings into one another, \ie\ they are topologically distinct. 
For this reason, a conserved charge such as $C$ in 
Eq.~(\ref{Eq:KinkCharge})\ 
is called a \emph{topological charge}, and the conservation law
a
\emph{topological conservation law}. 

This simple example has shown how a topological conservation law 
can divide the set of finite-energy solutions into 
several different sectors: $n=0$ (the vacuum);
$n=1$ (the kink); $n=-1$ (the anti-kink); etc.  
Their existence has far-reaching implications both at the
microscopic and cosmic scales.
Clearly, the distinct
topological sectors were 
made possible by the topology of the vacuum manifold
${\mathcal{M}}_0$: With the potential  
of Eq.~(\ref{Eq:DoubleWell})\ the vacuum manifold has disconnected
components.

\subsubsection{Higher-dimensional defects}
The simple (1+1)-dimensional 
kink has no direct analogue in higher dimensions.  There is
a
theorem due to Derrick \cite{Derrick:JMP64}
which states that in a scalar theory in with two spatial dimensions or more,
the only non-singular time-independent solutions of finite
energy are the ground states.

One can evade the dimensional restriction of Derrick's theorem 
by allowing time-dependence or by adding other fields, \eg\
gauge fields.  
The 't~Hooft--Polyakov monopole \cite{`tHooft:NPB74,Polyakov:JETPL74}
and the Nielsen--Olesen vortex-line \cite{NO:NPB73}
are examples of the latter type.
Time-dependent (non-topological or semi-topological)\ 
solitons have also been 
discovered 
\cite{Rosen:JMP68a,Rosen:JMP68b,%
LW:PRD74,FLS:PRD76,Coleman:NPB85}
(for reviews of 
boson stars and 
non-topological solitons,  
see
Refs.~\cite{LP:PRep92,Jetzer:PRep92}).

\subsection{Homotopy groups and defect classification}
Topological defects can be classified according to the
topology of the vacuum manifold.
Central to this classification scheme is the concept of
\emph{homotopy classes} and \emph{homotopy groups}.
It is defined as follows:
Let ${\mathcal{X}}$ and ${\mathcal{Y}}$ be two topological
spaces,  
and let $\Xi_0(x)$
and $\Xi_1 (x)$ 
be two continuous mappings from ${\mathcal{X}}$ to ${\mathcal{Y}}$. 
If $I$ is the unit 
interval $I=[0,\ 1]\subset \mathbb{R}$,
then the two functions  
$\Xi_0(x)$
and $\Xi_1 (x)$ 
are said to be homotopic if and only if
there exists
a continuous mapping $F: {\mathcal{X}}\otimes I\rightarrow {\mathcal{Y}}$
such that $F(x,0)=\Xi_0(x)$ and $F(x,1)=\Xi_1(x)$. The
function $F$ is called the
\emph{homotopy}. In other words,
two mappings are homotopic 
if there exists a continuous function which deforms one into the other.
In this case the two functions are said to be members of the same
\emph{homotopy class} $[\Xi_0]=[\Xi_1]$.

Consider now mappings $\Xi$ from the $n$-dimensional sphere $S^n$ 
to a vacuum manifold ${\mathcal{M}}_0$. 
Let us first consider $n=1$. Then each $\Xi$ represents 
a closed loop in ${\mathcal{M}}_0$ and each loop
can be placed in homotopy class $[\Xi]$. One can now
define a class multiplication by
\[
[\Xi_1] [\Xi_2] = [\Xi_1 \circ \Xi_2] 
\]
where loop multiplication is defined as
first going through one loop and then the other:
\[
\Xi_1 \circ \Xi_2 (t) = \left\{ 
\begin{array}{ll} 
\Xi_1 (2 t),           & t\in [0,\ 1/2] \\
\Xi_2 (2 t-1),\ \      & t\in [1/2,\ 1]. 
\end{array}
\right.
\]
The set of homotopy classes is a group under this
operation; the identity element consist of all loops
which can be contracted to a point and thus to the constant map $x$,
and the inverse is defined as $[\Xi]^{-1}\equiv [\Xi^{-1}]$. 
Since 
the base point $x$ 
on a connected space 
is movable through an isomorphism, we can omit it and denote 
this group $\pi_1 ({\mathcal{M}})$. It is called the 
first homotopy group 
of the manifold.  Similar groups can be defined for 
all
$n$-dimensional spheres that can be mapped onto the manifold.
We have seen that $\pi_1({\mathcal{M}})$ is non-trivial if there
are loops in the vacuum manifold which cannot be deformed
into a point;
$\pi_2({\mathcal{M}})$ is non-trivial if there are  
unshrinkable surfaces, and so on. 
The case $n=0$ corresponds to 
mappings of a point onto the manifold, and 
$\pi_0({\mathcal{M}})$ 
is non-trivial if and only if the manifold has disconnected components
as for example in the kink model discussed in Section~\ref{Sect:Kink}. 

The topological defects are therefore
classified according to the homotopy groups and are
shown in Table~\ref{Table:HomotopyClass}.

\begin{table}[htb] 
\caption{Classification of topological defects in three space dimensions.}
\label{Table:HomotopyClass}

{\vspace{1em}}

\centering{%
\begin{tabular}{l c c} \hline
Defect type  &\multicolumn{1}{l}{Dimension}&\multicolumn{1}{l}{Homotopy group}\\ 
\hline  
Domain walls & 2                           & $\pi_0({\mathcal{M}})$ \\
Strings      & 1                           & $\pi_1({\mathcal{M}})$ \\
Monopoles    & 0                           & $\pi_2({\mathcal{M}})$ \\
Textures     & --                          & $\pi_3({\mathcal{M}})$ \\ \hline 
\end{tabular}}
\end{table}

In addition, there are hybrid defects 
such as monopoles connected by strings \cite{LSW:NPB82,Vilenkin:NPB82}
or walls bounded by strings \cite{VE:PRL82,KLS:PRD82} (see also 
Refs.~\cite{Vilenkin:PRep85,VS:CUP94}). 

\subsection{Formation of topological defects}
\label{Sect:Formation}
Besides identifying and explicitly constructing 
defect solutions,
one would like to understand whether or not such structures could
form in the early Universe. To this end one needs theories
or scenarios for
defect formation. In this section we comment on
such scenarios.  

\subsubsection{The Kibble mechanism}
In general spontaneously broken symmetries are 
restored at high temperatures. According to the standard
model of the Universe, it began in a very hot state,
perhaps with temperatures close to the 
Planck temperature. 
It is believed that the 
effective potential of the Higgs field(s)
was different at such high temperatures and that
the ground state then was symmetric with respect to
the symmetries 
which we today see as broken symmetries 
\cite{Kirzhnits:JETPL72,KL:PL72,Weinberg:PRD74}. 

As the Universe
expanded and cooled the effective potential changed
and ended up with degenerate ground states
with broken symmetries. In different
regions of the Universe, the Higgs field(s) could settle down 
at different parts of the vacuum manifold, and then
depending on its topology (as discussed in the previous
sections) various types of topological defects can form.
This phenomenon is the Kibble mechanism \cite{Kibble:JPA76}
whose realization
depends only 
on the topology of the vacuum manifold and 
the assumption that the Universe went through a phase transition.

\subsubsection{Quantum creation} 
Quantum fields in curved space--time 
\cite{BD:CUP82}
can induce negative energy--mo\-men\-tum densities in the vacuum.
This allows particle creation even with 
a conserved stress energy $\langle T_{\mu\nu}\rangle$. 
Such
particle creation 
also takes place in 
the de~Sitter space--time 
\cite{GH:PRD77}, not only for particles, but also
for axionic domain walls \cite{LL:PLB90}
and other topological defects \cite{BGV:PRD91}.
As a result, topological defects 
with an energy scale not too far above the 
energy scale of inflation, can still
be present after inflation.
This could also happen if the defect field
$\phi$ is coupled to the inflation field 
responsible for inflation \cite{VOS:NPB87}.

Within the framework of quantum cosmology---where
the creation of the whole Universe is
envisaged  
as a quantum tunneling process 
from ``nothing'' 
\cite{Vilenkin:PLB82,Vilenkin:PRD83,Vilenkin:PRD84,GH:PRD90}%
---the Universe 
could 
in principle
be born with defects.
However, as recently pointed out by Gibbons \cite{Gibbons:96},
since their action is infinite,
universes with domains of negative cosmological constant
separated by supersymmetric domain walls can not be created
in this way. 

\subsubsection{False vacuum decay}

Another way of forming domain walls is through
false vacuum decay 
\cite{ZKO:JETP75,%
Coleman:Erice77}\@.\footnote{Early papers on vacuum decay include 
Refs.~\cite{LW:PRD74,%
VKO:SJNP75,%
Frampton:PRL76,%
Stone:PRD76,%
Stone:PLB77,%
Coleman:PRD77,CC:PRD77,Linde:PLB77,Linde:PLB80}.} 
In this case the scalar fields have  
non-degenerate minima, and there is a finite
probability of quantum tunneling from a false vacuum (a local minimum)\
into
the true vacuum (the global minimum):
Quantum fluctuations may spontaneously generate a ``bubble''
of the true vacuum\@.\footnote{Gravity makes the concept of
degenerate vacua somewhat more complicated. Decay out of a 
zero-energy minimum is suppressed by gravity \cite{CDeL:PRD80}
and two vacua with 
zero and negative energy-densities may therefore become degenerate
due to gravitational effects.}
If this bubble  has a radius larger than
a certain critical size, the bubble will expand rapidly and convert
the false vacuum into a true one. 
Such a true vacuum bubble  
is separated from the false vacuum by a domain wall. 
These tunneling bubble
walls
are \emph{not} topological defects, 
yet still of fundamental importance in a complete picture
addressing domain walls. 

\subsection{Domain walls as a particular type of topological defect}
 
Domain walls provide a special example of topological defects.
Therefore the review of the nature of topological defects, their 
cosmological implications  and  their dynamical formation, 
as spelled out in the above subsections,  applies to 
the domain walls as a special case. 

In particular,  domain walls are topological defects that can occur in the fundamental theory when  the vacuum manifold ${\mathcal{M}}_0$ has 
disconnected components, \ie\  
the homotopy group $\pi_0({\mathcal{M}})$ 
is non-trivial. An example  of  such a mapping  of a point onto the vacuum manifold ${\mathcal{M}}_0$ is  the kink model discussed in Section~\ref{Sect:Kink}. 

Dynamic formation of domain walls in 
the early universe may proceed, as discussed in 
Section~\ref{Sect:Formation}, as:
\begin{itemize}
\item
 a particular case of the Kibble mechanism,
\item
within the framework of quantum cosmology the domain walls could separate
different
universes, 
\item
domain walls would provide boundaries  of the 
false vacuum decay bubbles. 
\end{itemize}

Because of their extended structure,
the  dynamic effects of domain walls, including the gravitational ones, 
are most drastic.  Here we briefly summarize 
some of their  dynamical
effects. 

As discussed in  Section~\ref{Sect:CosmologicalImplications} domain 
walls that may form via Kibble mechanism, are incompatible with
the observed level of cosmic isotropy 
\cite{ZKO:JETP75},  unless  they  disappear at a 
sufficiently early stage. In the case of 
inflationary universe, the  domain walls  
can effective
disappear   by being inflated away \cite{Linde:PLB82,AS:PRL82},  
if the energy scale of inflation is below the 
scale of domain wall formation.  
Another 
mechanism  of getting rid of domain walls may  take place, 
if the walls are  bounded by strings. In this 
case, intersecting walls
bounded by strings 
cut holes in 
each other, and are also
chopped up by intersecting strings \cite{VE:PRL82}. 
In this way such walls 
may disappear 
before they dominate
the gravitational field of the universe. 

The vacuum decay bubbles 
are distinguished from 
those originating as results of phase transitions
by their smaller surface energy density. 
For vacuum decay bubbles, the liberated  energy 
of the false vacuum is 
converted into kinetic energy of the wall.
It could finally
be released as thermal energy after 
bubble collisions\@.\footnote{ 
The possibility that 
the vacuum energy is 
directly released as thermal energy of 
the medium inside the bubble, a process dubbed as
``vacuum burning'', 
was  
investigated in Refs.~\cite{BKT:PLB83b,BKT:JETP84}.} 
However, in the old inflationary model 
the phase transition could not be 
smoothly completed 
\cite{Guth:PRD81,GW:PRD81,GW:NPB83}.  
This is 
sometimes called  the ``graceful exit problem''. Moreover,  
the bubble collisions 
could not lead to sufficient 
thermalization of the wall energy \cite{HMS:PRD82,GW:NPB83}.
In 
Ref.~\cite{HMS:PRD82} it was concluded that 
bubble collisions could lead to formation of black holes, but
a more detailed analysis of 
the gravitational effects 
showed that black hole formation is impossible
in two-bubble collisions \cite{Chao:PRD83}. 
However, since the domain wall sets up a repulsive gravitational field,
one expects black hole pair creation to occur \cite{CCG:96}.

The graceful exit and thermalization problems 
were solved by the new inflationary scenario 
\cite{Linde:PLB82,AS:PRL82} 
and the chaotic inflationary model
\cite{Linde:PLB83}
in which inflation takes place during a slow-rollover 
transition. Here the observable universe comes from one inflationary bubble and
thermalization is due to dissipation of rapid scalar field oscillations
rather than bubble collisions. In these inflationary
models the thin wall approximation is
inapplicable. However, in the extended inflationary model \cite{LS:PRL89}
where gravity is described by an effective Brans--Dicke theory, 
the phase transition is first-order, and the bubbles
can again be treated in the
thin wall approximation.

The focus of this  Review  are the
gravitational aspects of domain 
walls\@.\footnote{For more general information about 
topological defects and their cosmological implications
we refer the readers to Ref.~\cite{VS:CUP94};
for a recent review on cosmic strings, see Ref.~\cite{HK:RPP95};
and
for a review on solitons in 
super\-string theory, see Ref.~\cite{DKL:PRep95}.
Earlier work on domain walls in global supersymmetry is found in
Refs.~\cite{AT:NPB91,CQR:PRL91}.} 
We shall concentrate on the 
su\-per\-sym\-me\-tric walls 
and their non-su\-per\-sym\-me\-tric generalizations within $N=1$ 
su\-per\-gra\-vi\-ty theory and effective  $N=1$ 
su\-per\-gra\-vi\-ty from su\-per\-strings, as a particular example.
For that purpose,  we shall review
the space--time symmetry assumptions,
the metric ansatz, 
and
the thin wall formalism  in Section~\ref{Sect:IsoWalls}.
In Section~\ref{Sect:SUSY_Embedding} the embedding  of such  walls in
in $N=1$ supergravity theory is
 given, with  vacuum domain walls  and dilatonic domain walls  studied 
Sections~\ref{Sect:VacuumWalls}  
and  \ref{Sect:DilatonicWalls}, respectively.
Connection of these walls 
to other topological defects  
as well as  implications for the domain walls within
 effective supergravity vacua from superstring theory are discussed 
in Section~\ref{Sect:ConnectionOTD}.

           \addtocounter{equation}{-\value{equation}}
%

%
           \section{Isotropic domain walls}
\label{Sect:IsoWalls}

A domain wall is a surface with a timelike
velocity vector.
As the wall propagates through space--time it
defines a set of events that
describe the wall's history. In analogy with a particle's
\emph{world line} 
and a string's \emph{world sheet}
or \emph{world tube} (for closed strings), 
the wall's history shall be referred to as
its \emph{world volume}.
A priori, the world volume may be spatially 
closed, semi-closed,  
or open depending on the 
topology of the wall.

\subsection{Domain wall symmetries} 
\label{Sec:DomainWallSymmetries}

An ideal domain wall is an
infinitely thin surface in which 
the tension in all directions 
equals the 
energy
density.  
The symmetries of the wall's
energy--momentum tensor
imply that it 
is boost invariant along its spatial directions. 
This property
is characteristic of a vacuum \cite{Gliner:JETP66}; 
an unpolarized vacuum has no
preferred frame 
\cite[page~367]{LPPT:Princeton75}. Accordingly, a domain wall is 
said to be vacuum-like. 
One consequence of this equation of state is that the energy-density
is \emph{constant}, even if the wall expands.  
Covariant energy--momentum conservation 
dictates that new vacuum energy is created 
in proportion to the change of volume.
This increase of energy in a fixed coordinate-volume
(as opposed to a proper volume)\
is equal to the positive work done by the
negative-pressure force on the surroundings, 
\ie\
$dU=-pdV$.
Exactly the same happens in the de~Sitter phase of
inflationary cosmologies. 
Another consequence is that the energy--momentum tensor  
of a domain wall must be isotropic and homogeneous.

\subsection{Induced space--time symmetries} 

Space--time need not have the same symmetries
as the energy--momentum tensor. 
The Kasner cosmology \cite{Kasner:AJM21}
is a very simple example of a symmetric 
energy--momentum tensor
in an an\-iso\-tro\-pic space--time; here the 
en\-er\-gy--mo\-men\-tum ten\-sor vanishes identically,
but the gravitational field is an\-i\-so\-tro\-pic.

Kasner-like solutions for domain walls are also 
known \cite{Tomita:PLB85,JS:96}, but
here we shall 
focus on the basic structure of 
domain wall field
configurations
without any additional complications coming from anisotropic
gravitational background fields.
Consequently,
we
\emph{assume} that
the gravitational field has the same high degree of symmetry as
the source.

\subsubsection{Metric ansatz}

It is
most convenient to describe the wall system
in the comoving coordinates,
\ie\ in the wall's rest frame.  
In this reference frame the
wall system is (locally) 
static, and its stress energy depends only on the
spatial distance from the wall surface.

First, we
assume 
that the spatial part of the metric
of the wall's world volume 
and of the two-dimensional spatial sections
``parallel'' to the wall
are \emph{homogeneous} and \emph{isotropic} in the 
\emph{comoving frame}. 
Homogeneity and isotropy reduce the
``parallel'' metric to the spatial part of a (2+1)-dimensional
\cite{GAK:GRG84}
Friedmann--Lema{\^{\i}}tre--Robertson--Walker (FLRW)
metric
\cite[page~412]{Weinberg:JWS72}.  
In the conventional
coordinates the line element has the form
\[
(ds_{\parallel})^2=R^2\left[(1-kr^2)^{-1}dr^2+r^2 d\phi^2\right] ,
\]
where the scale factor $R$ is constant on the surface.
This is a surface of constant curvature with  Ricci scalar 
equal to $2k/R^2$.

Since any non-zero $k$ can be absorbed into
the scale factor $R$, only the sign of $k$ is important.
We can therefore normalize $k$ to $0$, $+1$, or $-1$. 
This gives 
three possible wall geometries.  
If
$k=0$, 
the
metric $(ds_\parallel)^2$ can be transformed
to Cartesian coordinates
$(ds_{\parallel})^2 = R^{2} (dx^2+dy^2)$,
and the wall is planar. 
If $k=1$, then  the wall is
a bubble in which case
one may introduce $r=\sin\theta$
which 
gives the line-element
$(ds_{\parallel})^2=R^2 ( d\theta^2+\sin^2{\!\theta} d\phi^2)$.
If $k=-1$, 
then
the coordinate transformation
$r = \sinh\varrho$,
with $\varrho\geq 0 $, 
brings the line element to
$(ds_{\parallel})^{2} = R^2 (d\varrho^{2} +  \sinh^2\!\varrho \,d\phi^{2})$,
which 
corresponds to
a Gauss--B{\'{o}}lyai--Lobachevski
surface.  This non-compact
surface cannot be embedded in ordinary
3-dimensional Euclidean space 
\cite[page~5]{Weinberg:JWS72}.

\begin{figure}[hbt]
\centering{\epsfig{file=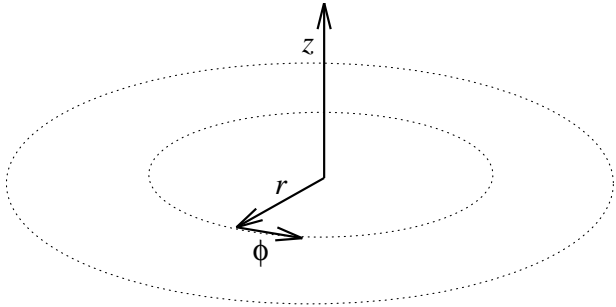,%
        height=4cm,%
        clip=}}

\caption{The spatial coordinates in the wall space--time; $z$ describes the
direction orthogonal to the wall surface.}
\label{Fig:Coord}
\end{figure}

Let us also assume that the two-dimensional space--time
sections ortho\-gonal to the wall
are \emph{static}
as observed in the  rest frame of the wall.
Hence, if $z$ denotes a coordinate describing the direction
transverse to the wall, and if $t$ represents the \emph{proper}
time as measured by observers at rest on the wall surface,
then
$g_{tz}=0$, and both $g_{tt}$ and $g_{zz}$ depend only on $z$.
By an appropriate rescaling of the $z$-coordinate
the 
orthogonal part of the metric can be written as%
\[
(ds_{\perp})^2=\e^{2 a(z)}\left(dt^2-dz^2\right). 
\]
With $ds^2\equiv (ds_{\perp})^2-(ds_{\parallel})^2$, we get
\[
ds^2=\e^{2 a(z)}\left(dt^2-dz^2\right)-R^2\left[(1-kr^2)^{-1}dr^2+r^2d\phi^2
\right] ,
\]
where $R=R(t,z)$. Without loss of generality, we 
can choose the origin of $z$ at the centre of the wall surface.
The range of $z$ is $z\in\langle -\infty,
\infty\rangle$, and the range of the other coordinates is
that of a FLRW cosmological model \cite[page~412]{Weinberg:JWS72}.
In Ref.~\cite{CGS:PRD93}
it was shown that Einstein's equations together with 
the requirement that each surface of constant $z$ has a
boost invariant extrinsic curvature, imply that 
$R(z,t)$ is separable. If we also demand that the
world volume has a non-singular and geodesically complete
metric, then the space--time metric 
can be written in the form \cite{CGS:PRD93} 
\begin{equation}
\begin{array}{l}
ds^2 = \e^{2 a(z)}\left\{ dt^2-dz^2-S^2(t)
\left[(1-kr^2)^{-1}dr^2+r^2 d\phi^2\right]\right\} \\ 
S(t)  = \left\{
   \begin{array}{lcl}
    1           & & k=0 \\ 
    \cosh \beta t &  &k=\beta^2
    \end{array}
    \right.
\end{array}
\label{Eq:Ansatz}  
\end{equation}
where $\beta$ is a positive real constant.
The functional form of $S(t)$ is determined by the boost-invariance symmetry;
the world volume is either a (2+1)-dimensional Minkowksi space--time or
a (2+1)-dimensional de~Sitter space. 
These two possibilities are characterized by different 
topologies; 
$\mathbb R^{3}$ for the Minkowski case and $\mathbb R \times \mathbb S^2$
for the de~Sitter case 
\cite{IS:PRD84,Ipser:PRD84,Gibbons:NPB93,CGS:PRL93}.
One could also think of 
anti--de~Sitter space with $S(t)=\sin\beta t$ as a third possibility, 
but this  
option is excluded since this metric would lead to a 
space--time 
with 
singularities
periodic 
in time.
Hence with the requirement that the world volume 
is non-singular, the wall must be either a static plane or
an accelerated bubble. 
Note that by relaxing some of our symmetry assumptions, one can
get more general solutions such as for example walls in Schwarzschild,
Schwarzschild--de~Sitter
\cite{BGG:PRD87,BKT:PRD87},  
or Reissner--Nordstr{\"o}m backgrounds. 
In this paper we shall, however, restrict ourselves to
domain wall space--times \emph{without} charges or
Schwarzschild masses.

\begin{figure}[hbt]

\centering{\epsfig{file=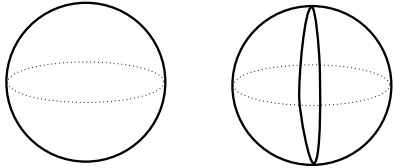,%
        height=2.2cm,%
        clip=}}

\caption{Two-dimensional analogue of the spatial section of the 
geodesically complete Vilenkin domain wall (domain wall
between Minkowski vacua) space--time. 
The whole spatial geometry is the closed surface (to the left),
and the domain string (the analogue of a
domain wall in three dimensions)\ 
is the dividing circle depicted on the right.
Note that angular circles decrease on both sides of the string:
both sides are inside the circular domain string.}  
\label{Fig:DomainString}
\end{figure}

Very often the line-element of Eq.~(\ref{Eq:Ansatz}) is written on the 
plane-symmetric \cite{Taub:AM51}
de~Sitter form with $S(t)=\e^{H t}$ 
and $k=0$
(see \eg,  
Refs.~\cite{Vilenkin:PLB83,Goetz:JMP90,%
Mukherjee:CQG93} and
\cite[page~282]{Peebles:Princeton93}).
It should be observed that the  
spatially-flat 
line-element is 
geodesically incomplete, and 
even though it is 
locally plane-symmetric,
it describes only a \emph{part} of a spherical bubble, 
cf.\
Refs.~\cite{IS:PRD84,Ipser:PRD84,Gibbons:NPB93,CGS:PRL93,KIF:PRD94}
and
\cite[page~380]{VS:CUP94}.
Nevertheless, recent papers 
\cite{WL:PRD95,PW:PRD95}  
still reveal some confusion about the 
physical interpretation of these solutions:    
in support of the view that the wall really
is planar, 
Wang and Letelier \cite{WL:PRD95} argue that 
the wall divides space in two and that none of these parts are
on the outside 
of a bubble.
This argument makes an unexpressed Euclidean 
assumption about the
spatial topology which imply that only planar walls can 
divide space in two equal parts. 
However, if space is closed, such a division can be made with
a spherical domain wall (see Fig.~\ref{Fig:DomainString}).  

While it is true that the high symmetry of de~Sitter space--time
allows for different time slicings, these 
are only different representations of the same
geometry. In other words,
using a coordinate system covering only part of
space--time does not change its physical nature.

\subsection{The thin wall formalism}
\label{Sect:ThinWallFormalism}

Far from a domain wall one expects the 
gravitational field to
be independent of the internal 
structure of the wall. In particular, the
large-scale structure of 
space--time or its 
topology  
should \emph{not} depend 
on 
such details. 
This expectation has been confirmed by studies of ``thick''
domain walls \cite{Widrow:PRD89b,Goetz:JMP90} 
where the metric 
approaches that of a
thin wall
at infinity. 
Therefore, in order to understand the global 
properties of domain wall space--times, it is sufficient to
study domain wall solutions in the thin wall approximation.
In this approximation the wall is regarded as infinitely thin,
with a $\delta$-function singularity in the 
energy-momentum tensor 
and the Einstein tensor.
Lichnerowicz's formalism 
\cite{Lichnerowicz:CR71,Taub:JMP80}
deals with distribution valued curvature tensors 
and can be applied to thin domain walls \cite{LW:JMP95}. 
However, the most used description 
of domain walls
is 
the Gauss--Codazzi formalism \cite[page~514]{MTW:Freeman73}. 
It
is a method of viewing a 
four-dimensional space--time as being sliced
up into three-dimensional hypersurfaces.
The proper junction conditions for such surfaces of discontinuity
were worked out by Israel \cite{Israel:INC66}. Later Israel's
formalism has been generalized also to include
lightlike surfaces \cite{CD:CQG87,BI:PRD91}  
and non-vacuum space--times \cite{BKT:PLB83c,Lake:PRD84,Maeda:GRG86}.
The $\delta$-function singularities of the singular layers 
correspond to  
step-function discontinuities  
in
the first-order derivative of the metric 
coefficients. 
The hight of 
the steps 
determines the components of the energy--momentum tensor
of the singular surface layer.
More precisely, 
in Israel's formalism
the 
energy--momentum 
of the surface 
is described by 
Lanczos' 
tensor
${S^i}_j$. This tensor is 
given by \cite{Israel:INC66}
\[
\kappa {S^i}_j\equiv -[{K^i}_j]^{-}+{\delta^i}_j [K]^{-},
\]
where 
$K_{ij}$ is the extrinsic curvature of the surface  
and
where $[\Omega]^{-}\equiv \Omega^{+}-\Omega^{-}$ 
signify the discontinuity at the wall. 
The indices $\{i\}$
label the 
components relative to a basis intrinsic to the world volume.
The extrinsic curvature 
describes how the normal vector in the embedding space
changes along the world volume. It
is given by the covariant derivative of
the spacelike 
unit normal vector. In the comoving frame 
of the metric ansatz (\ref{Eq:Ansatz}), the 
extrinsic curvature is
\[
K_{ij} = -\frac{\zeta}{2} e^{-a(z)} g_{ij,z}, 
\] 
where $\zeta$ is a sign factor determined by the 
orientation of the outer normal. 

Written in the perfect fluid form, the surface energy--momentum tensor  
takes the form
\[
\kappa {S^i}_j
= \sigma u^i u_j-\tau ( {\delta^i}_j + u^i u_j ).
\]
Here $u^i$ is the velocity vector of the surface, $\sigma$ is 
its rest mass density (mass per unit area), and 
$\tau$ is the wall's tension measured in the comoving frame. 
For a domain wall the tension equals the energy-density:
$\tau=\sigma$.

With the metric (\ref{Eq:Ansatz})\ the 
energy-density of a domain wall placed at $z=0$ becomes
\begin{equation}
\sigma = 2\zeta_1 \left. a' \right|_{0^{-}} 
       - 2\zeta_2 \left. a' \right|_{0^{+}}
\label{Eq:WallEnergy}
\end{equation}
where the $z$ coordinate has been oriented so that the vacuum of lowest
energy will be placed on the $z<0$ side and where (without loss of
generality)\ the coordinates
have been
normalized so that $a(0)=0$.
The sign factors $\zeta_k=\pm 1$ with $k\in \{1,2\}$
where the signs
are determined by directing the outward normal in the
direction of the matter source gradient.
We restrict ourselves to the case where the underlying
matter
source is a  
kink-like scalar field configuration  and then
$\zeta_1=\zeta_2=1$ \cite{CGS:PRD93}.

Following the above 
approach for the metric tensor field,
one
can derive analogous junction conditions for other fields.
For a scalar field the result is \cite{CS:PRD95}
\begin{equation} 
\left. \phi'\right|_{0^{+}} 
-\left. \phi'\right|_{0^{-}} 
=
\int^{0^{+}}_{0^{-}} e^{2 a} V' dz,  
\label{Eq:WallScalar} 
\end{equation}
where $V(\phi)$ is the effective potential for the scalar field.

\subsubsection{Lagrangian and field equations}

The starting point for the study of the supergravity domain walls is the
bosonic part of the action, $S\equiv \int{\mathcal{L}} \sqrt{-g} d^4x$,
for the space--time metric $g_{\mu\nu}$,
the matter field $\vartheta$ responsible for the formation of the wall
and the dilaton field $\phi$, which couples to the matter potential. In the 
Einstein frame the Lagrangian for the Einstein-dilaton-matter system is
\begin{equation}
{\mathcal{L}}=
-\frac{1}{2}R 
+ \partial_\mu \vartheta\partial^\mu \vartheta 
+ \partial_\mu \phi \partial^\mu \phi 
-V(\phi,\vartheta).
\end{equation}
The potential is of the form: 
\begin{align}
\label{Eq:Potential}
V(\phi,\vartheta)&=f(\phi) V_0 (\vartheta) +\widehat{V}(\phi).
\tag{\ref{Eq:Potential}a}
\label{Eq:Potentiala}
\displaybreak[3]\\ 
\intertext{Comparing with Eq.~(\ref{susypota}),
the functions in
the potential (\ref{Eq:Potentiala})\ are given 
in terms of the fields of $N=1$ supergravity:} 
f(\phi) &= \e^{2 \sqrt{\alpha}\phi},
\tag{\ref{Eq:Potential}b}
\displaybreak[3]\\ 
V_0(T,T^*) &= \e^{K_M} \left[ |D_T W|^2 K^{TT^*}- (3-\alpha)|W|^2 \right]. 
\tag{\ref{Eq:Potential}c} 
\end{align}  
\addtocounter{equation}{1}
The dilaton couples to the matter potential with the
function $f(\phi)$ which we shall call ``the dilaton
coupling''. For the sake of generality we include a
dilaton self-interaction term $\widehat V({\phi})$.  It is not 
present
in the original theory, but is believed to be
generated after dynamical 
supersymmetry breaking. This term is responsible for giving the
dilaton a mass.
In the thin wall approximation the matter field $\vartheta$ 
is frozen
outside the wall and its only contribution to the field equations are
constant potential terms $V_0 (\vartheta_1)$ and $V_0(\vartheta_2)$. 

The Einstein tensor for the metric (\ref{Eq:Ansatz})\ is
\begin{align}
\label{Eq:Einstein}
{G^z}_z &= 3 \left( \beta^2 - {a'}^2 \right) e^{-2 a},
\tag{\ref{Eq:Einstein}a}\\ 
{G^i}_i &= \left( \beta^2 - {a'}^2 -2 a''\right) e^{-2 a}. 
\tag{\ref{Eq:Einstein}b} 
\end{align}\addtocounter{equation}{1}%
The corresponding energy-momentum tensor is
\begin{align}
\label{Eq:EnergyMomentum}
{T^z}_z &= V(\phi) - {\phi'}^2 e^{-2 a}, 
\tag{\ref{Eq:EnergyMomentum}a}\\ 
{T^i}_i &= V(\phi) + {\phi'}^2 e^{-2 a},  
\tag{\ref{Eq:EnergyMomentum}b}\\ 
\end{align}\addtocounter{equation}{1}%
where $V(\phi)$ 
as defined in Eqs.~(\ref{Eq:Potential})\  
is the effective ``dilaton potential''
on either side of the wall. 
Einstein's field equations $G_{ij}= T_{ij}$
and the second Bianchi identity then lead to
\begin{eqnarray}
& e^{2a} V(\phi) + 2  {\phi'}^2 + 3 a'' =0,
\label{Eq:DynamicalEq}\\
& -e^{2a} V'(\phi) + 4  a' \phi' + 2 \phi'' =0,
\label{Eq:ConservationLaw}\\
& 3 \beta^2 - e^{2a} V(\phi) -3 {a'}^2  +  {\phi'}^2  =0.
\label{Eq:Constraint}
\end{eqnarray}
Note that the equation of motion for the scalar field
is identical to the energy--momentum conservation law 
(\ref{Eq:ConservationLaw}) and consequently there
are only two independent field equations. The constraint 
(\ref{Eq:Constraint})\
can be used to determine the boundary conditions.

\subsubsection{Junction conditions as boundary conditions}

In the case when the solution for the metric and the dilaton is
known on one side of the wall, Israel's 
matching condition (\ref{Eq:WallEnergy})\
and the dilaton matching condition (\ref{Eq:WallScalar})\
can be used to determine the boundary condition for the
metric and the dilaton on the other side of the wall.
Thus for walls 
with Minkowski space--time and a constant dilaton field on one side,
these conditions can be used to determine the solution on the other side.
For reflection symmetric walls, the boundary conditions are fixed
on both sides. Therefore, in both these cases one is 
able to find numerical solutions
to the field equations.

           \addtocounter{equation}{-\value{equation}}
%

%
           \section{Supersymmetric embedding}
\label{Sect:SUSY_Embedding}

Below we shall  
derive the Bogomol'nyi bound on the energy 
density and the  Killing spinor equations 
for supersymmetric (extreme)\ domain  
wall configurations. Our calculation will be based
on the  form of the  
Lagrangian spelled out in  Section~\ref{Sect:Supergravity}.

We shall primarily concentrate on 
su\-per\-sym\-me\-tric con\-fi\-gu\-ra\-tions,  
corresponding to extreme walls in\-ter\-po\-lat\-ing be\-tween 
su\-per\-sym\-me\-tric  
minima of the matter potential.

\subsection{Bogomol'nyi bound}
\label{Sect:BogomolnyiBound}

In order to derive the Killing
spinor equations and the Bogomol'nyi 
bounds for dilatonic domain walls, 
the technique of the generalized 
Israel--Nester--Witten form \cite{Witten:CMP81,Nester:PLA81}, applied to  
the study of supergravity walls  \cite{CGR:NPB92,CS:PRD95}, was used.  
We review the embedding   \cite{CS:PRD95}\footnote{Analogous procedures 
were followed in the derivation of the  Bogomol'nyi bounds for the mass of
the corresponding  charged  black holes  
\cite{GH:PLB82,GHHP:CMP83,GP:NPB84,KLOPP:PRD92,CY:NPB95}.}  into $N=1$ 
supergravity 
theory 
(see Section~\ref{Sect:Supergravity}  and specifically,  subsection \ref{BPL} 
for the bosonic part of the Lagrangian).  The   dilaton field $S$  and the 
matter field(s)  $T$,  responsible for the formation of the wall,  are assumed 
to  have a  general  separable  K\" ahler potential  (\ref{Eq:Kahlerpota}).  
Extreme walls with an exponential dilaton  coupling 
\cite{Cvetic:PRL93,Cvetic:PLB94}, as specified by the K\"ahler potential 
(\ref{Eq:kald}), and ordinary  
supergravity walls \cite{CGR:NPB92} (without dilaton)  are thus  special 
examples of such walls.
A generalization   of  the results  to more than one ``dilaton'' field
is straightforward,  as long as the dilatons   have no superpotential 
and  the   K\"ahler potential   
decouples from the matter fields responsible 
for wall formation\@.\footnote{
In superstring theory the additional ``dilatons''  may be identified
with  the compactification moduli, if the matter fields responsible for the 
formation of the wall do not couple to the moduli fields in the K\" ahler  
potential and the superpotential.  However,
matter fields do in general couple to the moduli fields.}

Since the extreme domain walls are planar and infinite,
we review a derivation  of the 
Bogomol'nyi bound for the  energy per unit area of the
wall.  
Note also 
that a precise definition of the energy density of the
wall is possible only in the thin wall 
approximation, namely, when the ``interior'' and the
``exterior'' regions of the wall 
are clearly separated.

We consider a generalized
Nester form \cite{Nester:PLA81}: 
\begin{equation} 
N^{\mu \nu} = \bar{\epsilon}\gamma^{\mu \nu \rho}
\widehat{\nabla}_{\rho} \epsilon 
\label{Eq:NesterForm} 
\end{equation}
where $\epsilon$ is a  Majorana spinor.
The supercovariant derivative is defined as
$\widehat\nabla_{\rho}\epsilon\equiv
\delta_{\epsilon}\psi_{\rho}$  and  
$\widehat\nabla_{\rho}=2\nabla_{\rho} + 
Q_{\rho}$, where $Q_{\rho}=
i \e^{K/2}
\left( \Re (W) +\gamma^5 \Im (W)\right)\gamma_{\rho}
 - \gamma^5 \Im (K_{T}\partial_{\rho}T) - \gamma^{5}
\Im (K_{S}\partial_{\rho}S)$ and
 $\nabla_{\mu}\epsilon = (\partial_{\mu}
  + \frac{1}{2}{\omega^{ab}}_{\mu}\sigma_{ab})
\epsilon
$; $\psi_{\rho}$ is the spin $\frac{3}{2}$ gravitino field. 
Therefore, the explicit  expression for Nester's form is:%
\begin{align}
N^{\mu\nu} & = 
 \bar\epsilon\gamma^{\mu\nu\rho}
\left[2\nabla_{\rho} + i \e^{K/2}
\left(\Re (W)+\gamma^{5}\Im (W)\right)\gamma_{\rho}\right.
\nonumber
\\ 
  & \quad \left. {}- \Im(K_{T}\partial_{\rho}T)\gamma^{5} -
 \Im(K_{S}\partial_{\rho}S)\gamma^{5}\right]\epsilon. 
\end{align}
Here $W$ is the superpotential (\ref{superpot})
and $K$ is the K\"ahler potential (\ref{Eq:Kahlerpot}). 
Stokes' theorem  ensures the following relationship:
\begin{equation}
\int_{\partial \Sigma} \! N^{\mu \nu} d\Sigma_{\mu \nu}
= 2\int_{\Sigma}\!\nabla_{\nu}N^{\mu \nu} d\Sigma_{\mu} 
\label{Eq:Stokes}
\end{equation}
where $\Sigma$ is a spacelike hypersurface.

After a lengthy calculation,\footnote{For details 
related to the derivation see  
Ref.~\cite{CS:PRD95} and appendices in 
Ref.~\cite{CGR:NPB92}.}  
the volume integral (\ref{Eq:Stokes})\ yields:
\begin{align}
&  \int  
\biggl[\, {\overline{\widehat\nabla_{\nu}\epsilon}}\,
\gamma^{\mu\nu\rho}{\hat\nabla_{\rho}\epsilon}  
+
K_{T  T^*}\overline{\delta_{\epsilon}\chi}  
\gamma^{\mu}\delta_{\epsilon}\chi
\nonumber \\ 
& \quad\quad {}
+ K_{S 
{S}^*} \overline{\delta _{\varepsilon}\eta} \gamma^{\mu}
\delta_{\varepsilon}\eta 
+(G^{\mu\nu}-T^{\mu\nu}){\overline{\epsilon}}\gamma_\nu\epsilon 
\, \biggr] \, d\Sigma_{\mu} 
   \ge 0,   
\label{volume}
\end{align}
where $\delta_{\epsilon}\chi$ and
$\delta_{\epsilon}\eta$
are the supersymmetry transformations of fermionic 
partners $\chi$ and $\eta$  to  the matter field $T$ and the 
dilaton field $S$, respectively; $T^{\mu\nu}$ 
is the energy-momentum 
tensor and $G^{\mu\nu}$ is the Einstein tensor. The first term in  
Eq.~(\ref{volume})\ is non-negative, provided 
the spinor $\epsilon$ satisfies the (modified)\ 
Witten condition, 
\ie\ 
${\mbox{\boldmath{$n$}}}\widehat{\mbox{\boldmath{$\nabla$}}}\epsilon  
=0$ 
({\boldmath{$n$}} is the four-vector  normal to
{\boldmath{$\Sigma$}}).
The 
K{\"a}hler metric coefficients $K_{T T^*}$ and $K_{S S^*}$ 
are positive definite,
and thus the second and 
the third terms in Eq.~(\ref{volume}) are non-negative
as well. The last term in  Eq.~(\ref{volume})\ is zero
due to Einstein's equations.   Thus, the  integrand in 
Eq.~(\ref{volume})
is always non-negative and it is zero  if and only if 
the supersymmetry transformations (\ref{Eq:deltapsi}),
(\ref{Eq:deltachi}), and
(\ref{Eq:deltaeta}) on the gravitino 
$\psi_{\rho}$ as well as on  
$\chi$ and  $\eta$  vanish, \ie\  
if the
configurations  are supersymmetric.

The surface integral of 
Nester's form  in Eq.~(\ref{Eq:Stokes})\ yields the  corresponding  
Bogomol'nyi
bound for the energy  associated with the  configuration. Such a  bound  
can be derived precisely 
only in the thin wall approximation, 
because  
the region inside the wall must be 
clearly separated from the region outside the wall in order 
for  its energy density to be well defined.

In this case the density of 
the  surface integral of Nester's form  in Eq.~(\ref{Eq:Stokes})\  
is  of the form: 
\begin{equation}
\left.\bar\epsilon_0\gamma^0\epsilon_0\sigma
+\bar\epsilon_0\gamma^{03}
\e^{K/2}\left[\Re(W)+\gamma^5\Im(W)\right]
\epsilon_0\right|_{0^-}^{0^+}. 
\label{surface}
\end{equation}
The spinor $\epsilon_0$ is
defined at the boundaries  $z=0^+$ and $z=0^-$ of the wall.   
In the first
term,  we have used the fact 
that for the thin wall the magnitude of the spinor 
components does not change. The first term  of the surface  
integral 
(\ref{surface})\ of the Nester's form (\ref{Eq:NesterForm})\  
can then be 
identified with the energy density of the wall. 
The second term 
corresponds to the  topological
charge density $C$ evaluated on  both sides of the wall. 
Positivity 
of the volume integral (\ref{volume})\ 
translates  through Eq.~(\ref{Eq:Stokes})\
into the corresponding Bogomol'nyi bound
for the energy density of  a thin wall:
\begin{equation}
\sigma \ge
|C|, 
\label{Eq:LocalBound}
\end{equation}
which is saturated if and only if 
the bosonic background is supersymmetric.

In the following subsection 
we shall derive the explicit  phase factors by which
the components  of the 
$\epsilon_0$ spinor  change at the wall
boundaries for the case of extreme solutions.
These phase factors   
will in turn allow us to 
obtain the  explicit form of $\sigma_{\text{ext}}=|C|$. 

Note that the existence of the  Bogomol'nyi bound (\ref{Eq:LocalBound}) crucially depends on the   fact that in  
Eq.~(\ref{volume})\ 
the spinor $\epsilon$ satisfies the (modified)\ 
Witten condition, 
\ie\ 
${\mbox{\boldmath{$n$}}}\widehat{\mbox{\boldmath{$\nabla$}}}\epsilon  
=0$ 
({\boldmath{$n$}} is the four-vector  normal to
{\boldmath{$\Sigma$}}). One example, where such a bound  
(\ref{Eq:LocalBound}) is satisfied \cite{CGR:NPB92}, is 
for  domain walls interpolating between the 
supersymmetric Minkowski vacuum and  the 
anti-de Sitter vacuum, which is either supersymmetric 
or with spontaneously broken supersymmetry. 
In general, however, the constraint  spinors 
is not satisfied, and  the bound is violated by the existence of 
 the ultra-extreme, 
false vacuum decay
bubbles. Thus, false vacuum decay bubbles  are \emph{not} topological defects. 

\subsection{Killing spinor equations}
\label{Sect:KillingSpinor}

We now write down  
explicit  Killing spinor equations 
\cite{CGR:NPB92,CS:PRD95}, 
\ie\ 
$\delta\psi_{\mu} = 
\delta\chi =\delta\eta=0$ (see Eqs.~(\ref{Eq:deltapsi}), (\ref{Eq:deltachi}),
and (\ref{Eq:deltaeta})), 
which are  
satisfied by supersymmetric, static
configurations.
With the metric ansatz (\ref{Eq:Ansatz})\
with $\beta =0$:
\[
ds^2= \e^{2a(z)}\left(dt^2-dz^2-dx^2-dy^2\right) 
\]
and $T(z)$ and $S(z)$ being functions only of $z$, 
and using the supersymmetry transformations 
specified in Section~\ref{Sect:Supergravity}, the Killing spinor
equations are of the form:
\begin{align}
\label{Eq:KillingSpinor}
\delta\psi_x  = & \left[- \gamma^{1}\gamma^{3}\partial_{z}a
- i \gamma^{1} \e^{(a+{K/2})}\left(\Re W+\gamma^5\Im W\right)
\right]\epsilon, 
\tag{\ref{Eq:KillingSpinor}a}
\displaybreak[3]\\ 
\delta\psi_y  = & \left[- \gamma^{2}\gamma^{3}\partial_{z}a
- i \gamma^{2} \e^{(a+K/2)}\left(\Re W+\gamma^5\Im W\right)
\right]\epsilon, 
\tag{\ref{Eq:KillingSpinor}b}
\displaybreak[3]\\ 
\delta\psi_z  = & \left[ 2\partial_{z}
 - i\gamma^{3} \e^{(a+K/2)}\left(\Re W
+\gamma^5\Im W\right) \right. 
\notag\\
 & \left. {} 
-\gamma^{5}\Im(K_{T}\partial_{z}T+
K_{S}\partial_{z}S)\right]\epsilon, 
\tag{\ref{Eq:KillingSpinor}c}
\displaybreak[3]\\ 
\delta\psi_t =& \left[\gamma^0\gamma^3 \partial_za
 +i \gamma^{0} \e^{(a+K/2)}\left(\Re W+\gamma^5\Im
W\right)\right]\epsilon, 
\tag{\ref{Eq:KillingSpinor}d}
\displaybreak[3]\\ 
\delta\chi =& -\sqrt2\left[ \e^{K/2}K^{T  T^*}
\left(\Re( D_{T}W)+\gamma^5\Im(  D_{T}W)\right) \right. 
\notag\\ 
  & \left. {}+ i  \e^a\left(\Re (\partial_{z}T )
+ \gamma^5(\partial_{z} T) \right)
 \gamma^{3} \right]\epsilon, 
\tag{\ref{Eq:KillingSpinor}e}
\displaybreak[3]\\ 
\delta\eta  = &  -\sqrt2 \left[ \e^{K/2}K^{S S^*}
\left(\Re( K_SW)+\gamma^5\Im(K_SW)\right) \right. 
\notag\\ 
 & \left. {} +i \e^a\left(\Re (\partial_{z}S )
+ \gamma^5(\partial_{z} S) \right)
 \gamma^{3}\right] \epsilon. 
\tag{\ref{Eq:KillingSpinor}f} 
\end{align}
\addtocounter{equation}{1}
We have assumed that  the Majorana spinor $\epsilon=(\epsilon_1,
\epsilon_2,\epsilon_2^*,-\epsilon_1^*)$  does not  depend on 
$x^{i}\in\{t,x,y\}$. Note that  
in Eqs.~(\ref{Eq:KillingSpinor})\ the K\" ahler  
potential
$K=K_{\text{dil}}(S,S^*)+K_{\text{matt}}(T,T^*)$ 
is separable  and $W=W(T)$,    
cf.\ Eq.~(\ref{Eq:SUSYLagrangian}). 
No Killing spinor exist for the anisotropic generalization with
$g_{tt}$, $g_{xx}$, and $g_{yy}$ 
all different \cite{JS:96,CGR:NPB92,Griffies:PHD}.

The vanishing of the above expressions  yields 
a set of
first-order 
differential 
equations\footnote{Equations~(\ref{Eq:KillingSpinor})\ set to
zero 
can be viewed as ``square roots'' of 
the corresponding Einstein and
Euler--Lagrange equations; 
they provide a particular solution of the equations
of motion which saturate the Bogomol'nyi bound (\ref{Eq:LocalBound}).} 
(known as the self-dual or 
Bogomol'nyi equations)\ for the metric
coefficient $a(z)$, 
$T(z)$ and $S(z)$
as well as the constraint on
the spinor $\epsilon$. The field equations are of the form: 
\begin{align} 
\label{boge}
0&=\Im \left(\partial _{z}T D_{T} \ln W\right), 
\tag{\ref{boge}a}
\label{boge1}
\displaybreak[3]\\
\partial_{z} T &=\zeta  \e^{(a+K/2)} 
 |W| K^{T{T^*}} 
D_{{T^*}} \ln W^*, 
\tag{\ref{boge}b}
\label{boge2}
\displaybreak[3]\\
\partial_{z}  a &= \zeta 
\e^{(a+ K/2)}|W|, 
\tag{\ref{boge}c}
\label{boge3}
\displaybreak[3]\\
\partial _{z}S &= -\zeta 
\e^{(a+K/2)}|W|K^{SS^*}K_{S^*}.
\tag{\ref{boge}d}
\label{boge4}
\end{align}
\addtocounter{equation}{1}
Here  $\zeta=\pm 1$ and 
it
can change sign only when $W$ crosses zero.  
There is another constraint
on the  ``field geodesic'' motion of the 
dilaton field, namely  $\Im(K_{
S}\partial_z S)=0$. 
However,
by multiplying Eq.~(\ref{boge4})\ 
by $K_{  S}$,  
this constraint is 
seen to be automatically satisfied. 
In this case the right-hand side 
of the equation is real, 
since $K^{SS^*}> 0$ is real and
$K_{S^*} =(K_{S})^*$. 

Equations (\ref{boge1})\ and (\ref{boge2})\
describe the evolution of the 
matter field $T = T(z)$ with $z$. 
The first equation is 
a ``field
geodesic'' equation,  
which  determines 
the path of the complex scalar field $T$  in the complex
plane  between the two 
minima $T_1$ and $T_2$ of the matter potential. 
Equation (\ref{boge2})\ governs the change of
the $T$ field with coordinate $z$ along this path.  

Equations (\ref{boge3})\ and (\ref{boge4})\
determine the evolution of the  
metric coefficient $a(z)$ and the
complex  field $S$.
These two equations
imply another  
interesting relation between  the  dilaton K\"
ahler potential $K_{\text{dil}}(S,S^*)$ and  $a(z)$:  
\[
2K^{SS^*}|K_{S}|^2\partial_za+\partial_zK_{\text{dil}} =0. 
\]

In addition, the Killing spinor 
equations (\ref{Eq:KillingSpinor})\   impose
a constraint on the phase of the Majorana spinor. Namely,
the solution for the Killing spinor component  is of the form
\begin{equation}
\epsilon_1=  \e^{i\theta}\epsilon_2^*=\mathcal{C} \e^{(a+i\theta)/2}, 
\label{Eq:KillSp}
\end{equation}
where the phase $\theta(z)$ satisfies 
\[
\partial_z\theta=-{\Im}(K_T\partial_zT).
\]
The constant $\mathcal{C}$ can be set to $\frac{1}{2}$ for  
the Majorana spinors normalized as
$\epsilon^\dagger\epsilon=1$. The constraint (\ref{Eq:KillSp})\
on the Killing
spinor $\epsilon$ in turn 
implies that the extreme configurations preserve
``$N=\frac{1}{2}$'' 
of the original $N=1$ supersymmetry.

The energy  density of the  wall (in the thin wall
approximation) is determined by setting
Eq.~(\ref{surface})\ to zero. With the explicit 
form for the Killing spinor
components (\ref{Eq:KillSp}), 
Eq.~(\ref{surface})\ yields:
\begin{align} 
\sigma_{\text{ext}} & = |C|  =  
2 \left|\left(\zeta \e^{K/2}W\right)_{z=0^+}-
\left(\zeta \e^{K/2} 
W\right)_{z=0^-}\right| \nonumber\\
& =  2\e^{K_{\text{dil}}(S_0,S_0^*)/2}\left|
\left(\zeta 
\e^{K_{\text{matt}}/2}
W\right)_{z=0^+}-\left(\zeta  
\e^{K_{\text{matt}}/2}
W\right)_{z=0^-}\right|. 
\label{sig}
\end{align}
Here the subscript $z=0^{\pm}$ refers to  either  side of the wall. 
Without loss of generality 
we have normalized $a(0)=0$ and set $S(0)=S_0$.

\paragraph*{Classification of extreme domain wall solutions:}
Solutions to the Bogomol'nyi equations (\ref{boge})\
fall into three types, 
depending  on whether 
$W(T)$ crosses zero
or not 
along the wall
trajectory.    

{\bf Type I} walls correspond to those 
where on one side of the wall, say for
$z>0$, $W(T_1)=0$. In this case the 
energy density of the wall is of the form:
$\sigma_{\text{ext}}=2 \left| \e^{K/ 
2}W\right|_{z=0^-}$, and 
the side of the wall with $z>0$
corresponds to the Minkowski space--time with a constant $S$.

{\bf Type II} walls 
correspond to the walls with $W(T)$ crossing zero
somewhere along the wall trajectory. 
At 
$W=0$ 
$\zeta$ changes sign. 
The energy density of the wall is specified by: 
$\sigma_{\text{ext}}=2 \left|  \e^{K/
2}W\right|_{z=0^+}+\left| \e^{K/2}W\right|_{z=0^-}$. 
Reflection symmetric
walls  fall into this class.

{\bf Type III} walls correspond to the walls where $W(T)\ne 0$ 
everywhere in 
the domain wall background. For this type
$\zeta$ does not change 
sign. The energy density of such walls is: 
$\sigma_{\text{ext}}=2 \left|\left|  \e^{K/ 
2}W\right|_{z=0^+}-\left|  \e^{K/2}
W\right|_{z=0^-}\right|$.

           \addtocounter{equation}{-\value{equation}}
%

%
           \section{Vacuum domain walls}
\label{Sect:VacuumWalls}

If each side of the domain wall corresponds to a vacuum,
that is, if on either side all the matter fields have
\emph{constant} expectation values, then all the local properties
of each side are Lorentz-invariant, and each side is a \emph{vacuum}. 
Such domain walls shall be called \emph{vacuum domain walls}. 

Vacuum domain walls can be classified according to the value of
their 
surface energy density $\sigma$, compared to the energy densities 
of the vacua outside the 
wall \cite{CGS:PRL93,CGS:PRD93}.
The three types are: (1) extreme walls with
$\sigma=\sigma_{\text{ext}}$ are planar, static walls. In this case
the gravitational mass of the wall is perfectly balanced by 
that of the exterior vacua;
(2) non-extreme walls with $\sigma=\sigma_{\text{non}} >
\sigma_{\text{ext}}$
corresponding to non-static bubbles
with two centres and (3) ultra-extreme walls with
$\sigma=\sigma_{\text{ultra}}<\sigma_{\text{ext}}$
representing expanding bubbles of false vacuum decay.

\subsection{Extreme vacuum walls}

The extreme vacuum domain wall solutions were discovered 
\cite{CGR:NPB92,CG:PLB92} 
in $N=1$ supergravity theory without a dilaton ($\alpha=0$). 
The walls represent regions interpolating between
isolated supersymmetric vacua of  the matter 
potential. 
The extreme solutions correspond to static
supersymmetric configurations saturating the
corresponding Bogomol'nyi bound.
The three possible types \cite{CG:PLB92}
of 
these extreme vacuum domain walls
are classified according to the nature of the
field path in the superpotential.
This aspect was discussed in Section~\ref{Sect:SUSY_Embedding}.
Here we shall analyse%
\footnote{This discussion 
is based on Refs.~\cite{CGS:PRD93,Griffies:PHD} 
where more details can be found.
It is included here for completeness.} 
the
space--times
induced by these extreme domain walls.  
We stress that for $\alpha=0$ the nature of the 
potential (\ref{susypota})  ensures that for supersymmetric minima, \ie\ 
those with $D_T W=0$, the value of 
the potential $V\le 0$. Thus the extreme walls
interpolate between anti--de~Sitter vacua or an anti--de~Sitter vacuum
and a Minkowski vacuum.

\subsubsection{The energy density of the extreme walls} 

In the thin wall approximation, the field equations
outside the vacuum domain wall
reduce to Einstein's vacuum equations with a non-positive
cosmological constant 
given by the supergravity potential (\ref{susypota})\
with $\alpha=0$. In a supersymmetric vacuum the K\"ahler covariant 
derivative of the superpotential vanish $D_T W=0$,
and thus the cosmological constant 
(here parameterized with 
parameter
$\chi$)\ is 
\begin{equation}
\Lambda \equiv V(\phi_0)= -3 \e^K |W|^2 = - 3 \chi^2. 
\label{Eq:ChiDefined}
\end{equation}  
In supergravity without a dilaton, 
supersymmetric vacua always 
have a non-positive $\Lambda$, but  
in order to include also the non-supersymmetric case,
we let $\chi$ take 
 a real positive 
value, 
if  the vacuum energy is negative, 
and conversely, if the vacuum has a positive energy-density,
then $\chi$ is positive imaginary. 

\begin{figure}[htb]
\centering{\epsfig{file=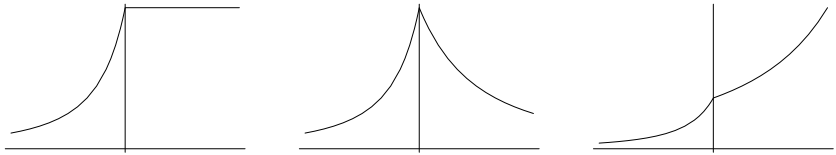,%
                   height=2.3cm,%
                   clip=}}
\caption{The metric conformal factor as function of the distance from the 
wall (at the origin)\ 
for extreme domain walls of 
Type I, II, and III, respectively .}
\label{Fig:TypeAll}
\end{figure}

Hence, the field equations (\ref{Eq:DynamicalEq})--(\ref{Eq:Constraint})\
reduce to
\[
{a'}^2= \e^{2 a} \chi^2, 
\]
which has the solution 
\begin{equation}
a=-\ln(1\pm\chi z) 
\label{Eq:ExtremeSolution}
\end{equation}
for a real $\chi$.
The integration constant has been absorbed into
a coordinate rescaling so that $a(0)$ is
normalized to unity.
To be definite, we place the 
vacuum with the most negative $\Lambda$ on the side $z<0$.
Then the three possible types of extreme domain walls are
characterized by the behaviour of the metric conformal factor 
(Fig.~\ref{Fig:TypeAll}).
For a Type I wall it is falling off from the wall on
the anti--de~Sitter side and is constant on the
Minkowski side. For a Type II wall it is falling off away from
the wall on both sides, and for a Type III wall it 
is increasing on the side with the least negative 
cosmological constant. 

The energy
density of an extreme wall is
\begin{equation}
\sigma = 2 (\chi_1 \pm \chi_2 ), 
\label{Eq:sigma_ext}
\end{equation}
with the plus sign for Type II walls 
and the minus sign for Type III.
Type I corresponds to $\chi_2=0$.
Linet's solution \cite{Linet:IJTP85} 
is a special case of a reflection symmetric wall of Type II.

One gets 
the same qualitative results 
by solving Eqs.~(\ref{boge}), which in the thin wall
approximation reduce to
Eq.~(\ref{boge3}) with all matter fields constant.
We would like to emphasize that the self-dual equations (\ref{boge})
are numerically tractable without 
using
the thin wall approximation (see Ref.~\cite{CGR:NPB92,CG:PLB92} and
the examples in Section \ref{Sect:ConnectionOTD}). 

\subsubsection{Gravitational mass for extreme walls}

Inertial observers on the
Minkowski side of the extreme Type I wall
experiences no gravitational effects.
One can understand this result by investigating
the
proper acceleration,\footnote{Hats denote tensor components
relative to an orthonormal tetrad frame.}
$g^{\hat{\mu}}$,
necessary to be a fiducial observer (an observer at a fixed spatial
position). 
This acceleration
is given by \cite{CG:PLB92}
\begin{equation}
   g^{\hat{\mu}}(z)= e^{-a} a'(z) {\delta^{\hat{\mu}}}_{\hat{z}}. 
   \label{Eq:ProperAcc}
\end{equation}
Hence, the metric conformal factor behaves
as a gravitational potential, and one can 
therefore
infer
the direction of the gravitational acceleration 
$\kappa^{\hat\mu}=-g^{\hat\mu}$
of test
particles
directly from Fig.~\ref{Fig:TypeAll}. 
For Type I and II walls it is directed \emph{away from}
the wall region. Hence, test particles
are repelled by the wall which is thus
exhibiting
``repulsive gravity''.

Clearly when $a(z)$ is constant as it is on the Minkowski side,
the acceleration is
zero, and no gravitational effects are felt.
On the anti--de~Sitter side 
the fiducial observers 
have a constant proper acceleration of magnitude $\chi$
which is 
\emph{half}
the surface mass
density of a Type I wall.

For the planar walls it is possible
to rewrite Einstein's field equations as an integral 
\cite{CDGS:PRL93,CGS:PRD93}
which can be
interpreted as a plane-symmetric 
analogue of Tolman's \cite{Tolman:PR30}
gravitational mass.
With help of
this concept
one can understand the ``gravitational forces''
and their sources. 
For example, in pure anti--de~Sitter space, 
the effective gravitational mass per
volume
is positive, thus indicating the attractive nature of the gravitational
field produced by a negative vacuum energy.
The converse holds for de~Sitter space,
which is a familiar result from inflationary
cosmology
where the repulsive nature of a 
positive vacuum energy drives a 
near 
exponential expansion
of the universe.

Tolman's formula for the gravitational mass
was originally derived
for a static spherically symmetric metric \cite{Tolman:PR30}.
This result has been adapted \cite{CDGS:PRL93,CGS:PRD93}
to
the case of static planar symmetry 
for which
the relevant
object is the gravitational mass per area.

In the derivation of Tolman's mass formula one focuses on the
\emph{generalized surface gravity},
that is, the gravitational
acceleration
as measured with \emph{standard}
rods and \emph{coordinate} clocks \cite{Gron:PRD85}
\[
\kappa^{i}(z) \equiv -\sqrt{g_{tt}} g^{\hat{\imath}}(z) 
\]
at any surface.
For a metric as given in Eq.~(\ref{Eq:Ansatz}), with $\beta=0$,
and with the proper acceleration (\ref{Eq:ProperAcc}), 
the
generalized surface gravity is
\begin{equation}
   \kappa^{i}(z)=-a'(z)\delta^{i}_{\; z}\, .
\label{Eq:k_result}
\end{equation}
In a spherically symmetric
four-dimensional space--time, one relates $\kappa^{r}$ to
the Tolman mass by defining  a mass $M$ in such a way that the Newtonian
force law is reproduced: $GM\equiv -r^2 \kappa^{r}$, and in the
three-dimensional case $G_{3}M_{3}\equiv -r\kappa^{r}$, where
$G_{3}$ is the (2+1)-dimensional Newton's constant 
\cite{Soleng:PS93}.
In the plane-symmetric case, considered here,
 we deal with
an essentially two-dimensional problem, 
and   the appropriate Newtonian
force law  implies $\Sigma\equiv -2\kappa^{z}$ where $\Sigma$ is the
\emph{gravitational mass per area} of the plane.
The factor of two is included because the gravitational acceleration is
half the mass per area
in the planar symmetric case 
(in the reflection symmetric case 
\cite{Vilenkin:PLB83,AG:PRD83}
one finds that the acceleration 
on both sides is a quarter of the mass
density).
Hence, Eq.~(\ref{Eq:k_result})\ leads to
\begin{equation}
\Sigma (z)= 2 a' .
\label{Eq:Sigma}
\end{equation}
For this equation to make sense in the wall case,
we rewrite the right hand side
in terms of an integral.
Starting from the Einstein tensor (\ref{Eq:Einstein}),
one finds  (in the static case $\beta=0$)
\begin{equation}
{G^{t}}_{t}-{G^{z}}_{z}-{G^{r}}_{r}-{G^{\phi}}_{\phi}=
e^{-4 a} ( 2 a' e^{2a})'.
\label{Eq:TolmanI}
\end{equation}
Using
$\sqrt{-g^{(4)}}=e^{4 a}$, $\sqrt{g^{(2)}}=e^{2 a}$,
Einstein's equations
${G^{\mu}}_{\nu} = {T^{\mu}}_{\nu}$, and
Eq.~(\ref{Eq:TolmanI}),
we can rewrite the right-hand side of Eq.~(\ref{Eq:Sigma})\
as 
\begin{equation}
   \Sigma (z) =  \frac{ \int_{-\infty}^{z}
   \sqrt{-g^{(4)}}dz'
   \left(
   {T^{t}}_{t}-{T^{z}}_{z}-{T^{r}}_{r}-{T^{\phi}}_{\phi}
   \right)
   \int dx dy }
   { \sqrt{g^{(2)}(z)} \int dx dy }
\label{Eq:Tolman}
\end{equation}
where we used
$a'(-\infty)=0$,  as is the case for both the
asymptotically Minkowski and anti--de~Sitter sides of the Type I and II
extremal walls.
The numerator on the right-hand side of
Eq.~(\ref{Eq:Tolman})\ 
is recognized as the Tolman mass of a static 
space--time \cite{Tolman:PR30}, and
the denumerator is the proper area.
Basically, this 
mass formula expresses the fact that
mass and energy are equivalent quantities in relativistic physics:
energy in the form of pressure contributes to the gravitational field
along with the mass density. On account of this, one can define
a gravitational mass density by $\rho_{g}\equiv\rho+3p$
for a perfect fluid in $3+1$ dimensions.
In the limit where we integrate from $z=-\infty$ to $z=\infty$, we get
$\Sigma(\infty)=0$. This means that
the total gravitational mass of
the Type I and II
extreme space--times is zero.
It should be noted that the Type III space--time is
causally identical to pure anti--de~Sitter. 
Therefore, the effective mass per volume of this
system is the relevant object; the effective mass
per area---as with pure anti--de~Sitter space---is infinite.

In the thin wall approximation, we can distinguish contributions to
the Tolman mass per area
due to the wall itself and due to the vacuum energy of
the adjacent
space--time.
In this case, a domain wall
has an effective  gravitational  mass per area, $\Sigma_{\text{wall}} =
{S^{t}}_{t} - {S^{r}}_{r } - {S^{\phi}}_{\phi}$, given by
$\Sigma_{\text{wall}} \equiv \sigma - 2\tau$.
Since the tension, $\tau$, is equal to the energy density, $\sigma$,
for a vacuum domain wall, 
we find $\Sigma_{\text{wall}} = -\sigma<0$.
By use of Eq.\ (\ref{Eq:sigma_ext}), one finds that $\sigma=2\chi$, which
yields
\[
\Sigma_{\text{wall}} = - 2 \chi.
\]
This negative gravitational mass per area for the wall,
with its \emph{repulsive}
gravity,
must be compensated by a positive gravitational
surface mass density  from the anti--de~Sitter space--time 
for there to
be no force on the Minkowski side.
This is precisely the case as we now show.
Again, taking into account 
the effect of vacuum pressure, $p_{v}=-\rho_{v}$,
the
gravitational mass density of anti--de~Sitter vacuum  is
\[
\rho_{g} =  \Lambda - 3\Lambda = 6\chi^{2} .
\]
Integrating out the $z$-direction from
$z= -\infty$ to the
position of the wall at
$z=0$
yields the
following mass per area for the anti--de~Sitter side of the wall:
\[
\Sigma_{\text{AdS}} = \lim_{ z  \rightarrow 0 }
\left[ {  \int_{-\infty}^{z} (6 \chi^{2})
\sqrt{-g^{(4)}}  dz\int dx dy \over \int \sqrt{g^{(2)}}dx dy  } \right]=
2\chi.
\]
Hence, as seen from the Minkowski side of the domain wall, 
there
are two gravitational surface mass densities on the
$z \le  0$ side. Firstly, there is a
\emph{negative} mass per area coming
from the domain wall: $\Sigma_{\text{wall}} = -2\chi$.
Secondly, there is a \emph{positive} integrated  mass per area coming from
anti--de~Sitter space itself: $\Sigma_{\text{AdS}} = 2\chi$,
which exactly cancels
that of the domain wall.

The analysis used for the extreme Type I wall can
also be applied to  the extreme Type II  wall.
When there is a Minkowski metric on one side,
the Killing time, $t$, corresponds to the
proper time of an observer infinitely far away from the wall
on this side. In the Type II case, one may use an observer sitting
in the center of the wall. Here too, there
is a frame where
all the connection coefficients vanish, the metric is Minkowskian,
and where the proper time of the observer is equal to the Killing time.
Thus, in the thin wall approximation,
one
finds the effective
mass per area of the two anti--de~Sitter 
sides to be $2(\chi_{1} + \chi_{2})$.
This positive effective mass is exactly cancelled by the negative
effective mass of the domain wall separating the two regions of 
anti--de~Sitter space.
Likewise, the general expression Eq.~(\ref{Eq:Tolman}) yields a
zero Tolman mass per area for the space--time.

Note that in the above calculations we have integrated along a
constant time slice $-\infty < z < \infty$. 
As we shall see below, 
there is a past and
future Cauchy horizon
for data placed on such a slice.
The above calculation implicitly assumes no
contribution to the effective mass arising from the
past of the past Cauchy horizon.  This is
consistent with the extensions of the space--time
beyond the Cauchy horizon considered in the next section.
It is also consistent with there
being a global balance of gravitational ``forces''.

\subsubsection{Global space--times of the extreme walls}

The three types of extreme domain walls realized
\cite{CGR:NPB92,CG:PLB92}
in four-dimensional ($N=1$)\ supergravity
theory  
are planar and static. 
The space--time metric (\ref{Eq:Ansatz}) induced by these walls
have
$\beta\equiv 0$
and
is conformally flat with conformal factor
$e^{a(z)}$ becoming unity on the Minkowski side of 
the Type I wall and
falling of as $(z \chi)^{-2}$ on the anti--de~Sitter side,
where $\Lambda = -3\chi^{2}$.
The Type II conformal factor
falls off as 
$(z\chi_{1})^{-2}$
and
$(z\chi_{2})^{-2}$
on the respective sides.
For the Type III wall,
the conformal factor
falls off in the same manner on one side of the wall, but has 
coordinate singularity
at a finite value of $z$ on 
the other side. This singularity represents
the affine boundary of the space--time.
In this section we present geodesically complete
extensions of the space--times for the 
Type I, II, and III extreme
domain walls.  The global space--times of
the Type I wall was first considered in Ref.~\cite{CDGS:PRL93}
and the Type II wall in Ref.~\cite{Gibbons:NPB93}.

For each of the walls, 
the space--time
must be extended 
across a
Cauchy horizon on the anti--de~Sitter side
\cite{CDGS:PRL93,Gibbons:NPB93}.
The Cauchy horizons occur on the nulls
at $|z|=\infty$ where
$a(z) = -\infty$,
\ie\ where the line element degenerates.
Although these nulls are an infinite
proper distance away, the geodesic distance is
finite.
This type of geometry is familiar from the extreme
black hole space--times 
\cite{Carter:PR66,Carter:PL66,HE:Cambridge73,Chandrasekhar:Oxford83}.
The comoving coordinates
must be extended across the Cauchy horizons
on these anti--de~Sitter sides.
The same need also
arises in pure anti--de~Sitter space.

A Cauchy horizon is the boundary of causal
evolution. Therefore, one has the possibility of making
identifications across the
Cauchy horizons which can introduce closed timelike curves (CTCs).
The possibility of CTCs is inherited from the
anti--de~Sitter portion of the space--time.
Identifications
of this type
are especially intriguing
for the Type I walls, as CTCs
could lead to supersymmetric
time-machine \cite{CDGS:PRL93}
for travelers leaving Minkowski space--time by
passing across the wall and
then re-emerging 
into the same Minkowski region
at an earlier time. 

There are three possible extensions across the
null Cauchy horizons:
\begin{enumerate}
\item
One can extend onto a new patch with the scalar field permanently
settled into its vacuum,
\ie\  beyond the Cauchy horizon
there is  a pure anti--de~Sitter vacuum.
\item
In the case of the Type II wall, one can 
shift the old diamond along the
null such that the new diamond
is oriented just as the old.  This extension
yields a new wall
as well as a jump in the cosmological constant at the Cauchy
horizon for non-reflection symmetric
walls.
\item
The old diamond can be reflected onto the new diamond
across the Cauchy horizon.
This extension leads to a new wall as well as a smooth
matching of the cosmological constant at the horizon.
\end{enumerate}

In the following, we concentrate on
geodesic extensions of the third kind.
One reason for doing so is that it yields
the most interesting causal structure for the
resulting space--times.
It is for the third approach that the
causal structure of the Type I and II space--times
exhibit a symmetric
lattice structure similar to those
first realized by the extensions
of Carter for the Kerr and
Reissner--Nordstr{\"{o}}m black holes \cite{Carter:PR66,Carter:PL66}.
The extension
for the Type I wall
realizes the identical causal structure as the
extreme Kerr black hole along its symmetry
axis \cite{HE:Cambridge73,Chandrasekhar:Oxford83,Carter:PR66}.
\begin{figure}[htb]
\centering{\epsfig{file=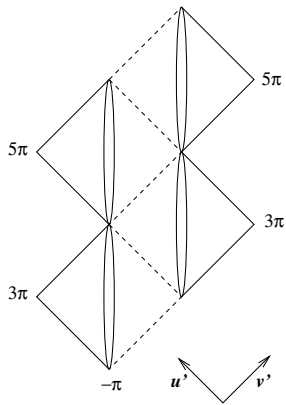,%
                   height=5.3cm,%
                   clip=}}
\caption{Conformal diagram of the 
extreme Type I domain wall. The walls
are represented by the double timelike 
arcs splitting the diamonds. The central regions with 
Cauchy horizons (represented by dashed nulls)\
consist of anti-de~Sitter patches, and the outer
regions are semi-infinite Minkowski regions. 
The vertices are an infinite affine distance away from
all other points.
The complete 
extension 
is 
an infinite lattice chain which 
continues forever towards the past and the future.} 
\label{Fig:ExtremeI}
\end{figure} 
\begin{figure}[htb]
\centering{\epsfig{file=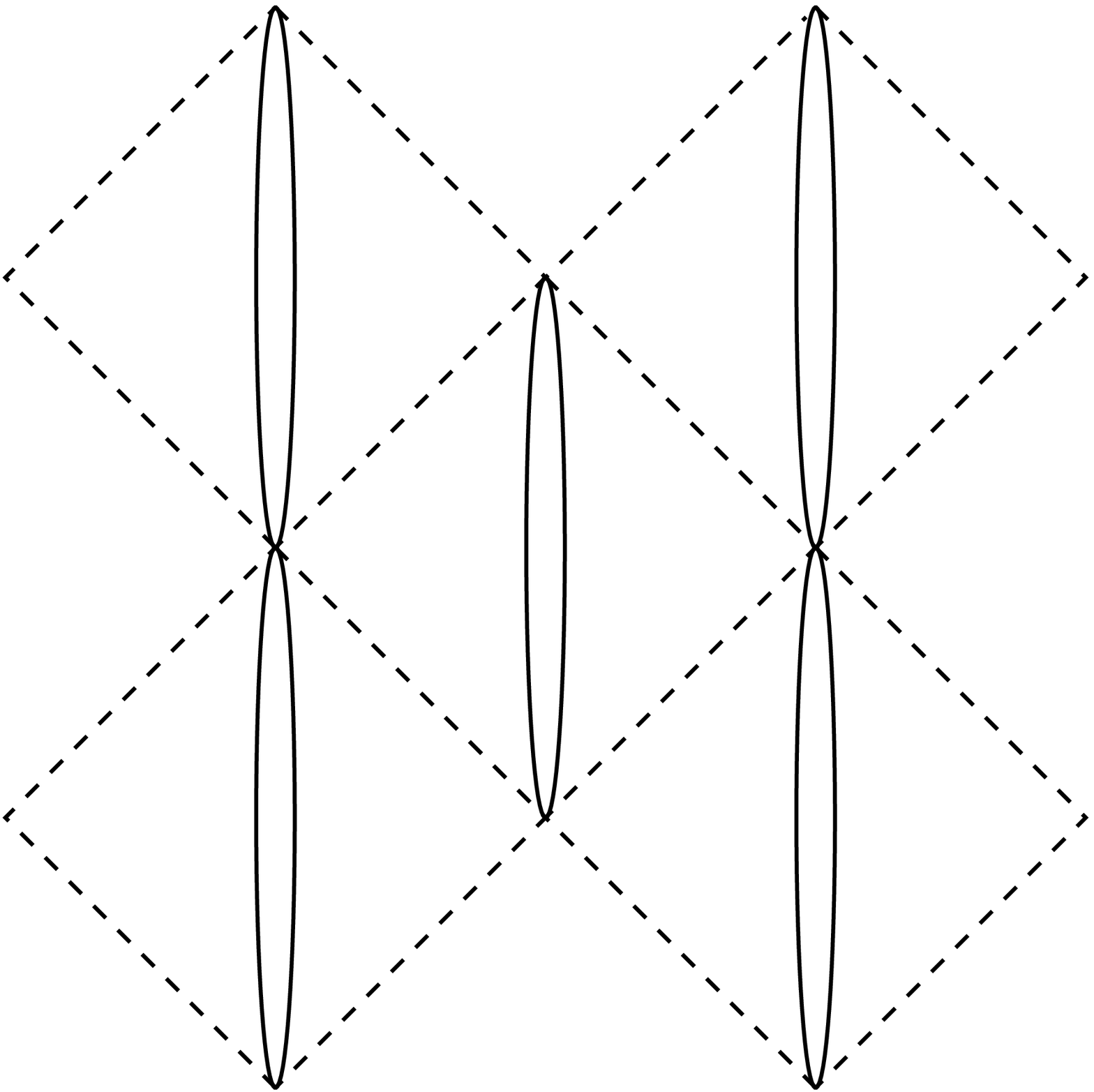,%
                   height=4.24cm,
                   clip=}}
\caption{Conformal diagram of the extreme Type II domain wall. 
Both sides of the walls are anti--de~Sitter spaces.
The complete extension covers the whole plane.} 
\label{Fig:ExtremeII}
\end{figure} 
Finally, it is through the infinite lattice
for the Type I and II space--times
that one eliminates the
timelike boundary of the covering space of
anti--de~Sitter space
in exchange for a countably infinite
number, $\aleph_0$, of isolated vertex 
points an infinite
affine distance away from any interior 
point
(see Figs.~\ref{Fig:ExtremeI} and \ref{Fig:ExtremeII}).
For example, the Cauchy problem for
the Type I space--time can be specified by
prescribing initial data on one constant
time slice in an AdS$_{4}$ region
and freely choosing
boundary data on past null infinity of the
countably infinite
number of adjacent Minkowski spaces 
(see Fig.~\ref{Fig:ExtremeI}).
In contrast, for the anti--de~Sitter covering space,
the Cauchy
problem is defined only after prescribing
an infinite amount of boundary
data
which has to be \emph{self-consistent}
with the specified initial
data \cite{AIS:PRD78,BF:AP82}.

\begin{figure}[htb]
\centering{\epsfig{file=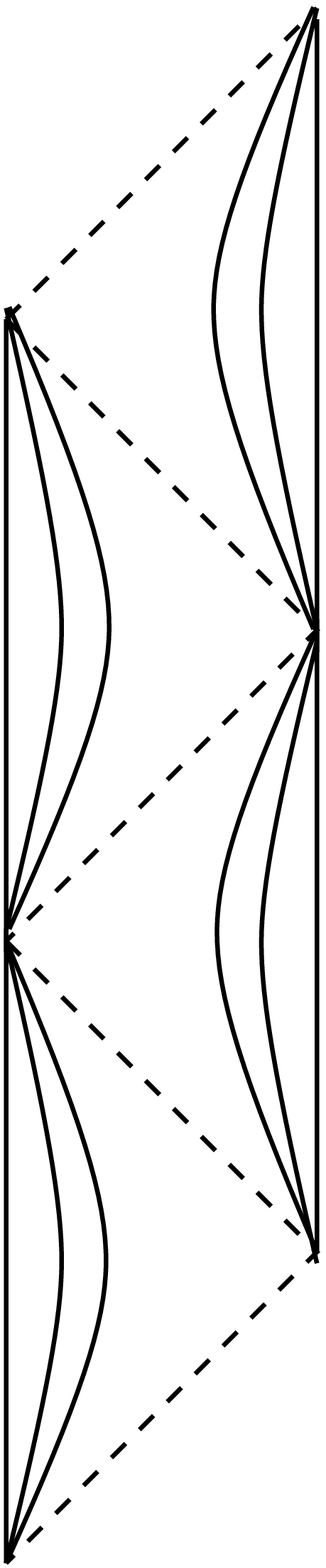,%
                   height=5.3cm,%
                   clip=}}
\caption{Conformal diagram of the extreme Type III domain wall. 
Both sides of the walls are anti-de~Sitter spaces. The outer
regions are limited by
timelike affine boundaries. The complete extension is an infinite lattice
which continues forever
towards the future and the past.} 
\label{Fig:ExtremeIII}
\end{figure} 

The three types of extreme space--times,
constructed from the third
kind of geodesic extension
described above,
have the conformal diagrams shown in
Figs.~\ref{Fig:ExtremeI},
\ref{Fig:ExtremeII}, and  \ref{Fig:ExtremeIII}.
In each of the figures,
the $x$- and $y$-coordinates are suppressed;
therefore, each point represents an infinite plane.
The compact null coordinates
$u'$ and $v'$ defined by
$\tan(u'/2)= (t - z)\chi$ 
and 
$\tan(v'/2)= (t + z)\chi$ 
define the axes.
These coordinates can be
extended
smoothly
across the
Cauchy horizons (denoted by the dashed nulls)\ 
separating the diamonds
on the anti--de~Sitter side.
This fact is seen explicitly by
writing the (1+1)-dimensional line element near the horizon as
$ds^{2} = (z\chi)^{-2}(dt^{2} - dz^{2}) =
\{ \chi \sin[1/2(u' - v')] \}^{-2} du' dv'$
which has a smooth extension across the
null specified by $u'=\pi$ and $-\pi < v' < \pi$
as well as across all the other Cauchy horizons.
Thus, the null $(u',v')$-coordinates
provide an atlas for describing
the global space--time.
Note that the full (3+1)-dimensional
metric has coordinate
singularities in the $x$- or $y$-directions
at the Cauchy horizons.

The extension chosen for the Type I wall \cite{CDGS:PRL93}
in Fig.~\ref{Fig:ExtremeI} possesses
the same causal structure as the extreme Kerr black
hole along its symmetry axis \cite{HE:Cambridge73,Chandrasekhar:Oxford83,%
Carter:PR66}.
The extension chosen
for the Type II wall in Fig.~\ref{Fig:ExtremeII}
tiles the whole plane with a lattice
of walls.  For the Type III wall
shown in Fig.~\ref{Fig:ExtremeIII},
the conformal factor 
diverges
at some finite coordinate \cite{CG:PLB92}.
This is the
edge of the space--time.
As a result, the extension of the 
Type III wall is causally the same as
that of pure anti--de~Sitter space.

For each of the extensions, the vertices are special
points \cite{Gibbons:NPB93} which 
are an infinite affine distance away from all other
points; \ie\ they represent an infinite conformal
compression.
This is  analogous to the situation in
the extreme  Reissner--Nordstr\" om
and Kerr black holes \cite{HE:Cambridge73,Chandrasekhar:Oxford83,%
Carter:PR66,Carter:PL66}.

The conformal diagram of a space--time
containing a Type I or II extreme domain wall centered at  $z=0$
should be compared to that of pure anti--de~Sitter space. 
The timelike affine infinity of anti--de~Sitter space 
is smoothed out by the wall, which allows
for another space--time region across the boundary of
pure anti--de~Sitter space.
In this sense, the
wall is located at spatial infinity.  
It has therefore 
been speculated \cite{Gibbons:NPB93} that the
extreme walls are related to the
\emph{``Membrane at the End of the Universe''}
in supermembrane
theory \cite{BDPS:PLB87,BDPS:PLB89}.

\subsection{Non- and ultra-extreme vacuum walls}
\label{Sect:non_and_ultra_vacuum_walls}

As we have seen, the extreme domain walls are characterized by an
exact cancellation 
of the gravitational contributions of the vacua and the
walls. 
By increasing or decreasing the energy density of the wall
and thus disturbing this
balance, supersymmetry is broken,
and the walls become non-static. These solutions are 
not planar; the walls have spherical topology (if their
world volumes are
assumed to be geodesically complete). See Section~\ref{Sect:IsoWalls}
for details about the metric. 

By increasing the 
wall's energy so that 
$\sigma=\sigma_{\text{non}}>\sigma_{\text{ext}}$, 
its repulsive gravitational effect increases,
and 
\emph{a priori} 
one could 
expect that the total effect of an anti--de~Sitter bubble surrounded
by such a \emph{non-extreme} 
domain wall bubble would be an example of a finite body with 
\emph{negative gravitational mass} which is known to have 
unusual physical properties 
\cite[Prob.~13.20]{Price:AJP93,Bonnor:GRG89,LPPT:Princeton75},
but   
this phenomenon is avoided here; both sides of the wall 
are on the inside of the wall; it is a two-centered bubble.
Hence, one can never be on the outside of the negative mass system.
In Ref.~\cite{CGS:PRD93} a \emph{Positive Mass Conjecture} 
was formulated saying that there is no 
singularity free solution of Einstein's 
field equations
with matter sources (not including the vacuum)\
obeying the weak energy condition 
equations for which 
an exterior observer can see a negative mass object.
A global monopole \cite{BV:PRL89,HL:PRD90,SL:CQG91}
would appear to be a counter example,
but in this case the Goldstone fields 
extend to infinity which means that these 
object are extended sources and all observers must be inside
the system. 
 
Geometrically we define the inside of the spherical
domain wall to be the side on which 
the radius of curvature $R\equiv e^{a(z)} S(t)/\beta$
(see Eqs.~(\ref{Eq:Ansatz}) for the definition of $S(t)$)\
of concentric shells
decreases as one goes away from the wall, and the outside is the side
where it increases with distance. As we shall see in the next
subsection, this local definition agrees with the global
picture. 

A
domain wall with energy density less than the extreme wall
$\sigma=\sigma_{\text{ultra}}<\sigma_{\text{ext}}$, 
is referred to as an ultra-extreme domain wall. As for the
non-extreme case, the wall has the topology of a 
sphere, but now only the side with the smallest vacuum
energy is on the inside; this is an ordinary one-centered
bubble. It corresponds to a vacuum decay bubble.

When $\beta\neq 0$ the vacuum Einstein equations reduce to
(cf.\ Eq.~(\ref{Eq:Constraint}))
\begin{equation}
{a'}^2 - \beta^2 =  e^{2a} \chi^2 
\label{Eq:NonUltraEinstein} 
\end{equation}
where $\chi$ is defined in Eq.~(\ref{Eq:ChiDefined}),
\ie\ $\chi^2=-\Lambda/3$ where the cosmological constant
$\Lambda=V(\phi_0)$.

The solutions are
\begin{equation}
a(z)=\left\{\begin{array}{lcl} 
     - \ln[ \beta \sinh(\beta z-\beta z')/\chi ]   
     & {\text{for}}&  V(\phi_0)<0,\\ 
      \pm \beta z 
     & {\text{for}}& V(\phi_0)=0,\\ 
     - \ln[ \beta \cosh(\beta z-\beta z'')/\chi ]   
     & {\text{for}}&  V(\phi_0)>0, 
            \end{array} \right.
\end{equation}
where there are two solutions for each of the
 integration constants $z'$ and $z''$ determined by the
normalization condition $a(0)=0$.
They are denoted by $z'_{\pm}$ and $z''_{\pm}$ and 
are given by
\[
e^{2\beta z'_{+}}=e^{-2\beta z'_{-}}=  1 +  \frac{2\beta^2}{\chi^2} 
+ \frac{2\beta}{\chi^2} \left(\chi^2+\beta^2\right)^{1/2}  
\geq 1
\]
(with equality in the extreme limit)\
and
\[
e^{2\beta z''_{+}}=e^{-2\beta z''_{-}}=  -1 -  \frac{2\beta^2}{\chi^2} 
+ \frac{2\beta}{\chi^2} \left(\chi^2+\beta^2\right)^{1/2}  
> 1. 
\]
{}From the last equation it is clear that there is no extreme 
limit $\beta\rightarrow 0$ in the de~Sitter case.

The energy density of a non-extreme domain wall is
\begin{equation}
\sigma_{\text{non}}= 
2\left( \chi_1^2+\beta^2 \right)^{1/2}
+ 2\left( \chi_2^2+\beta^2 \right)^{1/2}.   
\label{Eq:sigma_non}
\end{equation}
The ``planar'' reflection symmetric
wall discussed by Vilenkin \cite{Vilenkin:PLB83}
and by Ipser and Sikivie \cite{IS:PRD84} 
is a reflection symmetric non-extreme wall  
with vanishing vacuum energy. 

An ultra-extreme wall has 
\begin{equation}
\sigma_{\text{ultra}}= 
2\left( \chi_1^2+\beta^2 \right)^{1/2}
- 2\left( \chi_2^2+\beta^2 \right)^{1/2},  
\label{Eq:sigma_ultra}
\end{equation}
where it is understood that the smallest cosmological constant
(the most negative one)\ is on the left side.
The ultra-extreme bubbles are the 
tunneling bubble \cite{BKT:PLB83a}
of false vacuum decay \cite{Coleman:Erice77,CDeL:PRD80}.
See Refs.~\cite{BKT:PLB83a,Sato:PTP86,BGG:PRD87,BKT:PRD87}
for discussions of false vacuum decay also involving de~Sitter
space. 

The parameter $\chi$ is real for anti--de~Sitter space
and imaginary for de~Sitter space. 
Since the energy-density must be 
real, 
the cosmological constant is limited by $\beta^2\geq \chi^2$ in the
de~Sitter case \cite{Sato:PTP86}, again 
confirming that there is no extreme limit in this
case.  
Yet, the limiting case $\beta^2=\chi^2_{\text{deS}}$ 
is special
because for this value of $\beta$
the gravitational acceleration of fiducial
observers vanish near the wall.
Hence, just as for the extreme Type I
(anti-de~Sitter--Minkowski wall),
an anti-de~Sitter--de~Sitter wall 
with $\beta^2=\chi_{\text{deS}}^2$ exactly cancels
the gravitational mass of the anti-de~Sitter vacuum.

\begin{figure}[htb]
\centering{\epsfig{file=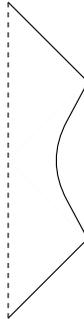,%
                   height=4.24cm,%
                   clip=}}
\caption{Conformal diagram for the Minkowski side 
inside
a non-\ or ultra-extreme bubble. This diagram 
represents the $(t,r)$-plane in compactified
coordinates. The curved line is the world volume of the wall,
which follows radial Rindler motion.
The dotted nulls are the Rindler horizons. 
The stippled line on the left is the center of the bubble.} 
\label{Fig:M4InsideNonUltra}
\end{figure} 

\begin{figure}[htb]
\centering{\epsfig{file=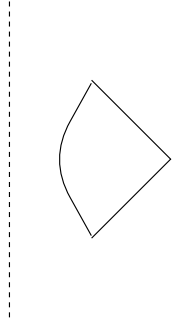,%
                   height=4.24cm,%
                   clip=}}
\caption{Conformal diagram for the Minkowski side 
outside
an ultra-extreme bubble. This diagram 
represents the $(t,r)$-plane in compactified
coordinates. The curved line is the world volume of the wall,
which follows radial Rindler motion.
The stippled line on the left is the center of the bubble. 
There are no bubbles with two outsides, so if the
interior 
is a vacuum, it
must be filled with anti--de~Sitter space.} 
\label{Fig:M4OutsideUltra}
\end{figure} 

\afterpage{\clearpage}

\subsubsection[Global structure]{Global structure
of non- and ultra-extreme domain wall space--times}

For these domain walls, the hyper-space of the 
world volume 
is a $(2+1)$-dimensional de~Sitter space. The global 
structure of the space--time is well known; 
it can be embedded in a $(3+1)$-dimensional Minkowski 
space where it takes the form of a hyperboloid (see, \eg\ 
Ref.~\cite{HE:Cambridge73,EG:IJMPD95} 
for the analogous case of $(3+1)$-dimensional de~Sitter space).
Hence, if the inside of the wall is a portion
of Minkowski space, it must be the inside of this 
hyperboloid (Fig.~\ref{Fig:M4InsideNonUltra}).
It is then clear that the wall, which is a spacelike section of the
hyperboloid,
is a sphere with a constant acceleration in the radial direction. 
In its rest frame one has Rindler horizons
at $|t\pm z|=\infty$. 
The extension across these  
horizons 
is unique.
Conversely, if the Minkowski space is on the outside,
it must
be the outside the de~Sitter hyperboloid (Fig.~\ref{Fig:M4OutsideUltra}). 
Timelike observers with insufficient acceleration will
eventually be hit by the ultra-extreme wall. 

\begin{figure}[htb]
\centering{\epsfig{file=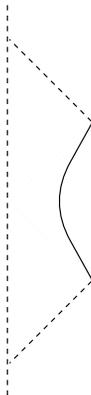,%
                   height=5.3cm,%
                   clip=}}
\caption{Conformal diagram for the anti--de~Sitter
side 
inside a non-\ or
ultra-extreme bubble. 
The curved line is the world volume of the wall,
which follows radial accelerated motion.
The straight timelike edge on the right side is the
timelike affine boundary of anti--de~Sitter space. 
The stippled line on the left is the center of the bubble.} 
\label{Fig:AdS4Inside}
\end{figure}

For anti--de~Sitter space--time, 
the Einstein universe coordinates \cite{HE:Cambridge73}
define a frame where the bubbles have kinematic 
properties
analogous to those seen by inertial 
observers in Minkowski space (see Refs.~\cite{CGS:PRD93,Griffies:PHD}
for details).
Non-extreme bubbles accelerate away from observers
on both sides, and therefore we get a horizon 
of the Rindler type
also in anti--de~Sitter space. 
Consequently, anti--de~Sitter insides of non-\ or ultra-extreme 
walls
have a similar 
conformal diagram as the corresponding Minkowski
interior, but instead of having a finite diagram one now
has Cauchy horizons (Fig.~\ref{Fig:AdS4Inside})\
and the possibility to 
extend to an infinite chain.

\begin{figure}[htb]
\centering{\epsfig{file=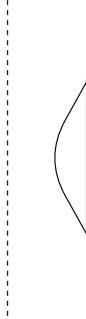,%
                   height=4.24cm,%
                   clip=}}
\caption{Conformal diagram for the anti--de~Sitter
side 
outside
an ultra-extreme bubble. This diagram 
represents the $(t,r)$-plane in compactified
coordinates. The curved line is the world volume of the wall,
which follows radial Rindler motion.
The straight timelike edge on the right side is the
timelike affine boundary of anti--de~Sitter space. 
The stippled line on the left is the center of the bubble. 
There are no bubbles with two outsides, so if the
interior 
is a vacuum, it
must be filled with anti--de~Sitter space with a
more negative vacuum energy density.} 
\label{Fig:AdS4OutsideUltra}
\end{figure}

An
ultra-extreme
bubble 
accelerates towards  outside observers in anti--de~Sitter space,
and thus, with exception of the affine
boundaries which are 
null in the Minkowski case and timelike for anti--de~Sitter space,
the causal structure 
(Fig.~\ref{Fig:AdS4OutsideUltra})\ 
is similar to that of the Minkowski exterior. 

\begin{figure}[htb]
\centering{\epsfig{file=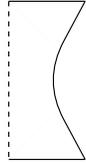,%
                    height=2.14cm,%
                   clip=}}
\caption{De~Sitter 
side 
inside a non-\ or
ultra-extreme bubble. 
The curved line represents the world volume of the wall. 
The stippled timelike line on the left-hand side is the 
coordinate singularity $\psi=0$ of de~Sitter space--time
\protect\cite[page~127]{HE:Cambridge73}. 
Note that in contrast to anti--de~Sitter and Minkowski space--time,
de~Sitter space has a spacelike infinity for 
timelike and null lines.}
\label{Fig:dS4Inside}
\end{figure}

\begin{figure}[htb]
\centering{\epsfig{file=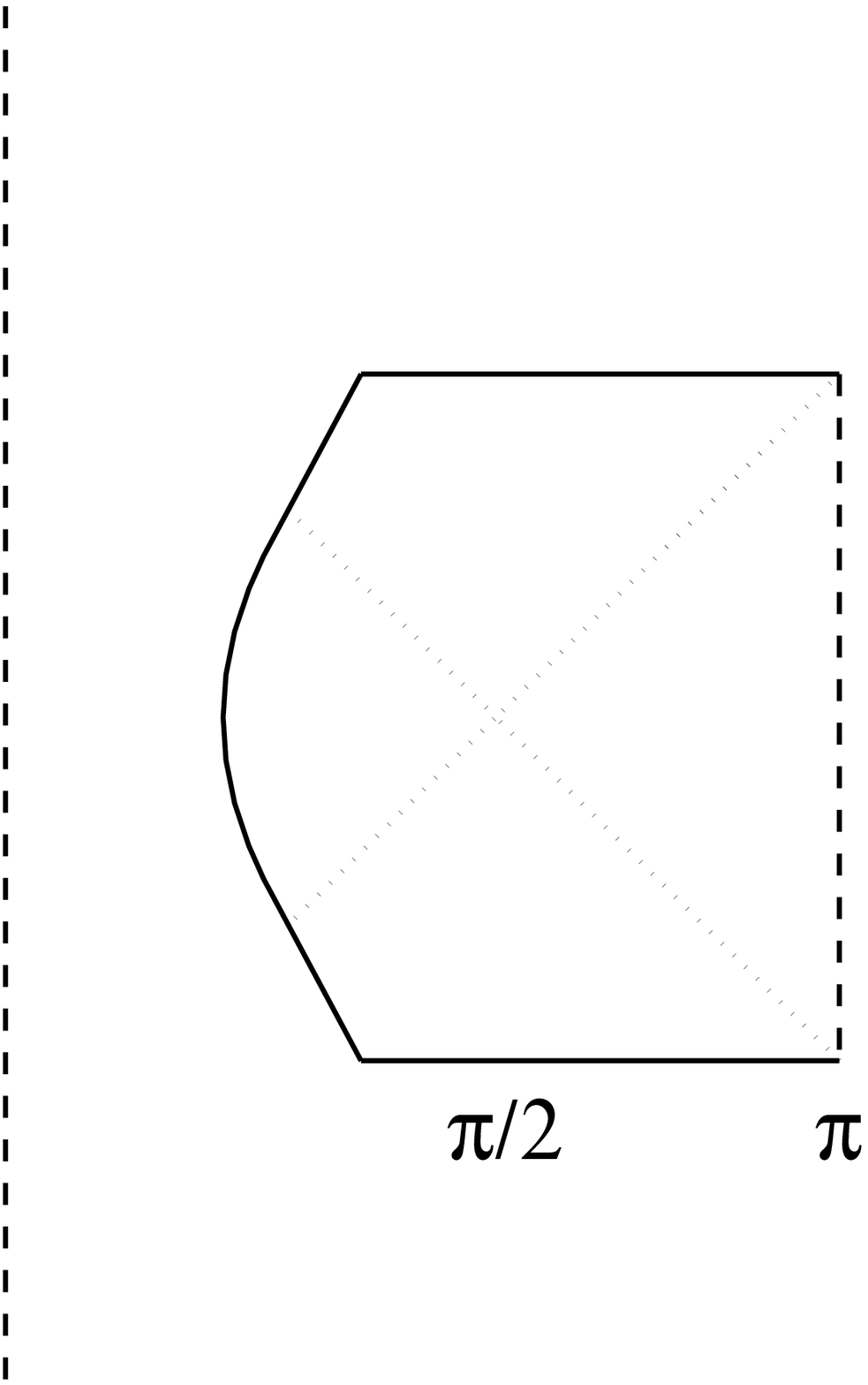,%
                   height=4.24cm,%
                   clip=}}
\caption{De~Sitter
space outside
an ultra-extreme bubble with
$\chi^2<\beta^2$. 
The curved line is the world volume of the wall. 
The stippled timelike line on the right-hand side is the 
coordinate singularity $\psi=\pi$ of de~Sitter space--time
\protect\cite[page~127]{HE:Cambridge73}. 
The stippled
line one the left-hand side is the centre of the
bubble on the \emph{other}
side of the wall.} 
\label{Fig:dS4OutsideUltra}
\end{figure}

\begin{figure}[htb]
\centering{\epsfig{file=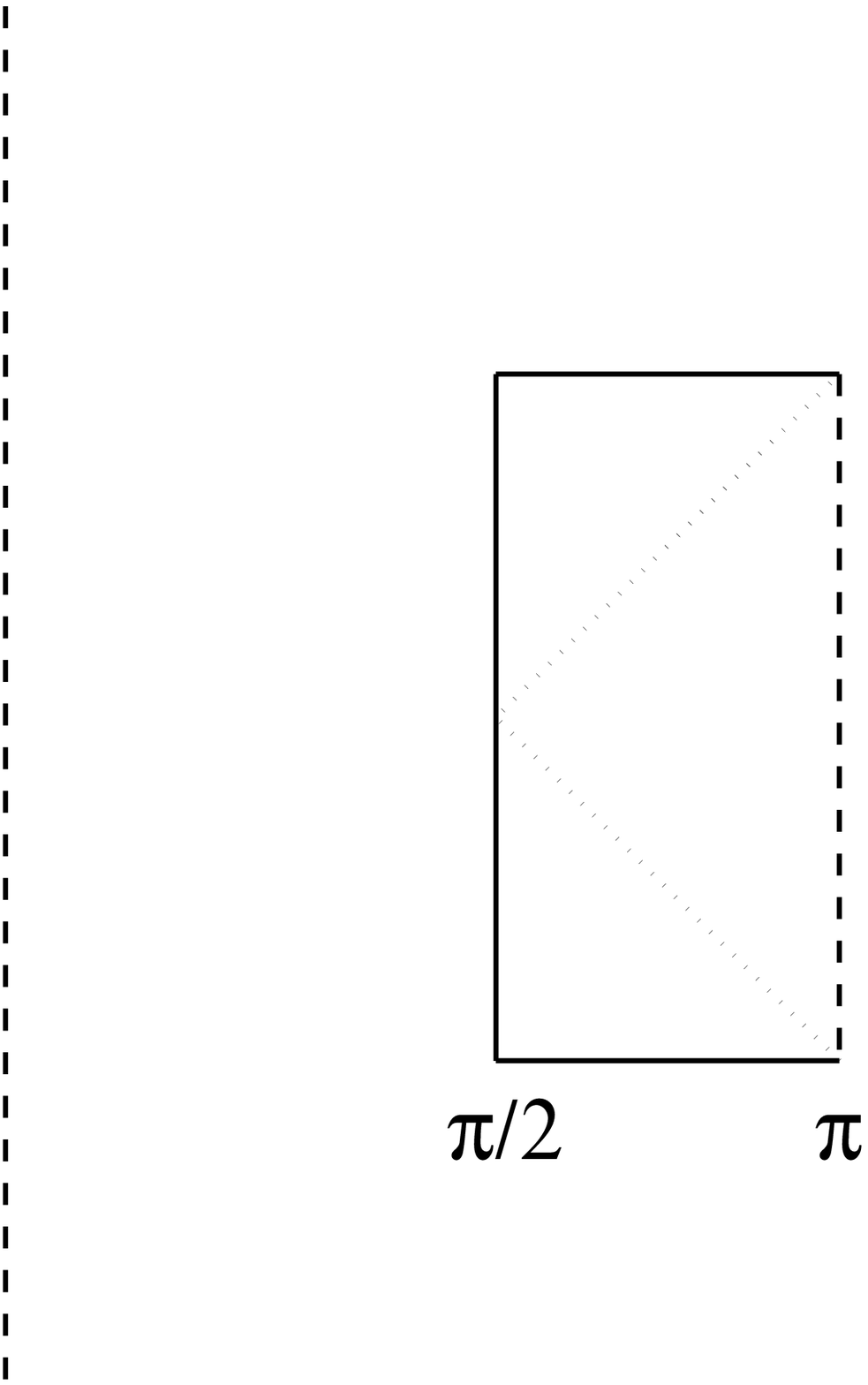,%
                   height=4.24cm,%
                   clip=}}
\caption{The de~Sitter
side 
outside
an ultra-extreme bubble with $\beta^2=\chi^2$. 
The timelike line at $\psi=\pi/2$ on the left 
is the world volume of the wall. 
The stippled timelike line on the right-hand side is the 
coordinate singularity $\psi=\pi$ of de~Sitter space--time
\protect\cite[page~127]{HE:Cambridge73}. 
The stippled
line on the left-hand side is the centre of the
bubble on the \emph{other}
side of the wall.} 
\label{Fig:dS4OutsideEks}
\end{figure} 

The de~Sitter 
inside of a non-\ or ultra-extreme
bubble corresponds to the part of de~Sitter space with the
metric
conformal factor
monotonously increasing toward the wall. The wall
surface
accelerates away from test particles 
(Fig.~\ref{Fig:dS4Inside}). 
In de~Sitter space outside 
an ultra-extreme bubble 
with $\chi^<\beta^2$, the wall
accelerates toward inertial observers, but only those close
to the wall are hit by it
(Fig.~\ref{Fig:dS4OutsideUltra}). 
If $\chi^2=\beta^2$, then
the accelerated expansion  
of the bubble is exactly cancelled by the self-expansion
of de~Sitter space, and the wall appears to be static 
(Fig.~\ref{Fig:dS4OutsideEks}). 

To get the complete space--time one must glue the two sides
of the wall. In the non-extreme case one must glue two 
bubble interiors; by gluing together two copies of 	 
the Minkowski interiors of Fig.~\ref{Fig:M4InsideNonUltra}
by identifying points from each of the 
diagrams along the walls world volume, one gets the maximal
extension of Vilenkin's \cite{Vilenkin:PLB83}
Minkowski--Minkowski wall.
This two-centered 
space--time is an example of a inhomogeneous closed universe with
eternal expansion.

Note again that the domain walls  with  de~Sitter space-times 
involve  vacua  
where the supersymmetry is (spontaneously) broken.   
Such vacuum domain walls always fall into a 
class of non-extreme 
($\mathbb{Z}_2$-symmetric walls) or ultra-extreme walls 
(false vacuum decay walls).  For more details 
about 
vacuum domain walls with 
de~Sitter space-times,  we refer the reader
to Ref.~\cite{BKT:PRD87}.

\subsection{Stability}

There is a large literature on the instability of Cauchy horizons
in Kerr and Reissner--Nordstr{\"o}m black holes \cite{SP:IJTP73,%
GSNS:PRD79,GNSS:PRD79,MZS:PRD79,NS:JETP80,Hiscock:PLA81,%
PI:PRD90,Ori:PRL91,BP:CQG92,MM:CQG92,Yurtsever:CQG93,KR:GRG93}.
In the generic case, 
such horizons are turned into singularities and
one could therefore suspect that the Cauchy 
horizons of the domain walls
have the same kind of instability. This issue has been
analyzed by Helliwell and Konkowski \cite{HK:PRD95} 
who 
considered the effect of adding 
infalling and/or outgoing null dust to the space--time. 
They
concluded that---%
except for one world line 
where 
a
scalar curvature
singularity is formed---%
the Cauchy horizons of
the anti--de~Sitter
domain wall space--times and vacuum bubbles
are stable under 
such perturbations. 
Further research on the stability of vacuum domain wall
space--times
is needed. 

The physics of vacuum  
domain walls also has a bearing upon the
stability of supersymmetric vacua. Their stability 
against decay into other
supersymmetric vacua
was 
first
shown perturbatively in the
gravitational constant expansion
\cite{Weinberg:PRL82}
and then by a full non-perturbative 
calculation in Ref.~\cite{CGR:NPB93}.
In addition, supersymmetric Minkowski vacua are known to be
absolutely stable 
\cite{CGR:NPB92}.

           \addtocounter{equation}{-\value{equation}}
%

%
           \section{Dilatonic domain walls}
\label{Sect:DilatonicWalls}

The existence of dilatons is a generic feature
of  unifying theories, including 
certain classes of supergravity theories, Kaluza--Klein  
theories, 
and effective theories from superstrings. 
The term `dilaton' is here used as a generic name for a  scalar
field without self-interactions that couples to the matter
sources, \ie\ the  potential of scalar
matter  fields as well as the kinetic 
energy of gauge fields.
Through this coupling 
it modulates the overall strength of such interactions (see
Section~\ref{Sect:Supergravity} for the $N=1$ supergravity case).
In the low-energy effective action of  super\-string theory, the  
dilaton
plays an essential r{\^o}le 
for the ``scale-factor duality'' \cite{Veneziano:PLB91}, 
which has been taken as an indication of a ``dual
pre-big-bang'' phase 
as a possible alternative to the
initial singularity of the standard cosmological model 
\cite{GV:APP93}.  The 
dilaton is believed to play a crucial r{\^o}le in
dynamical supersymmetry 
breaking\footnote{For a review see, \eg\ Ref.~\cite{Quevedo:95}.}
as well.
Moreover, in theories with  dilaton field(s)  
topological defects in general,
and black holes in particular, have a space--time structure that is
drastically changed compared to the non-dilatonic ones. 
Namely, since
the dilaton couples to the matter
sources, \eg\ the  charge of the 
black hole, or it modulates the strength
of
the matter interactions, it in turn 
changes the  nature of the space--time.
In the past, charged  dilatonic black holes have been studied
extensively (for review see  Ref.~\cite{Horowitz:Trieste92} and  
references
therein).
It is therefore of considerable interest to
generalize 
the vacuum domain wall solutions 
by including the dilaton, thus
addressing 
the
nature of space--time in the domain wall background with a
varying dilaton field. 

Such configurations may be of specific
interest in the study of  
domain walls in the early universe as they may
arise in fundamental 
theories that include the  dilaton, in particular in
an effective theory from  superstrings.  In addition,  the nature of
ultra-extreme dilatonic  domain walls, which describe 
false vacuum decay,  
in  
basic theories  that contain one (or more) dilaton fields
is
of importance. 

In the present  section we 
investigate 
dilatonic
domain walls.   First, within  $N=1$ supergravity 
coupled to a linear supermultiplet,
we give the  extreme dilatonic
solutions  to the case  with  an  arbitrary
separable  dilaton K\" ahler potential (\ref{Eq:Kahlerpota}).
These  solutions satisfy 
Killing spinor equations (\ref{Eq:KillingSpinor})  for the static, supersymmetric walls.
Since the form of the separable 
K{\"a}hler potential is kept arbitrary, the analysis applies
to  special cases  such as: $N=1$ 
supergravity theory coupled to a linear supermultiplet  
(see Section~\ref{BPL}) where
the 
dilaton $\phi$ corresponds 
to the scalar component of the  linear
multiplet and couples with an exponential coupling 
$e^{2\sqrt\alpha\phi}$ to the
potential of the matter scalar fields 
(see Sections~\ref{BPL} and \ref{Sect:Lagtopdef}) as
well as  examples of  the  
self-dual
case, where   the K{\"a}hler potential  has  
an extremum for a finite dilaton value.
We also comment on the  effects of a dilaton mass, 
which in basic theory can be
induced as a non-perturbative effect. Such a 
mass (or any other attractive self-interaction)\
does not alter the space--time  sufficiently  to remove
the naked singularity.

The first set of extreme  dilatonic 
domain wall solutions \cite{Cvetic:PRL93,CY:PRD95,Cvetic:PLB94}  
is reviewed in Section~\ref{Sect:Explicit}.
Their space--time structures  depend crucially
on the value of the coupling $\alpha$ 
of the dilaton to the matter potential (\ref{susypota}).
For $\alpha\le 1$ there is a
planar null singularity, while for $\alpha>1$ the 
singularity is \emph{naked}.  

The major part of this section involves a review of the space--time   
for  non-extreme  and
ultra-extreme  dilatonic domain walls  
in the thin wall
approximation (Section~\ref{Sect:Non/Ultra}). 

These walls are  generalizations
of the non- and ultra-extreme
vacuum domain walls discussed in Section~\ref{Sect:VacuumWalls}, but 
now, only numerical solutions have been obtained \cite{CS:PRD95}.
For this reason,  the analysis can be done
only for walls for which the boundary 
conditions  for the dilaton 
field and the metric can be specified uniquely at the
 wall surface. 
Nonetheless, such cases include the physically interesting
example of  
Type~I walls, which interpolate 
between Minkowski space--time with a constant
dilaton  value and  
a new type of space--time with varying dilaton, 
as well as reflection-symmetric
(non-extreme)\ walls.

The nature of the  space--time for 
non- and ultra-extreme dilatonic domain walls, 
which  possesses naked singularities,
poses serious
constraints on the phenomenological viability of 
theories with dilaton
fields, including a large class of $N=1$ supergravity theories
as well as the perturbative
effective
low energy
theory from superstrings.

Interestingly, the
space--time induced in the dilatonic domain 
wall backgrounds can  also be related to 
certain cosmological solutions by 
a complex coordinate transformation where 
$z$ is replaced with a cosmic time coordinate
and where the potential changes sign (Section~\ref{Sect:Cosmology}).

\subsection{Extreme dilatonic walls}
\label{Sect:Explicit}

In Section~\ref{Sect:SUSY_Embedding}
we spelled out the formalism  for an embedding of   
extreme domain walls into the corresponding tree level $N=1$  
supergravity theory as specified in  Section~\ref{Sect:Supergravity}.
Extreme domain walls are static, planar configurations
interpolating between \emph{supersymmetric} minima of
the corresponding supergravity potential. Such configurations satisfy  
the
Killing spinor  equations (\ref{Eq:KillingSpinor}) for \emph{any thickness} of the  
wall, and the
energy density  of the wall saturates the corresponding Bogomol'nyi  
bound.
In the thin wall approximation  the Einstein-dilaton system
outside the wall is described  by the
formalism spelled out in Section~\ref{Sect:IsoWalls}
with the non-extremality parameter $\beta
=0$.  

The  $N=1$
supergravity Lagrangian, described in Section~%
\ref{BPL} contains the 
(gauge neutral)\
chiral-superfield $\mathcal{T}$,  
whose scalar component
$T$ is  responsible for the formation of the wall. In addition,  
there is a chiral superfield $\mathcal{S}$,
which has no superpotential and whose 
K{\"a}hler potential decouples 
from the one of $\mathcal{T}$.  
In turn, the scalar component $S$ of the chiral 
superfield $\mathcal{S}$ 
acts as the dilaton
field, which couples to and thereby modulates the
effect of the matter potential. 

The vanishing of the Killing spinor 
equations (\ref{Eq:KillingSpinor})\
for the   
above bosonic field configuration yields 
first-order 
differential 
equations for the metric
coefficient $a(z)$ 
and the matter fields
$T(z)$ and $S(z)$
as well as the constraint (\ref{Eq:KillSp})\ on
the spinor $\epsilon$, as spelled out in Section~\ref{Sect:SUSY_Embedding}.
In the following   we shall  concentrate 
on the explicit form of the extreme   
solutions  for different special cases in the thin wall 
approximation\@.\footnote{Explicit numerical solutions of 
Eqs.~(\ref{boge})\ 
for  a wall of any thickness  
have  the same qualitative features outside the
wall region.}

We shall first  summarize%
\footnote{Special cases with $\alpha=1$,   
parameterizing the dilaton coupling in an effective theory from  
superstring were found in Ref.~\cite{Cvetic:PRL93}, and $\alpha=2,  
3,\ldots$, motivated  by no-scale supergravity theories, were 
studied in  
Ref.~\cite{CY:PRD95}.}
the   results for  extreme  walls
with a K{\"a}hler potential 
(\ref{Eq:kald})
for $S$:
$K_{\text{dil}}=-\alpha\ln (S+S^*)$  
\cite{Cvetic:PLB94}.
Then, we shall study extreme domain walls with self-dual 
$K_{\text{dil}}$,  \ie\   
$\left. K_{S}\right|
{\mbox{\raisebox{-1.0ex}{{\tiny{$\!S'\!$}}}}} 
=0$ 
for some $S'$. An example 
of the latter class 
corresponds to a solution of a
theory  with  a strong--weak (dilaton)\
coupling symmetry, \ie\ 
$\mathrm{SL}(2,\mathbb{Z})$ invariance of
the dilaton coupling.
Section~\ref{Sect:ExExp} 
presents the extreme solutions in theories with
an
exponential dilaton coupling. Their physical properties such as
the Hawking 
temperature associated
with the horizons and the
gravitational mass of the singularities are 
also discussed in Section~\ref{Sect:Temperature}.
Section~\ref{Sect:Self-dual} 
comments on the
self-dual extreme solutions.

\subsubsection{Extreme solutions for exponential dilaton coupling}
\label{Sect:ExExp}

Let us first consider the gravitational properties of extreme
super\-symmetric domain walls  
with $K_{\text{dil}}=-\alpha\ln(S+S^*)$, which has been
worked out in Refs.~\cite{Cvetic:PRL93,Cvetic:PLB94}.
The energy density of the
wall is of the
form:
\begin{align}
\sigma_{\text{ext}} &=  2^{1-{\alpha \over 2}}
\e^{\sqrt\alpha
\phi_{0}}\left| \left(\e^{{K_{\text{matt}}} \over 2}
W\right)_{z=0^+} \pm \left(\e^{{K_{\text{matt}}} \over 2}
W\right)_{z=0^-}\right| \nonumber\\
&\equiv 2(\chi_1\pm\chi_2) 
\label{Eq:sigp}
\end{align}
where we have chosen the boundary condition for  
$a(0)=0$ and $\phi(0)=\phi_0$.
The sum
corresponds to the Type~II wall
and the difference to the Type~III wall.
It is understood that the coordinates are chosen so that 
$\chi_2\leq \chi_1$.  With the  choice
\[
\chi_{1,2}=2\left|\e^{K/2}W\right|_{z=0^\mp}, 
\]
the solution of Eqs.~(\ref{boge})\
are satisfied with the following  choice for 
the sign factor $\zeta$: 
on the $z<0$ side,  $\zeta=1$, while on the 
side where $z>0$, $\zeta=-1$ for 
Type~II walls and   $\zeta= 1$ for Type~III walls. 
Type~I corresponds to the case
where $|W|$ is zero on one side of the wall
and nonzero on the other.

The  Killing spinor 
equations (\ref{Eq:KillingSpinor})\footnote{See 
Refs.~\cite{CS:PRD95,Cvetic:PLB94}  
for the explicit form of Bogomol'nyi equations (\ref{boge})\
applied to  
this particular case.} 
outside the wall region, \ie\ when
$\partial_z T\sim 0$, 
are the same as those of
the second-order equations (\ref{Eq:DynamicalEq})--(\ref{Eq:Constraint})\ 
with  $\beta=0$ and  the effective
dilaton potential  of the type of Eq.~(\ref{Eq:Potential}), 
where  
\[
f(\phi)= \e^{2\sqrt{\alpha}\phi}, \ \
V_0(\tau_{1,2})=
-(3-\alpha)2^{-\alpha}\left.\left( 
\e^{K_{\text{matt}}}|W|^2\right)\right|_{T_{1,2}},\ \
\widehat V (\phi)=0
\]
on either side of the wall. 

The  value of parameters $\chi_{1,2}$  in
Eq.~(\ref{Eq:sigp})\ on either
side of the wall is related to $V_0(\tau_{1,2})$ in the 
following way
\begin{align}
\chi_{1,2} &\equiv 2^{-{\alpha\over 2}} \e^{\sqrt\alpha
\phi_0}\left( \e^{{K_{\text{matt}}\over 2}}|W|\right)|_{T_{1,2}} \nonumber\\
& =  
\e^{\sqrt\alpha
\phi_0}\sqrt{-V_0(\tau_{1,2})/(3-\alpha)}. 
\label{Eq:chidetermined}
\end{align}
Note that $\alpha=3$
corresponds to  the point where $V_0$ changes sign.

\begin{figure}[hbt]
\centering{\epsfig{file=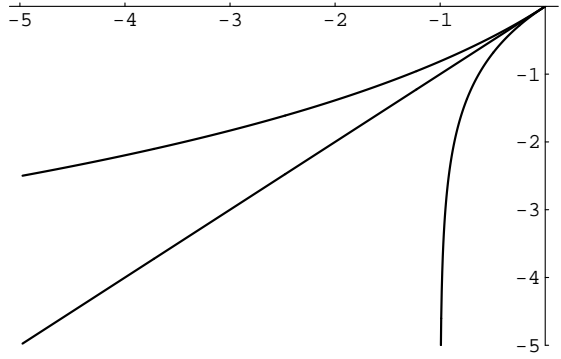,%
                   height=5cm,%
                   clip=}}
\caption{The metric function $a(z)$ versus $z$ in units of $\chi$ for
extreme solutions with $\alpha=0.5$, $\alpha=1$, and $\alpha=2$,
respectively. Solutions in models with $\alpha>1$ has a naked singularity
at a finite value of $z$.}
\label{Fig:Ext_A}
\end{figure}

The explicit 
solution on either side of the wall is of the form  
\begin{equation}
 a = \left\{
     \begin{array}{rl}
 \chi_1 z, &   \;
 \ \ z<0\\ 
\mp \chi_2 z,     &   \;
\ \ z>0  
     \end{array}
     \right.
\label{stringextreme}
\end{equation}
if $\alpha=1$ and
\begin{equation}
a= \left\{
     \begin{array}{ll}
(\alpha-1)^{-1}\,\ln [1+\chi_1 (\alpha-1 )z ], &  
\ \  z<0\\ 
(\alpha-1)^{-1}\,\ln [1\mp\chi_2 (\alpha-1 )z ], & 
\ \  z>0 
     \end{array}
     \right.
\label{Eq:extremesolutions}
\end{equation}
if $\alpha\neq 1$.
The upper and lower signs  of the solutions 
(\ref{Eq:extremesolutions})\ 
correspond to the Type~II and 
Type~III solutions, respectively. 
Type~I corresponds to the special case with $\chi_2=0$, \ie\  
those are
solutions with Minkowski space--time  ($a=0$)\
 and a  constant dilaton on  the $z>0$
side of the wall.
The Bogomol'nyi equations also imply that
\[
\phi = -\sqrt{\alpha}a
\]
everywhere in the domain wall background.  Consequently, 
these solutions are represented by  straight lines in the
$(a',\phi')$ phase diagram.
For $\alpha>1$ the domain walls have a naked (planar)\
singularity  at
$z=-1/[\chi_1(\alpha-1)]$  and for Type~II 
walls at $z=1/[\chi_2(\alpha-1)]$
as well.  For $\alpha \le 1$ 
the singularity becomes null, \ie\ it
occurs at $z=-\infty$ and for 
Type~II walls at $z=\infty$ as well. Note that
for $\alpha< 1$ Type~III walls have a coordinate singularity  at
$z=1/[\chi_2(1-\alpha)]$.  
Thus, extreme walls with the 
``stringy'' coupling $\alpha=1$ act as an intermediary 
between the extreme dilatonic walls 
with naked singularities and those with 
singularities covered by a horizon.
The behaviour of $a$ for different values of $\alpha$ 
is plotted in Fig.~\ref{Fig:Ext_A}. 
The  case  with 
$\alpha = 2$ had earlier been found in the 
guise of a static plane-symmetric space--time
with a
conformally coupled
scalar field \cite{VS:PRD83,AVS:JMP83}. 
In
Ref.~\cite{GS:PLA92} this space--time 
was shown to be induced by a domain wall.

\subsubsection{Temperature and gravitational mass per area}
\label{Sect:Temperature}

Static domain wall configurations 
with   space--time singularities are only possible
if there is an exact cancellation of the contributions to the
gravitational mass coming from the wall, the dilaton field,
and from the singularity. 

In  the case of  an
extreme Type~I wall, \ie\   
a
static dilatonic wall with a non-zero vacuum energy on
one side (say, $z<0$  and $\chi_1\ne 0$)\
and a Minkowski space on the other
($z>0$ and $\chi_2=0$),  one can  
employ the 
concept of gravitational mass per area 
\cite{CGS:PRD93}  
as 
discussed in Section~\ref{Sect:VacuumWalls}.
It 
corresponds to  a plane-symmetric version of
Tolman's \cite{Tolman:PR30} gravitational mass in the
spherically symmetric case.       
The contribution from all sources outside the horizon 
(or the naked singularity)\
can be 
expressed as
\[
\Sigma (\infty) = \left.2 a'\right|
{\mbox{\raisebox{-1.2ex}{{\tiny{$\!\infty$}}}}} 
 - \left.2 a'\right|
{\mbox{\raisebox{-1.2ex}{{$\!$\tiny{horizon}}}}}. 
\]
On the Minkowski side $a'\equiv da/dz=0$, and so 
$\left.a'\right|
{\mbox{\raisebox{-1.2ex}{{\tiny{$\!\infty$}}}}}=0$. 
Hence, $\Sigma$ is determined by the value of $a'$ 
at the horizon: 
\begin{itemize}
\item
For $\alpha<1$, 
the gravitational mass per area 
outside the horizon vanishes. 
Since the total mass vanishes,
there is also no mass beyond the horizon.
Accordingly, 
the 
Hawking temperature 
associated 
with this horizon also vanishes \cite{CS:PRD95}.

\item
For $\alpha=1$, the gravitational mass per area
from sources
outside the
horizon is negative: $\Sigma (\infty)= -2\chi_1$. In order
to have a vanishing total mass,
the mass of the singularity must be 
$\Sigma_{\text{singularity}}  = 2\chi_1$, but  
a proper mathematical description would 
require a distribution-valued 
metric.
We shall not pursue this issue further here,
but we note that a 
source at the singularity of the 
Schwarzschild metric
has recently been identified in terms of a 
distributional energy-momentum tensor
\cite{BN:CQG93}. 
The Hawking temperature for the extreme
Type~I wall with $\alpha=1$ is finite: 
$T=\chi_1/2\pi$ \cite{Cvetic:PLB94}. 

\item
For $\alpha>1$, the mass per area outside the singularity 
is negative and infinite. Accordingly, there must be an 
infinite positive mass in the singularity, which  is then 
a singularity even in  
a distributional sense. This naked
singularity also has an infinite Hawking temperature.
\end{itemize}

\begin{figure}[htb]
\centering{\epsfig{file=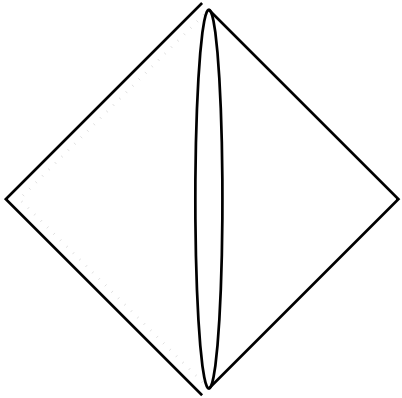,%
                   height=5.3cm,%
                   clip=}}
\caption{Penrose--Carter diagram in the $(z,t)$ plane for an extreme
Type~I dilatonic domain wall with $0<\alpha\leq 1$.
The thin arc corresponds to the wall separating the semi-infinite
Minkowski space--time and the space--time with a varying dilaton field
and the null singularity which coincides with the horizon (the double line).}
\label{Fig:DilExtremeI} 
\end{figure}

\begin{figure}[htb]
\centering{\epsfig{file=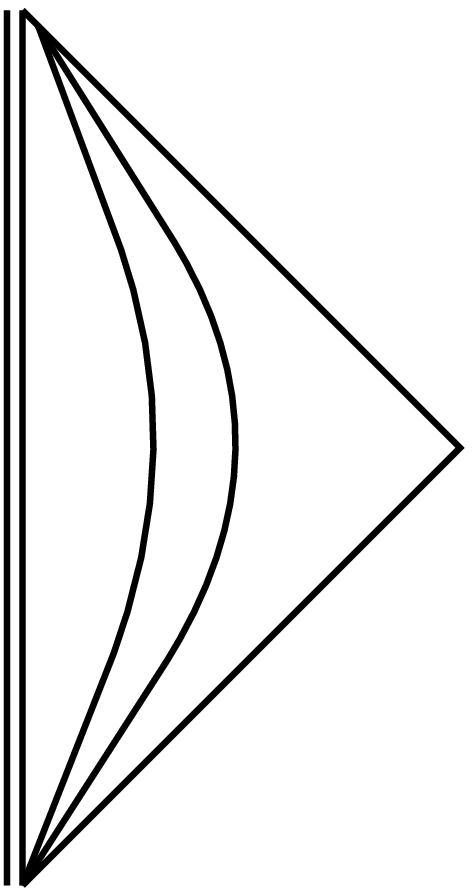,%
                   height=5.3cm,%
                   clip=}}
\caption{Penrose--Carter diagram in the $(z,t)$ plane for an extreme
Type~I dilatonic domain wall with $\alpha> 1$.
The arcs correspond to the wall separating the semi-infinite
Minkowski space--time and the space--time with a varying dilaton field
and the naked planar singularity (the double line).}
\label{Fig:DilExtremeNI} 
\end{figure}

\subsubsection{Self-dual dilaton coupling}
\label{Sect:Self-dual}
We now address  the case of  a
self-dual dilaton K{\"a}hler potential  
$K_{\text{dil}}$, namely,
 $K_{\text{dil}}$  has an extremum 
($K_S|_{S'}\equiv\partial K_{\text{dil}}/\partial S|_{S'}=0$) for
some  finite $S =S'$.   In the thin wall approximation, 
the 
Bogomol'nyi  equations (\ref{boge})\   
are of course satisfied with the  
choice $S=S'$ 
outside the  
wall region. 

In order for  such a    solution to be  
stable, one would have to  
show   that
as one moves down the wall, 
the dilaton
reaches  the point $S=S'$, 
and that from then on it would remain constant,   
\ie\ the point  $S=S'$ is an attractive fixed point. Such 
solutions would in turn reduce to singularity-free space--times of  
vacuum
domain walls.

However,  such  a class of extreme solutions is shown \cite{CS:PRD95}  
to be unstable within $N=1$ supergravity theory. In  
particular,  if at $z\sim 0$, 
$S(0)= S'+\Delta(0)$, with $\Delta(0)$ being an 
infinitesimal perturbation from the 
self-dual point, $S=S'$, then  
$\Delta(z)$ grows indefinitely as $z\rightarrow
-\infty$ and thus the solution with a 
constant dilaton outside the wall region is 
not dynamically stable \cite{CS:PRD95}.

\subsection{Non- and ultra-extreme solutions}
\label{Sect:Non/Ultra}
In this section we shall analyse 
the non-extreme and ultra-extreme
solutions. These are  
solutions that are not supersymmetric. They
correspond to the domain wall backgrounds  with moving wall boundaries.
Unlike extreme solutions, which have 
supersymmetric embeddings and 
where solutions can be given 
for any thickness of the wall, 
 non- and
ultra-extreme solutions  have only been
obtained  \cite{CS:PRD95}
in the thin wall approximation, employing 
 the formalism spelled
out in  Section~\ref{Sect:ThinWallFormalism}.

In the following subsections we 
review  the non-extreme solutions for
the exponential dilaton potential,  
as well as the case with a self-dual
dilaton potential. We 
shall also add a mass term  for the dilaton
field.  Since only   numerical solutions for non-extreme  walls
have been 
found \cite{CS:PRD95},  
we confine  
the analysis to the non-extreme Type~I walls (with
Minkowski space--time on one side of the wall)\
and reflection symmetric
solutions,  for which the boundary conditions  on either
side of the wall can be specified uniquely. 
In subsection~\ref{Sect:M4-dil} the boundary conditions
for the non- and ultra-extreme Type~I walls
are written down explicitly, and the field equations are reduced
to a first-order system. The results of the numerical integrations
are presented and discussed. Section~\ref{Sect:Reflection}
contains solutions for the reflection-symmetric cases,
and in Section~%
\ref{Sect:Self-Dual}
it is pointed out that
the singularity-free self-dual 
dilatonic domain walls are dynamically unstable. 
Finally, in Section~%
\ref{Sect:Self-interaction} 
the effects of dilaton self-interactions are studied.

\subsubsection{Walls with Minkowski space--time on one side} 
\label{Sect:M4-dil}

We now consider the 
case where the   
dilaton 
potential outside the wall region  has the  form 
specified in Eq.~(\ref{Eq:Potential})\
with $f= e^{2\sqrt\alpha \phi}$ 
and $\widehat V(\phi)=0$. 
Let the wall be non- or ultra-extreme 
with a non-vanishing $V_{0}$ and a running dilaton on one side
($z<0$)\ and a 
Minkowski space with $V_{0}=0$ and 
a constant dilaton on the other 
($z>0$).
According to the results of 
Section~%
\ref{Sect:ThinWallFormalism}, the boundary conditions 
are 
\begin{equation}
\left. \begin{array}{ll}
\left.\phi'\right|
{\mbox{\raisebox{-1.2ex}{{\tiny{$\!0^{-}$}}}}} 
= -\frac{1}{2}\sqrt{\alpha}\sigma,\ \ & 
\left.a'\right|
{\mbox{\raisebox{-1.2ex}{{\tiny{$\!0^{-}$}}}}} 
=  \frac{1}{2}\sigma -\beta \\
\left.\phi'\right|
{\mbox{\raisebox{-1.2ex}{{\tiny{$\!0^{+}$}}}}} 
 = 0\, \ \ &
\left.a'\right|
{\mbox{\raisebox{-1.2ex}{{\tiny{$\!0^{+}$}}}}} 
 = -\beta .
       \end{array}
\right.
\label{Eq:BoundaryM}
\end{equation}
$\beta<0$ and $\beta>0$
represent an ultra-extreme and a non-extreme wall, respectively.   
Without loss
of generality  we have also chosen $\phi|
{\mbox{\raisebox{-1.2ex}{{\tiny{$\!0$}}}}}=0$ and normalized the metric
coefficient $a|
{\mbox{\raisebox{-1.2ex}{{\tiny{$\!0$}}}}}=0$. The choice $\phi|
{\mbox{\raisebox{-1.2ex}{{\tiny{$\!0$}}}}}=\phi_0\ne 0$ would  
correspond to
the rescaling  $V_0\rightarrow \e^{2\sqrt\alpha\phi_0}V_0$.

At the boundary $z=0^{-}$, 
Eq.~(\ref{Eq:Constraint})\ gives
\begin{equation}
V_{0}- 3\beta\sigma + (3-\alpha) \left(\frac{\sigma}{2}\right)^2 =0.
\end{equation} 
For $\alpha\neq 3$ the above equation reduces to
\[
\sigma = \frac{2}{3-\alpha}\left[ \sqrt{ 9 \beta^2 +
(3-\alpha)^2\chi_1^2}+3\beta \right], 
\]
where we have used the fact that 
$\chi_1^2=-V_0(\tau_1)/(3-\alpha)$, as found in  
Eq.~(\ref{Eq:chidetermined}).
In the case $\alpha=0$, $\beta \neq 0$, 
and $\chi_1^2=|V_0|/3$, 
one recovers the result from the 
non- and ultra-extreme anti-de~Sitter--Minkowski 
walls  without a dilaton, 
and if $\alpha = 0$, $\beta>0$, and $\chi_1=0$, 
one recovers the 
dilaton-free
non-extreme Minkowski--Minkowski walls \cite{Vilenkin:PLB83,CGS:PRD93}.

For $\alpha=3$, one finds 
\[
V_{0}=3\beta\sigma, 
\]
which indicates that in the non-supersymmetric 
case with
$\alpha=3$, unlike in the supersymmetric case where $V_0=0$,
the potential 
itself has a non-zero value. 

\begin{figure}[hbt]
\centering{\epsfig{file=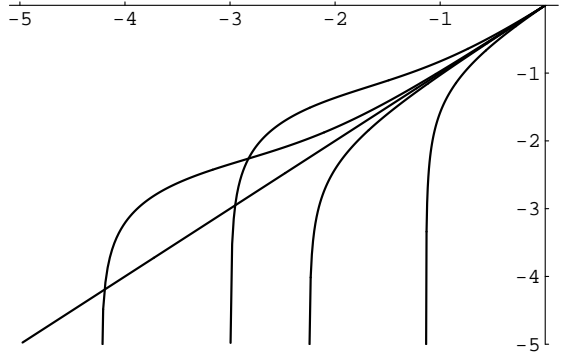,%
                   height=5cm,%
                   clip=}}
\caption{The metric function $a(z)$ versus $z$ in units of $\chi$ 
with
with $\alpha=1$ for different values of $\beta$.
Starting from the left at the bottom of the figure where
$a=-5$, the curves correspond to 
$\beta=0$, $\beta=-0.01$, $\beta=-0.1$, $\beta=0.01$
and $\beta=0.1$.} 
\label{Fig:Non_A}
\end{figure}

\begin{figure}[hbt]
\centering{\epsfig{file=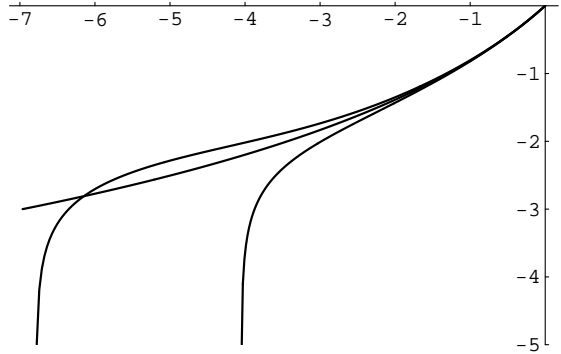,%
                   height=5cm,%
                   clip=}}
\caption{The metric function $a(z)$ versus $z$ in units of $\chi$ 
with
with $\alpha=1/2$.
The curve starting in the middle
corresponds to the extreme solution.
The non-extreme case with $\beta=0.01$ becomes singular shortly after $x=-4$.
The third curve corresponds to the ultra-extreme case with $\beta=-0.01$. It also
ends in a singularity.}
\label{Fig:Non_A2}
\end{figure}

The field equations  have been integrated numerically (for further
details see  
Ref.~\cite{CS:PRD95}). 
The  conformal factor goes to zero faster than in the extreme
space--times \emph{both} in the non- and 
ultra-extreme cases (see Figs.~\ref{Fig:Non_A} and
\ref{Fig:Non_A2}). This is the case for \emph{all}
values of $\alpha$ as long as $\beta\ne 0$. 
As illustrated in Fig.~\ref{Fig:Non_A}, 
when  $|\beta|$ is increased, the  conformal factor decreases even  
faster.
\emph{Such domain walls  thus always exhibit naked singularities.}

In order to  understand  these surprising  results, 
it is instructive to
look at the evolution of the dilaton.
The extreme solutions with $0<\alpha<3$ are characterized by a delicate 
balancing of the ``kinetic'' and ``potential'' energies.
As soon as the supersymmetry is broken, \ie\  
$\beta\ne 0$, the dilaton
speeds away along  its potential;
the kinetic energy eventually  becomes dominant in both cases and is  
thus  
responsible for the appearance of a naked singularity.

\begin{figure}[hbt]
\centering{\epsfig{file=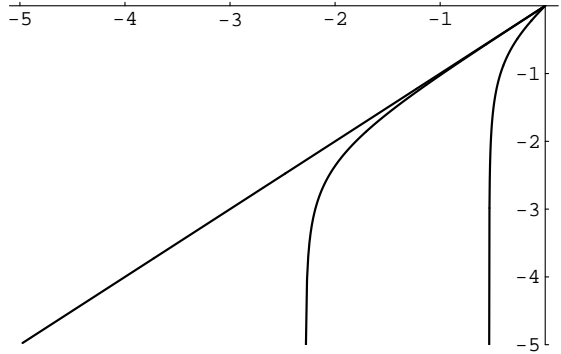,%
                   height=5cm,%
                   clip=}}
\caption{The metric function $a(z)$ versus $z$ in units of $\chi$ for
a reflection-symmetric
wall with $\alpha=1$. The straight line corresponds to the extreme solution.
The other two curves represent non-extreme walls with $\beta=0.1$ 
and $\beta=1$.} 
\label{Fig:Non_Ref_A}
\end{figure}

\subsubsection{Reflection-symmetric walls} 
\label{Sect:Reflection}

Now we consider  a reflection-symmetric 
non-extreme wall (a special case of Type~II
walls)\
with non-vanishing $V_{0}$ and a 
running dilaton\@.\footnote{Of course, there  
are no reflection-symmetric ultra-extreme walls.} 
The potential is taken to be that of 
Eq.~(\ref{Eq:Potential})\ with $f(\phi)=e^{2\sqrt{\alpha}\phi}$
and $\widehat V(\phi)=0$. 
According to Section~\ref{Sect:ThinWallFormalism}, the boundary conditions 
are 
\[
\left. \begin{array}{l}
\left.\phi'\right|
{\mbox{\raisebox{-1.2ex}{{\tiny{$\!0^{-}$}}}}} 
=
-\left.\phi'\right|
{\mbox{\raisebox{-1.2ex}{{\tiny{$\!0^{+}$}}}}} 
 = -\frac{1}{4}\sqrt{\alpha}\sigma,\\  
\left.a'\right|
{\mbox{\raisebox{-1.2ex}{{\tiny{$\!0^{-}$}}}}} 
=
-\left. a'\right|
{\mbox{\raisebox{-1.2ex}{{\tiny{$\!0^{+}$}}}}} 
=
 \frac{1}{4}\sigma . 
       \end{array}
\right.
\]
With these boundary conditions,
Eq.~(\ref{Eq:Constraint})\ gives
\[
-3 \beta^2 
+
V_{0} + (3-\alpha) \left(\frac{\sigma}{4}\right)^2 =0.
\]
If $\alpha\neq 3$, we find by use of
Eq.~(\ref{Eq:chidetermined})\ that 
\[
\sigma = 4\sqrt{ \chi_1^2 + 3\, (3-\alpha )^{-1}\beta^2}.
\]
For $\alpha=0$, this expression reduces to the one found for
reflection-symmetric vacuum domain walls \cite{CGS:PRD93}.
If $\alpha=3$, then $\sigma$ remains undetermined from this 
expression, but
\[
V_{0}= 3\beta^2. 
\]
This result again indicates that in  the non-supersymmetric case  with
$\alpha=3$, the potential
itself is modified to be non-zero. 
Thus, these walls are different from reflection-symmetric
domain walls in the background of a
Zel'dovich fluid \cite{Vuille:PRD91}  or, which is equivalent, 
domain walls minimally 
coupled to a scalar field with no effective potential  
\cite{Wang:PLB92}.
Yet, in all these cases one encounters
naked singularities.
The boundary 
conditions for reflection-symmetric non-extreme walls
are 
different from those
of the wall adjacent to Minkowski space. In this case the
singularity is further away from the wall (compare
Figs.~\ref{Fig:Non_A} and 
\ref{Fig:Non_Ref_A}); however,   as for the Type~I
non-extreme walls, the
reflection-symmetric solutions always have naked singularities as well.

\subsubsection{Self-dual dilaton coupling}
\label{Sect:Self-Dual}
We now review the case when the  dilaton coupling $f(\phi)$ is  
self-dual. Namely,  for a finite
$\phi=\phi'$, $f(\phi)$ satisfies:
\begin{equation}
\left.{{\partial f}\over{\partial \phi}}\right|
{\mbox{\raisebox{-1.7ex}{{\tiny{$\!\phi'$}}}}} 
=0 
\label{fAnsatz}
\end{equation}
and that $\widehat V(\phi)=0$ (see Eq.~(\ref{Eq:Potential})). 

Note that the equation for  the dilaton 
is of the form (see Eq.~(\ref{Eq:ConservationLaw}))\ 
\[
2{\phi}'' + 4 a '\phi'
+  e^{2a}\,{{\partial f(\phi)}\over{\partial\phi}} V_{0}=0,  
\]
with $\phi'=d\phi/dz$.  With  the boundary conditions 
$\left.\phi\right|
{\mbox{\raisebox{-1.2ex}{{\tiny{$\!0$}}}}}=\phi_0$ and 
$\left.\phi'\right|
{\mbox{\raisebox{-1.2ex}{{\tiny{$\!0$}}}}} 
= 0$, 
the solution of the field equations corresponds 
to a dilaton frozen at $\phi=\phi_0$.  
The global space--times
of these solutions are then identical to those
of the vacuum domain walls 
\cite{CGS:PRD93,CDGS:PRL93}.

When addressing the stability of the constant dilaton
$\phi(z)=\phi_0$ solutions for extreme, non- and ultra-extreme  
solutions,
one adds to 
$\phi(0)=\phi_0$   
an
infinitesimal virtual displacement $\delta(0)$. 
Supersymmetric
embedding of an extreme solution  with a constant  
(complex)\ dilaton  $S=S'$  is 
\emph{always} unstable.
One  can also  show an  
instability of such solutions  by   
solving  the corresponding differential equation for 
$\delta$,
directly, 
\ie\
without reference to  the effective  dilaton
potential restricted by $N=1$ supergravity 
theory\@.\footnote{The extreme solutions with a real dilaton field $\phi$
can be viewed as corresponding to  
a special supersymmetric embedding, which renders
the imaginary part of the complex field $S$ constant.} 
Thus, the extreme self-dual
solutions with a frozen 
dilaton are always dynamically unstable.
The origin of  the instability of
extreme  solutions  
may be related to an infinite extent of such planar
configurations.

On the other hand,  the  non-extreme  and ultra-extreme solutions
($\beta\ne 0$)  turn out  to be    
\emph{stable} under this perturbation \cite{CS:PRD95},
and may thus be ``phenomenologically'' stable.

\subsubsection{Domain walls 
with a  massive dilaton}
\label{Sect:Self-interaction}

Let us now consider the situation where, 
supersymmetry breaking
due to non-perturbative effects, 
introduces an additional self-interaction term
in the dilaton potential. In general such a term is
of a complicated form. 
One might hope that  a dilaton self-interaction term, providing 
a mass for the dilaton, could stabilize the system
and keep the dilaton from running away. 
For the sake of simplicity we
consider  an additional self-interaction potential $\widehat V$ of
the form
\[
\widehat V = \lambda^2 \chi_1^2 \sinh^2 \omega \phi, 
\]
where $\lambda$ and $\omega$ are real constants.  
Note that in the cosmological 
picture, this  potential has the opposite sign.

Since $V(0)=V'(0)=0$, the boundary  conditions for the 
equations of motion at $z = 0^{-}$ for a wall 
adjacent to Minkowski space, remain as
in Eq.~(\ref{Eq:BoundaryM}).

In the non-extreme case adding  a mass term
forces  the dilaton to
 ``run'' in a way which causes the   appearance  
of
the naked singularity.
In the ultra-extreme case 
with
a dilaton self-interaction potential,   
the dilaton  also  produces
a naked singularity, in general. In the latter case the appearance of
the naked singularity  can only be avoided by a fine-tuned  
self-interaction 
potential whose fine-tuning would have to depend also on $\beta$. 

The same qualitative features take place 
in the reflection-symmetric case.

\subsection{Correspondence with a cosmological model} 
\label{Sect:Cosmology}

In this section 
we point out that the  Einstein--dilaton system
outside the  wall is  
equivalent to that of an Einstein--dilaton 
Friedmann--Lema{\^{\i}}tre--Robertson--Walker (FLRW)\ cosmology.
Formally, one can flip the wall into a spacelike hyperspace 
by
a complex coordinate transformation:
\begin{equation}
z\rightarrow \eta,\ \ \ \cosh\beta t\rightarrow i \beta r. 
\label{Eq:tocosmologyform} 
\end{equation}
If one regards the new coordinates
as real, then
$\eta$ becomes the conformal time and $r$ a spatial coordinate in a 
metric with opposite signature: $(-,+,+,+)$. Changing the 
sign of the metric implies a change of sign of the curvature 
scalar, 
$R$, and of all kinetic
energy terms in the Lagrangian (\ref{Eq:SUSYLagrangian}). 
Because the overall sign of the total Lagrangian is arbitrary,
one can change back
the sign of the metric,
if one at the same time changes the sign of
the potential $V(\phi)$.  Thus, the complex 
coordinate transformation (\ref{Eq:tocosmologyform})\ 
\emph{maps the domain wall system onto  a cosmological model 
having a potential with the opposite sign.}
As a result, 
the line-element takes the form 
\[
ds^2=  e^{2 a(\eta)}\left[ d\eta^2 - \frac{dr^2}{1+\beta^2 r^2} - 
r^2 d\Omega_{2}^{2}\right].     
\]
This is a FLRW line-element where  
the spatial curvature is $k=-\beta^2$.

The equivalence of the Einstein-dilaton system outside the wall  with  
the
dilaton-FLRW  cosmology  (by using the  
coordinate transformations 
(\ref{Eq:tocosmologyform}),
as well as identifying
$V(\phi)\rightarrow -V_{\text{c}}(\phi)$ and 
$\beta^2\rightarrow -k$) proves useful, because 
it allows us to 
carry over results from the corresponding  cosmological studies. 
In the cosmological  picture the domain wall
is a spacelike hyper-space. It could be
interpreted as representing  a  
phase transition taking place
simultaneously throughout the whole universe.
In our case, the boundary conditions at this hyper-space are fixed
by the 
boundary conditions
of the domain wall. 
In addition,  it is  useful to 
compare the Einstein-dilaton system outside the wall with the 
evolution  of corresponding well-known
perfect fluid
cosmologies and to compute the corresponding effective equation of  
state
for the dilaton. 
In terms of a perfect fluid description,
the energy-momentum tensor  is of the form:
\begin{equation}
\left. \begin{array}{lcl}
T^{\eta}_{\;\; \eta} &=&
  V_{\text{c}}(\phi) +
   {{{{{\dot \phi}}^2}} {{{e}^{-2\,a}}}}\\
T^{i}_{\;\, i} &=&
 V_{\text{c}}(\phi)- 
  {{{{{\dot \phi}}^2}} {{{e}^{-2\,a}}}},  
      \end{array}
\right.
\label{Eq:Tmunu}
\end{equation}
where $\eta$ is the conformal time and  
$\dot \phi=d\phi/d\eta$. 
Here the index $i$
refers to the three
spatial coordinates. 
Note that in the energy--momentum tensor (\ref{Eq:Tmunu})\ the sign of
$V_{\text{c}}(\phi)$ is 
reversed with respect to the potential of Eq.~(\ref{Eq:Potential}).
The expressions 
(\ref{Eq:Tmunu})\
correspond to an energy
density
\begin{equation}
\rho\equiv T^\eta_{\;\;\eta}=
\dot{\phi}^2 e^{-2 a}  
+
V_{\text{c}}(\phi) 
\label{rho}
\end{equation}
and a pressure
\begin{equation}
p\equiv -T^i_{\;i} = 
\dot{\phi}^2 e^{-2 a}   
-V_{\text{c}}(\phi)
\label{pressure}
\end{equation}
of a perfect fluid with a four-velocity 
$u^{\mu}=e^{-a}\delta^{\mu}_{\;\;\eta}$.
It is conventional to parameterize the equation 
of state of a perfect fluid
by the $\gamma$-parameter: $p = (\gamma-1)\rho$.  
The following values of $\gamma$ are singled out:
$\gamma=0$ corresponds to the equation of state of a 
cosmological constant;   
$\gamma=2/3$ is the equation of state of a cloud of randomly oriented
strings;  
$\gamma=1$ represents dust (non-relativistic cloud of particles); 
$\gamma=4/3$ is radiation (ultra-relativistic matter);   
and 
$\gamma=2$ corresponds to a Zel'dovich fluid (maximally stiff matter).  
All physical equations of state are confined to the range  
$0\le\gamma\le 2$.
This  is
also the range covered by a minimally coupled scalar field $\phi$:  
\begin{equation}
\gamma =  \frac{2 \dot\phi^2 e^{-2a}}{\dot\phi^2 e^{-2 a}   
+V_{\text{c}}(\phi)}. 
\end{equation} 
It has $\gamma=2$
if the kinetic energy dominates, and $\gamma =0$ if the potential
energy dominates.
Matter satisfying an equation of state with $\gamma <1$ 
has negative pressure.  If $\gamma <2/3$, then the 
repulsive gravitational effect of the negative pressure is
greater than the attractive gravitational effect
of the energy density. Matter obeying such 
an equation of state is therefore 
a source of repulsive gravity. An effective equation of state of this 
kind is a necessary ingredient in
inflationary universe models.

\subsubsection{Domain wall space-time as a cosmological solution}
\label{Sect:Correspondence}

The equivalence  of  the domain wall solutions on either side of the
wall  and a  class of cosmological solutions implies that
we are able to relate the  
above solutions to known solutions of inflationary cosmology
with exponential potentials \cite{Barrow:PLB87}, which were
later 
generalized to higher-dimensional
FLRW cosmologies \cite{BB:NPB88}. 
Properties  of general (extreme  and non-extreme)\ 
scalar field cosmological models and their 
corresponding phase diagrams were studied  in
Ref.~\cite{Halliwell:PLB87}\@.\footnote{Note that
in the cosmological picture, the extreme Type~I  vacuum domain wall
becomes a flat inflationary universe where the inflation 
(the wall-forming scalar field in the original picture)\ 
rolls down the inflation potential with just the right speed so that 
it
stops at a local maximum 
corresponding to a vanishing cosmological 
constant. At this point the universe also stops expanding.}

Note that after the substitution 
$z\rightarrow \eta$ and $V_0\rightarrow -V_{0 \text{c}}$
the Type~II solutions  
and Type~III solutions (on the $z>0$ side)\ correspond to
to contracting and expanding 
cosmological solutions,
respectively. For cosmological models 
$\chi_{1,2}^2\equiv V_{0\text{c}}/(3-\alpha)$.
The value $\alpha=3$ corresponds to the point 
where $V_{\text{c}0}$ changes sign
from positive (for $\alpha<3$) to negative (for $\alpha>3$). 
Since the  extreme solutions are characterized by
$\phi = -\sqrt{\alpha}a
$ they  are represented by  straight lines in the
$(\dot a,\dot\phi)$ phase diagram \cite{Halliwell:PLB87}.

\subsubsection{Cosmological horizons and domain wall event horizons} 

We shall now relate the nature of the cosmological horizons to  
the
event horizons in the domain wall background.
If we write the cosmological line element in the standard form
\begin{equation}
ds^2 = d\tau^2 - R^2(\tau)\, \left(\frac{dr^2}{1+\beta^2 r^2} + r^2  
d\Omega_{2}^2\right),
\end{equation}
then the \emph{convergence} of the integral
\begin{equation}
I=\int^{\tau_{1}}_{\tau_{0}}\! \frac{d\tau }{R(\tau)} 
=\int^{\eta_{1}}_{\eta_{0}}\! d\eta 
\end{equation}
in the limit $\tau_{1}\rightarrow \tau_{\text{max}}$
is a necessary and sufficient condition for the existence of
a cosmological event horizon \cite{Weinberg:JWS72}. 
Note that
the complex rotation to the domain wall space--time interchanges  
a space dimension with the time
dimension.
Because of this, 
the sufficient and necessary condition for having 
an event horizon in the domain wall space--time is 
that
\begin{equation}
I=\int^{\eta_{\text{max}}}_{\eta_{0}}d\eta = \eta_{\text{max}}-\eta_{0}
\end{equation}
\emph{diverges.}
In other words, if there is a singularity at finite $\eta$, then
this singularity is naked.

\begin{figure}[hbt]
\centering{\epsfig{file=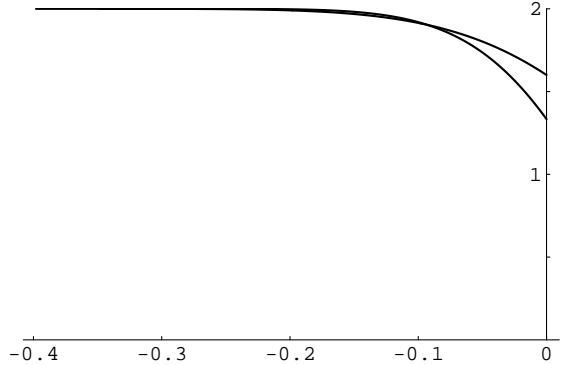,%
                   height=5cm,%
                   clip=}}
\caption{The ``equation of state''  parameter $\gamma$ versus $z$ in units of $\chi$ 
for
extreme solutions with $\alpha=4$ (upper curve) and $\alpha=6$.
For $0\leq \alpha \leq 3$, the equations of state are straight lines
$\gamma=2\alpha/3$.}
\label{Fig:Ext_Gamma}
\end{figure}

\subsubsection{Equation of state}
For $0\leq \alpha\leq 3$, 
the equation of state
is given by
\begin{equation}
\gamma = \frac{2\alpha}{3}. 
\end{equation}
The ``stringy'' value, $\alpha=1$, is therefore the border line 
between the solutions corresponding to attractive and repulsive
equations of state in the cosmological picture. 
In the domain wall system, this is the dividing line of domain walls
with naked singularities ($\alpha >1$), and domain walls
with  
the singularity hidden behind a horizon ($\alpha < 1$).

For $\alpha>3$ the equation of state is time-dependent 
(see Fig.~\ref{Fig:Ext_Gamma}).  
It approaches $\gamma=2$ near the singularity.

           \addtocounter{equation}{-\value{equation}}
%

%
           \section{The r\^ole of supergravity domain walls in basic theory}
\label{Sect:ConnectionOTD}

The purpose of this chapter is two-fold. First, we 
emphasize a connection of supergravity 
domain walls to other topological defects of 
four-dimensional supergravity theories. 
Specifically, a complementary nature of dilatonic 
Type I extreme domain walls and certain extreme 
black holes, some of them appearing as 
Bogomol'nyi--Prasad--Sommerfield 
(BPS)
saturated states 
\cite{Bogomolnyi:SJNP76,PS:PRL75,CPNS:PRD77} 
 of $N=4$ (or $N=8$) superstring vacua, are summarized. 
Second,  we 
illuminate how (and when) such domain walls can arise in 
fundamental  theories, like 
the $N=1$ effective supergravity  
theory  of superstring vacua. 
In particular we shall emphasize  different nature of the domain walls
 when the perturbative as well as non-perturbative effects
in superstring theory are included.

\subsection{Connection to  topological defects in superstring theory}
\label{Sect:connection}

The study of topological defects  in superstring theory is 
an important topic extensively discussed in the literature. 
For  recent reviews see, \eg\   
Refs.~\cite{CHS:Trieste91,DKL:PRep95}. 
Recently, it has been recognized  that supersymmetric solitons, 
also referred to as BPS-%
saturated states 
play a 
crucial r\^ole \cite{HT:NPB95,Witten:NPB95} 
in establishing non-perturbative dualities  in string theory. 
Namely, BPS-saturated states are  non-perturbative 
configurations with the  minimal energy in its class, 
and in the case when  the corresponding string 
vacuum has a large enough supersymmetry, \ie\ 
$N\ge 2$, the energy of the BPS-saturated states may not receive 
quantum (loop) corrections. In this case the  
expression for the mass of such solitons can be 
trusted not only in the weak coupling, but also 
in the strong coupling regime.  Since such 
BPS-saturated states, along with the 
perturbative string excitations, 
contribute to the  full spectrum of the theory,  
they provide a nontrivial test  to establish equivalence 
(at the  spectrum 
level)  of certain  strongly coupled and  
(dual) weakly coupled string vacua.

The majority of BPS-saturated states in 
super\-string theory, which have been 
studied in the literature, 
corresponds to (charged) p-brane configurations 
 in various dimensions  ($d \ge 4$)  and  
su\-per\-string vacua with 
su\-per\-sym\-me\-try  $N\ge 2$.\footnote{For a 
recent review, see, \eg\  Ref.~\cite{Townsend:95} 
and references therein.}  
Such configurations therefore  need  not bear 
direct connection to the supergravity walls, 
which were addressed  in previous Sections as  nontrivial con\-fi\-gu\-rations 
in four-dimensions   
within $N=1$ supergravity theories. 
It is however interesting that  certain  
p-brane solutions in (p+2)-dimensions  may still possess
similar features as  extreme dilatonic domain walls. 
In Refs.~\cite{DK:NPB94,DKMR:NPB94} a special example of 
four-dimensional su\-per\-sym\-me\-tric domain walls 
of toroidally compactified string theory was found, 
whose extreme limit \cite{DK:NPB94}
corresponds to a specific
example of dilatonic domain walls,
with the r{\^o}le of
the cosmological constant being played by the constant gauge field
strength. 

Interestingly, certain  p-brane solutions in dimensions  higher than four 
possess \cite{GT:PRL93,DGT:PLB94}
the same interesting feature as  extreme four-dimensional 
domain walls; namely,  they  interpolate between different types of  (higher 
dimensional) supersymmetric vacua.

In the following we   shall also  see 
that there is an intriguing  
complementarity between  the  
space--time structure of certain extreme   
black holes  and certain extreme  
Type I domain walls as described in 
the following subsection \ref{Sect:WBHComplement}. 
Some of these these extreme black holes  
appear as  special  cases of BPS-saturated  
black holes of $N=4$ (or $N=8$)  superstring vacua as 
will be shown in the subsequent subsection \ref{Sect:BPSBH}.

\subsubsection{Extreme domain wall and black hole complementarity}
\label{Sect:WBHComplement}

Interestingly, Type I supersymmetric (extreme)\  
domain 
walls  in the $(x,y)$ plane of 
four-dimensional  $N=1$  supergravity 
theories with a general dilaton coupling $\alpha > 0$ (see Sections~\ref{BPL}
and \ref{Sect:Lagtopdef}) 
have the same global space--time 
structure  in the $(t,z)$ slice  as 
of the extreme magnetically charged black
holes  with the  coupling $1/\alpha$   in the
($t,r)$ hyperspace 
\cite{Cvetic:PLB94}. 

The origin of this complementarity lies in the nature of the $N=1$ 
supergravity Lagrangian (\ref{Eq:SUSYLagrangian}) with  
one $\mathrm{U}(1)$ gauge superfield 
$\mathcal{W}$, a gauge neutral chiral 
matter superfield $\mathcal{T}$ (with 
nonzero 
superpotential) and  one  linear supermultiplet  
rewritten in terms of a gauge 
neutral chiral superfield $\mathcal{S}$. 
Recall (see Section~\ref{BPL}), 
that in this case the theory  is specified by  
the gauge coupling function 
$f_{a b}=\delta_{a b}\mathcal{S}$,   superpotential 
$W=W_{\text{matt}}(\mathcal{T})$ and a
 a separable 
K\"ahler potential $K = K_{\text{matt}} (\mathcal{T}, \mathcal{T}^*) + 
K_{\text{dil}}(\mathcal{S}, \mathcal{S}^* )$
where
$K_{\text{dil}}(\mathcal{S} , \mathcal{S}^* )= -\alpha 
\ln(\mathcal{S},\mathcal{S}^*)$ and $S=\e^{-2\phi/\sqrt\alpha}+iA$
Eqs.~(\ref{gaugefunc})--(\ref{Eq:Kahlerpot}). 

The black holes 
with  a general dilaton coupling \cite{GM:NPB88,HW:NPB92}
are  spherically symmetric solutions of the 
theory (\ref{Eq:SUSYLagrangian})\  with 
the  matter fields $T$ turned off,  
\ie\ $V\equiv
0$, however, with non-zero  $\mathrm{U}(1)$ 
gauge fields $F_{\mu\nu}\ne 0$. 
Note that  now the kinetic energy term 
for the gauge field is of the 
form: ${1\over 4}\e^{-2\phi/\sqrt{\alpha}}F_{\mu\nu}F^{\mu\nu}$. 
This particular coupling arises in the 
$N=1$ supergravity Lagrangian with the 
$\mathrm{U}(1)$ Yang--Mills gauge field 
and a general linear supermultiplet (see Eq.~(\ref{Eq:SUSYLagrangian}).  It is, however,  not  
known whether such  a term  with a  
general coupling $\alpha$  of the dilaton 
to the gauge fields has an embedding 
into  supergravity Lagrangian  with $N\ge 2$. 
We shall however see later that for special values of 
$\alpha$, the black hole Lagrangian has an embedding into 
a consistently truncated  $N=4$ (as well as $N=8$)  
Lagrangian of string vacua. In  this case it 
can be shown that the corresponding extreme 
black holes  are supersymmetric and 
thus have the minimum energy  in their class, 
\ie\ they are BPS-saturated states. We shall 
defer the discussion of these states to Section~\ref{Sect:BPSBH}.

\begin{figure}[htb] 
\centering{\subfigure[{}]{\epsfig{file=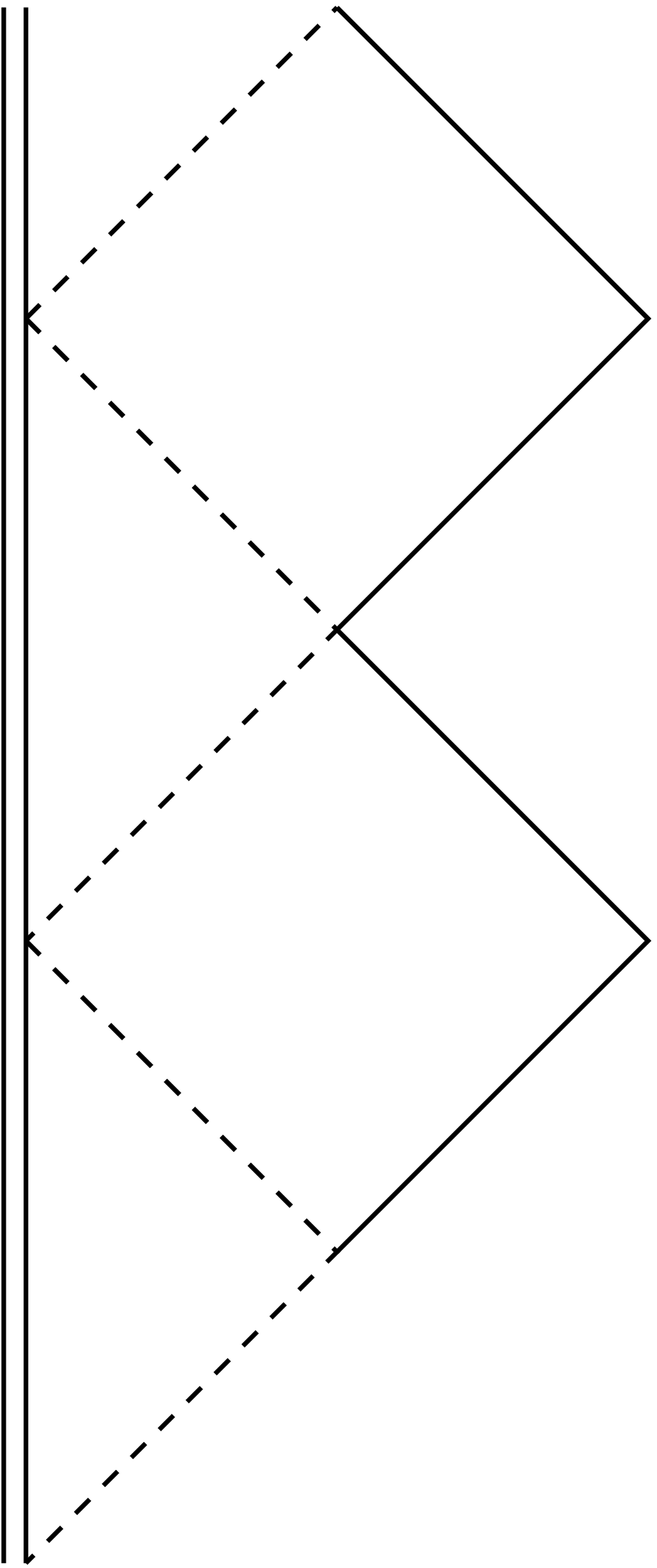,%
                   height=5.75cm,%
                   clip=}}
                   \quad\quad 
           \subfigure[{}]{\epsfig{file=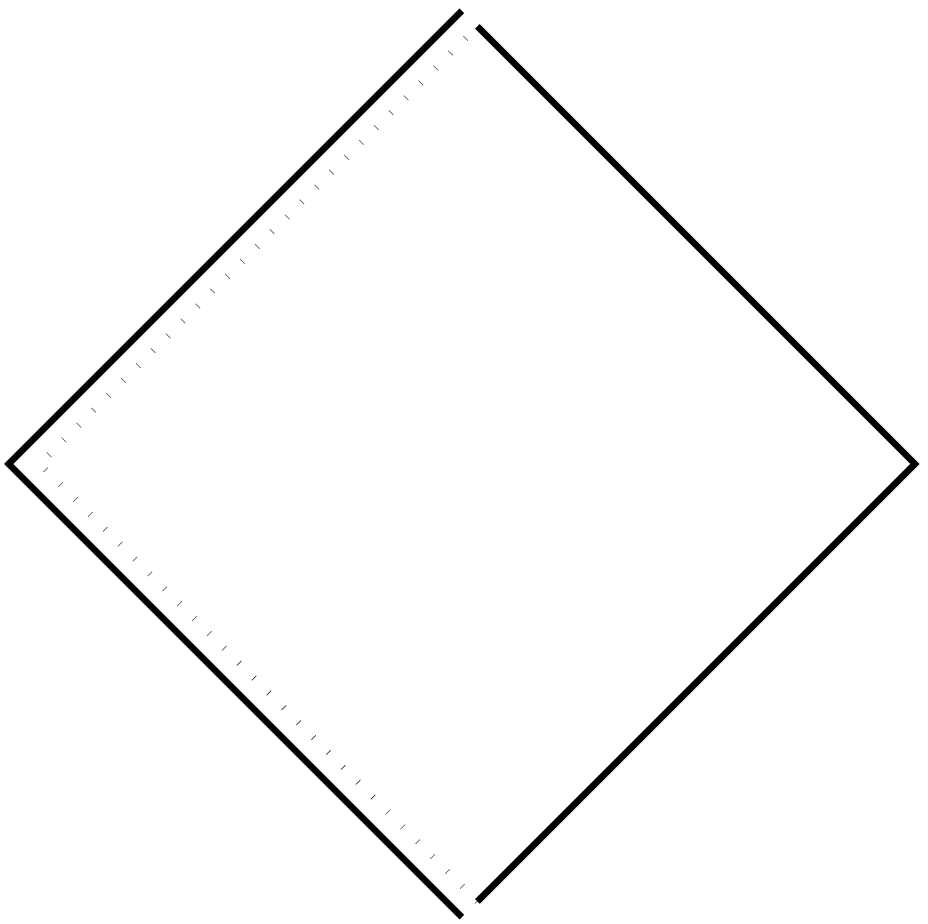,%
                   height=2.3cm,%
                   clip=}}
                   \quad\quad
           \subfigure[{}]{\epsfig{file=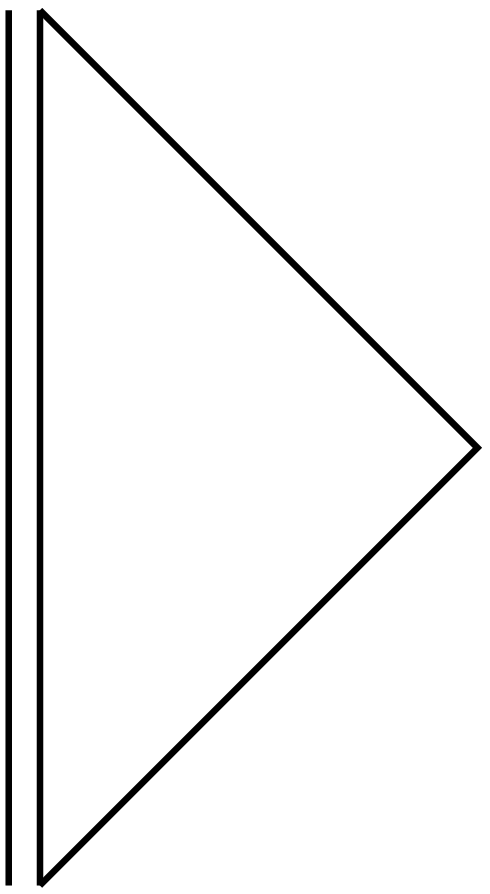,%
                   height=2.3cm,%
                   clip=}}
                   }
\caption{
Penrose--Carter 
diagrams 
for the extreme  magnetically  charged black holes 
with
$\alpha=\infty$, $\infty>\alpha\ge 1$, and $1>\alpha\ge 0$,
respectively.
The double lines corresponds to  the timelike, null  
and naked singularity.
} 
\label{Fig:BHoleandNaked}
\end{figure}

The  (Einstein frame)\ metric  for the extreme  magnetically charged 
solution  with the general dilaton coupling is of the form
\cite{GM:NPB88,GHS:PRD91,HW:NPB92}:\footnote{%
The corresponding electrically charged 
black holes have the same  
Einstein
frame metric, however, the dilaton solution is related to the  
corresponding
magnetic one by the  transformation $\phi\rightarrow -\phi$.}
\begin{align}  
\label{Eq:metbh} 
ds^2 & = \lambda(r)dt^2 - \lambda(r)^{-1}dr^2 -  
R(r)d\Omega_2^2,
\tag{\ref{Eq:metbh}a}
\label{Eq:metbha}
\displaybreak[3]\\
\intertext{with:} 
\lambda(r)&=\left(1-{r_0\over r}\right)^{{2\alpha\over {1+\alpha}}},\  
\ R(r)=r^2
\left(1-{r_0\over r}\right)^{{2\over {1+\alpha}}}.
\tag{\ref{Eq:metbh}b}
\end{align}\addtocounter{equation}{1}%
The dilaton field $\phi$ and the magnetic field are of the form:
\[
\e^{2\phi/\sqrt\alpha}=\left(1-{r_0\over r}\right)^{-{2\over  
{1+\alpha}}}, 
\ \  F_{\theta\phi}= P\sin\theta .
\]
Here $P$ is the magnetic charge of the black hole, 
$r_0^2=P^2({{1+\alpha}\over \alpha})$ and the mass $M$ of the black  
hole is
\[
M= |P| \left({\alpha \over {1+\alpha}}\right)^{1\over 2}. 
\]
The global space--time structure (and the related
thermal properties)\ of the extreme magnetically charged
dilatonic black holes  bear  striking similarities
to the one  of the domain wall configurations, however,  
now the
role of $\alpha$ is inverted: 
\begin{itemize}
\item
$\alpha=\infty$   corresponds to the  case, where the dilaton field does not
couple to the gauge fields. The solution therefore corresponds to the  
extreme  Reissner--Nordstr\"om black hole, 
which has a timelike singularity
at $r=0$ and where $r=r_0$ corresponds to a Cauchy horizon.  
Its  
global space--time  structure (see Figure~\ref{Fig:BHoleandNaked}a)\ 
in the $(r,t)$ direction  is
\emph{the same} as the one  of 
the Type I supergravity  walls  in the $(z,t)$ direction
(cf.\ Figure~\ref{Fig:ExtremeI}). 
In the latter case, however,
the  timelike singularity is replaced by the  wall. 
The corresponding 
black hole
Hawking temperature vanishes.
\item
$\infty>\alpha>1$ corresponds to  solutions with  
the curvature singularity at
$r=r_0$. Timelike radial geodesics
reach $r=r_0$ in  infinite time.
Therefore $r=r_0$
corresponds to a null singularity (see Figure~\ref{Fig:BHoleandNaked}b). 
The temperature vanishes.
\item
$\alpha=1$ corresponds to the stringy extreme magnetically charged
black hole  with  a null singularity at $r=r_0$ 
 (see Figure~\ref{Fig:BHoleandNaked}b) however,
 the temperature is $T= M/8\pi$. 
\item
$\alpha<1$, corresponds to solutions, where the  singularity at $r=r_0$  is
reached  by a radial geodesics in a
\emph{finite} proper time. Thus,  the singularity is \emph{naked} 
(see Figure~\ref{Fig:BHoleandNaked}c), and
the temperature $T$ is infinite.
\end{itemize}
Thus, the extreme magnetically charged
stringy dilatonic black hole with $\alpha=1$ and  $T=M/(8\pi)$  serve as a
 dividing line \cite{GM:NPB88} between 
extreme charged dilatonic black holes with $\alpha>1$ and $T=0$
and the naked singularities with $\alpha<1$ and $T=\infty$.  

Therefore the global space--time in the
$(t,z)$ slice  for  extreme walls with coupling $\alpha$
is the same as the one  in the $(t,r)$ slice for  extreme magnetically
charged black holes  with coupling $1/\alpha$ 
(see Figure~\ref{Fig:BHoleandNaked}).
Between the two solutions 
the dilaton coupling
$\alpha$ is
inverted, while  the 
r\^ole of $W_{\text{matt}}(T)$ on one side of the wall  and 
the magnetic charge $P$
of the black hole are interchanged.

Near the singularity, the   metric (\ref{Eq:Ansatz})\ in the $(t,z)$
slice of the wall  with the coupling $\alpha$
is the same as the metric (\ref{Eq:metbh})\  in the $(t,r)$ slice of the   
black 
hole with the coupling $1/\alpha$. Namely, in the region
$r-r_0\equiv\rho \rightarrow 0^+$, the  
coordinates $(t,\rho)$  of the black hole with the coupling
$\alpha$  and  the coordinates $(t,z)$ of the wall with the
coupling $\tilde \alpha\equiv1/\alpha$ are related  in the following  
way:
\begin{equation} 
\frac{\rho}{r_0}=
\left\{ \begin{array}{ll} 
\left[1-\frac{1}{2} 
(\tilde\alpha-1)\sigma_{\text{ext}} |z|\right]^
{\frac{\tilde\alpha+1}{\tilde\alpha(\tilde\alpha-1)}} & \quad 
\text{for}\quad \alpha\ne 1 
\\
\e^{-\sigma_{\text{ext}}|z|}& \quad  
\text{for}\quad  
\alpha=1 
       \end{array}
\right.
\end{equation}
 where  
$r_0=2/[(1+\tilde\alpha)\sigma_{\text{ext}}]$
and $\tilde\alpha\equiv 1/\alpha$.

Near the singularity the
dilaton blows up  in both cases, however, unlike the
two-dimensional
metric slices, the  coordinate dependence of
the dilaton near the singularity is \emph{different} in either case. 
This fact is also reflected in the different form of the
corresponding two-dimensional effective actions.

The  complementarity  between the 
global space--time structure of the extreme dilatonic
domain walls with coupling $\alpha$ and 
extreme charged dilatonic black holes
with coupling $1/\alpha$ can be  
traced back to  the nature of the coupling 
$e^{2\sqrt\alpha\phi}$ of
the dilaton to the matter potential (the source for the wall)\
and the
complementary coupling $e^{-2\phi/\sqrt\alpha}$ of
the dilaton to the  gauge kinetic energy  (the source of
the charge of the black hole).

The complementarity 
($\alpha\leftrightarrow 1/\alpha$)\ between  the  
extreme wall and extreme charged
black hole solutions is a generalization of  
the one found \cite{CDGS:PRL93} 
 between  extreme 
vacuum domain walls ($\alpha=0$)\ and ordinary extreme black holes 
($\alpha=\infty$). Interestingly, only for the 
$N=1$ supergravity with the
coupling $\alpha=1$, 
which corresponds to an effective tree level theory 
from super\-strings, both extreme dilatonic walls \emph{and} 
extreme  charged
dilatonic black holes are 
void of  naked singularities.
Such a  complementarity seems to exist only in the case of 
\emph{extreme} configurations.
In the case of the  non-\ or ultra-extreme configurations the 
complementarity is not carried over. For one thing, the domain walls are not 
static configurations anymore, while non- and ultra-extreme black holes 
remain static configurations. Nevertheless, the connection between the 
space--times of domain walls and black holes deserves further study.

In the following subsection we shall summarize 
results about supersymmetric embedding of  
some of the above extreme black holes within $N=4$ superstring vacua.

\subsubsection{Relationship to the BPS-saturated black holes of 
               $N=4$ superstring vacua}
\label{Sect:BPSBH}
Effective $N=4$ superstring vacua  in four dimensions  
can be  parameterized in terms of  massless fields 
of heterotic string compactified in six-torus  
$T^6$ (and equivalently, due to string-string duality, 
in terms of massless fields of  
type IIA string theory compactified on 
$K3\times T^2$ where $K3$ 
is the two- complex-dimensional Calabi--Yau manifold)\@.\footnote{An analogous embedding exists also in the case  of $N=8$ superstring vacua, parameterized in terms of the massless fields of  toroidally compactified Type IIA superstring.}
For a review see  Ref.~\cite{Sen:IJMPA94}. 

The  bosonic part of the 
 effective $N=4$  Lagrangian is parameterized in terms of  the graviton, 28
$\mathrm{U}(1)$ gauge fields and 134 scalar fields,  
with complex field $S$, parameterizing 
the gauge coupling and 132 scalar-moduli, 
parameterizing the  six-torus of the 
compactified space. The solutions of the theory
possess the  $T$-duality $\mathrm{O}(6,22)$ 
symmetry, associated with the symmetries 
of the compactified space,  and $S$-duality
$\mathrm{SL}(2,\mathbb{R})$  relating the 
strong and weak couplings of string vacua.
The general BPS-saturated spherically symmetric  
solutions of this effective Lagrangian, 
parameterized by 28 electric and 28 magnetic charges 
have been obtained in Ref.~\cite{CT:95,CY:PRD96,CT:PRD96},  
while general
non-extreme solutions 
(compatible with the corresponding Bogomol'nyi bound) 
have been constructed in Ref.~\cite{CY:95}\@.\footnote{%
See also Ref.~\cite{CY:95} for the references for 
certain special solutions of this effective theory.}

Here we will quote only  a special example  of the 
BPS-saturated  spherically symmetric static solutions, 
which are parameterized by 
two electric charges $Q_1$ and $Q_2$ of the  
respective Kaluza--Klein and the `two-form' 
$\mathrm{U}(1)$ gauge fields  associated with 
the first compactified direction of the six-torus, and 
two magnetic charges $P_1$, $P_2$  of the  
respective Kaluza--Klein and the `two-form' $\mathrm{U}(1)$ 
gauge fields  associated with the second  
compactified direction of the six-torus first obtained in 
Ref.~\cite{CY:PRD96}\@.\footnote{The special case
with $Q_1=Q_2$ and $P_1=P_2$ was
obtained in Ref.~\cite{KLOPP:PRD92}.} 
In this case the  space--time metric (\ref{Eq:metbha}), 
the dilaton field $\Phi$ and the two 
scalar fields $g_{11}$ and $g_{22}$, \ie\ the respective moduli 
for the circles of the first and the second 
compactified direction, are of the form:
\begin{equation}
\lambda={{r^2}\over{\left[(r+Q_1)(r+Q_2)(r+P_1)(r+P_2)\right]^{1\over 2}}},
\ \  R\lambda =r^2, 
\label{meto} 
\end{equation}
\begin{equation}
\e^{2\Phi}=\left[{{(r+P_1)(r+P_2)}\over{(r+Q_1)(r+Q_2)}}\right]^{1\over 2}
, \ \ g_{11}={{r+P_2}\over{r+P_1}},  \ \ 
g_{22}={{r+Q_1}\over{r+Q_2}}, 
  \label{modo}
\end{equation}
with the ADM  mass:
\begin{equation}
M={1\over 4}(Q_1+Q_2+P_1+P_2).
\label{massBPS}
\end{equation}

Interestingly, for special values of  
the above four charge assignments the solution 
reduces to the  solution (of the consistently) 
truncated $N=4$ bosonic Lagrangian with only one 
$\mathrm{U}(1)$ gauge field,  coupled to one scalar field 
field $\phi$ with the specific value of 
the coupling $\alpha$. These non-zero 
values of the  coupling $\alpha$ and the corresponding 
non-zero  charge assignments  
are  of the following form \cite{CY:PRD96,DLR:NPB96}:
\begin{alignat}{2} 
\label{diffal}
\alpha &=\infty, & \qquad    P_1&=P_2=Q_1=Q_2\equiv P,
\tag{\ref{diffal}a}
\displaybreak[3]\\ 
\alpha &=3,      & \qquad    P_1&=P_2=Q_1\equiv {2\over \sqrt3} P,       
\tag{\ref{diffal}b}
\displaybreak[3]\\ 
\alpha &=1,      & \qquad    P_1&=P_2\equiv {\sqrt 2}P,
\tag{\ref{diffal}c}
\displaybreak[3]\\ 
\alpha &={1\over 3} & \qquad P_1&\equiv {\sqrt3\over 2}P.
\tag{\ref{diffal}d}
\end{alignat}\addtocounter{equation}{1}%
With the  charge assignments (\ref{diffal}),   
Eqs.~(\ref{meto}) and (\ref{modo}) 
reproduce the  solution for the extreme magnetically 
charged black holes (discussed in the 
previous Section)  whose   value of  
$\alpha$  is also given in (\ref{diffal}),  while the 
 the radial coordinates  related as
$r\to r+P\sqrt{(1+\alpha)/\alpha}$. 
 Thus, within $N=4$ superstring vacua  
there are BPS-saturated black hole solutions 
with four specific charge assignments. Each of these assignments 
falls within the four specific classes of the 
scalar couplings $\alpha$, \ie\  
$\alpha=\infty$, $\alpha= 3>1$, 
$\alpha=1$, and $\alpha=1/ 3<1$,  each of them 
with  its distinct space--time structure 
(see Figure~\ref{Fig:BHoleandNaked}) 
and thermal properties, which are complementary 
to the corresponding extreme Type I domain walls.

\subsection{Domain walls within $N=1$ superstring vacua}
\label{DWstrings}

In this Section we would like to discuss  
how the  supergravity domain wall configurations,  
arising within general effective $N=1$  
supergravity  Lagrangian  (described in  
Section~\ref{Sect:Supergravity}), could arise within  effective 
$N=1$ super\-gra\-vi\-ty of   su\-per\-string 
vacua, with or with\-out inclusion of non-per\-tur\-ba\-tive effects. 
For a review of the  structure and in particular  
constraints on the field dependence of the couplings for the
$N=1$ effective Lagrangian from  superstring theory, 
see Ref.~\cite{KL:NPB95} (see also Ref.~\cite{Quevedo:96}.).

\subsubsection%
[Walls without inclusion of   
non-perturbative effects]%
{Stringy domain walls without in\-clu\-sion of 
non-per\-tur\-ba\-tive  effects}

\label{withoutnonpert}

Perturbative $N=1$  supersymmetric four-dimensional string vacua 
are specified by the $N=1$ supergravity 
Lagrangian (\ref{Eq:SUSYLagrangian})  of Section~\ref{Sect:Supergravity} 
 where the  linear multiplet $\mathcal{S}$  
has a tree level K\"ahler potential (\ref{Eq:kald}) 
with $\alpha=1$. 
The matter chiral superfields split into  moduli 
$\mathcal{T}_i^{\text{mod}}$,
 parameterizing the symmetries of the 
 compactified space, which do not have any  
self-interaction in the superpotential (\ref{superpot}), 
and the matter  chiral superfields $\mathcal{T}_j^{\text{matt}}$ 
which are in general charged under 
the gauge group and    
have non-trivial superpotential $W$.  
In certain cases, if some of the moduli $\mathcal{T}_i^{\text{mod}}$,
do not couple to the (gauge neutral)  
matter fields which can  acquire non-zero 
vacuum expectation value (and form the topological defect),  
these moduli could act as   
effective   ``dilaton-type'' fields with a separable 
K\"ahler potential of the form
(\ref{Eq:Kahlerpot}). In some cases the 
toroidal moduli  would play the r\^oles of dilaton-type 
fields with the separable K\"ahler potential 
$-\alpha\ln(\mathcal{T}_i^{\text{mod}}+\mathcal{T}_i^{\text{mod}*})$ with
 $\alpha=1,2,3$. 

In principle, the superpotential for  the  matter fields can 
possess  discrete symmetries.
Sometimes they are consequences of discrete symmetries of the 
compactified space, \ie\ Calabi--Yau compactified space.  
Therefore  one can
 (at least) in principle allow for  the  existence of isolated 
minima of the matter potential, and thus for the  existence of 
dilatonic domain  walls discussed in Section~\ref{Sect:DilatonicWalls}  
with an effective coupling 
$\alpha$  which  assumes a discrete values $\ge 1$.

Thus, it is in principle possible that perturbative string vacua 
allow  for  the existence of dilatonic  
domain walls with  effective discrete 
values of $\alpha\ge 1$ \cite{Cvetic:PRL93,CY:PRD95}  
at  energy scales 
that  
are larger  
than the  
energy scale 
where the 
non-perturbative effects, \eg\ gaugino condensation,   
take place. However, we should point out that  the 
existence of isolated  
superstring vacua with non-zero  vacuum expectation 
of the (neutral) matter fields  
would be an indication of the perturbative 
\emph{instability} of such string vacua,  
which is  an unlikely possibility\@.%
\footnote{For the study of domain walls 
(with fixed dilaton  values)  due to  
spontaneously broken discrete symmetry 
 of the compactified Calabi--Yau 
manifolds see Refs.~\cite{CK:PRL85,CR:PLB87}.}

\subsubsection[Walls with inclusion of non-perturbative effect]%
{Domain walls with inclusion of non-perturbative effects}
\label{withnonpert}

A more likely possibility is that 
specific supergravity walls  exist  
as solutions within $N=1$ superstring vacua  after 
the (non-perturbative) supersymmetry breaking effects are taken into 
account. However, at present
the non-perturbative effects in string theory 
are not well understood. One scenario is based on 
gaugino condensation in the hidden sector 
of the string theory, which would in turn 
provide a dilaton dependent, and due to 
genus-one threshold corrections,  also a 
moduli dependent superpotential\@.\footnote{For a recent 
review of the status of non-perturbative 
effects due to  gaugino condensation 
see for example Ref.~\cite{Quevedo:95}.
For another non-perturbative phenomenon
within Liouville (non-critical)
string theory, which induce a periodic
non-perturbative potential see Ref.~\cite{OT:PLB91}.}  
In this case it is believed that  
the dilaton could be  
stabilized and  the   non-perturbatively 
induced superpotential for the (toroidal) moduli respects
a discrete non-compact  (generalized) 
target space duality symmetry 
also referred to as $T$-duality, \eg\ the 
$\mathrm{SL}(2, \mathbb{Z})$, which is the 
symmetry associated with the 
toroidally compactified space of the string theory. 
$\mathrm{SL}(2,\mathbb{Z})$ is specified by:
\begin{equation}
T^{\text{mod}}\to {{aT^{\text{mod}}-ib}\over{icT^{\text{mod}}+d}}, \ \   
ad-bc=1,\ \ 
a,b,c,d\in \mathbb{Z},
\label{SLTZ}
\end{equation}
where the real part of $T^{\text{mod}}$ specifies 
the radius of the compactified space.
Non-perturbative potentials for the modulus field $T^{\text{mod}}$, 
respecting $\mathrm{SL}(2,\mathbb{Z})$ symmetry,
a discrete noncompact symmetry, were 
extensively classified  in Ref.~\cite{CFILQ:NPB91}.

The  physics of  the moduli fields is an 
intriguing generalization of the well known 
axion physics  introduced to solve the strong CP problem in QCD\@.%
\footnote{For a review see \cite{Kim:PRep87}.} 
Namely, perturbatively the effective 
Lagrangian for  the modulus field $T^{\text{mod}}$, 
only,  possesses
a non-compact symmetry $\mathrm{SL}(2,\mathbb{R})$, 
which is broken down to its discrete subgroup $\mathrm{SL}(2,\mathbb{Z})$. 
Thus, prior to the non-perturbative effects taking place, 
the modulus Lagrangian allows for the existence of  
``stringy'' cosmic strings \cite{GSVY:NPB90}. 

In general,  non-perturbative moduli potentials for the  moduli fields, 
respecting $\mathrm{SL}(2,\mathbb{Z})$ symmetry,   allow for discrete 
isolated vacua, and  thus for the existence of  supergravity 
vacuum domain walls of the type discussed in Section~\ref{Sect:VacuumWalls}. 
In the cosmological context such domain walls  
are bounded \cite{CQR:PRL91} by stringy cosmic strings. 
Further cosmological implications of such domain walls were 
studied in  Ref.~\cite{CD:PLB92}.

\begin{figure}[htb] 
\centering{\epsfig{file=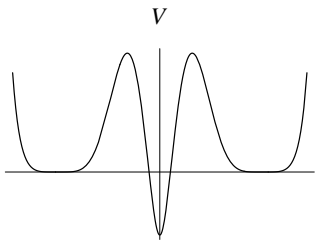,%
                   height=3.3cm,%
                   clip=}}
\caption{
Modular invariant potential along the geodesic 
$T(z)=\e^{i\Phi(z)}$ for the K\"ahler potential and superpotential
of Eq.~(\protect\ref{Eq:kahlsuper}).}
\label{Fig:ModInvariantV}
\end{figure}

\begin{figure}[htb] 
\centering{\subfigure[{}]{\epsfig{file=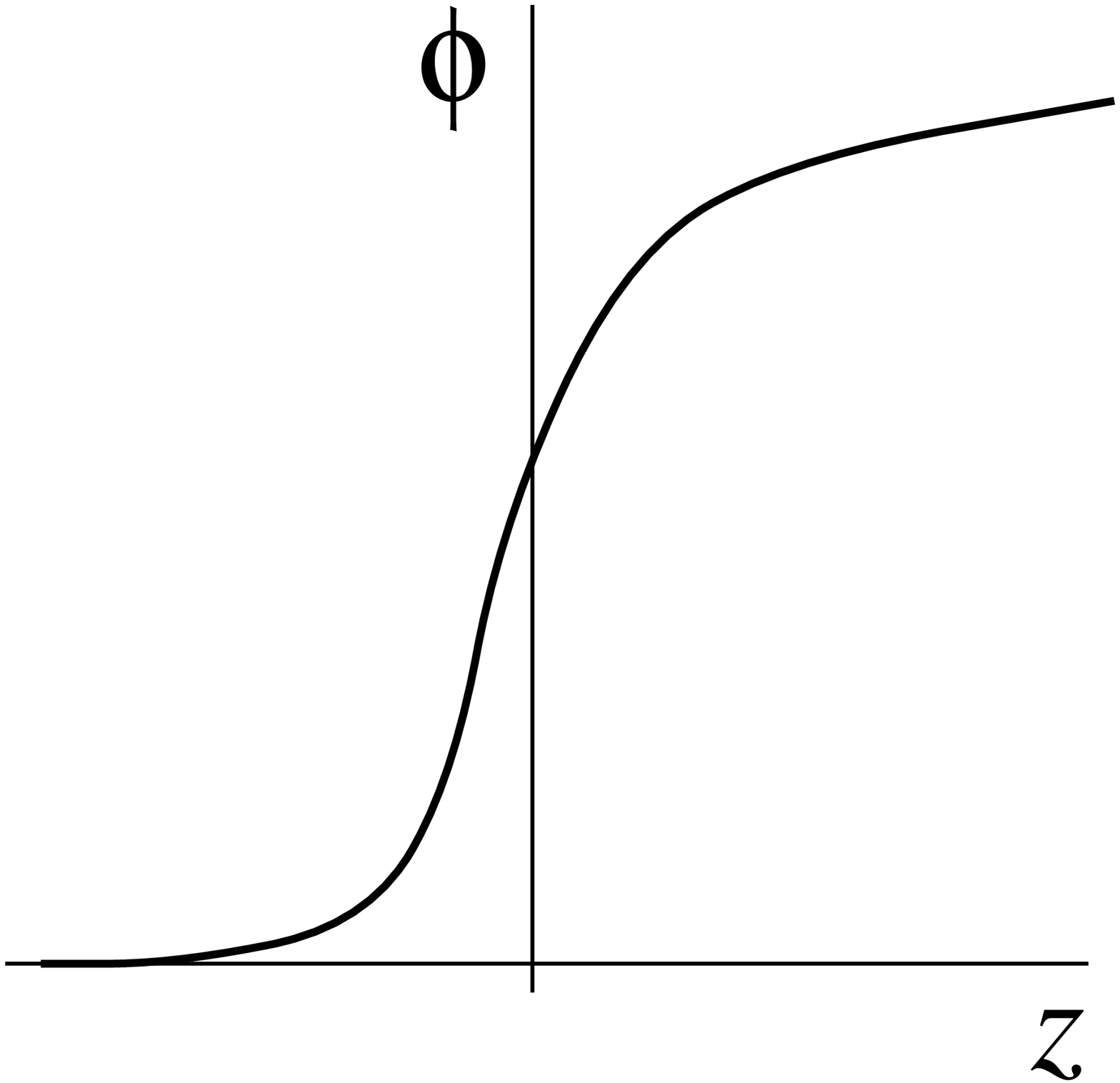,%
                   height=3.3cm,%
                   clip=}}
                   \quad\quad
           \subfigure[{}]{\epsfig{file=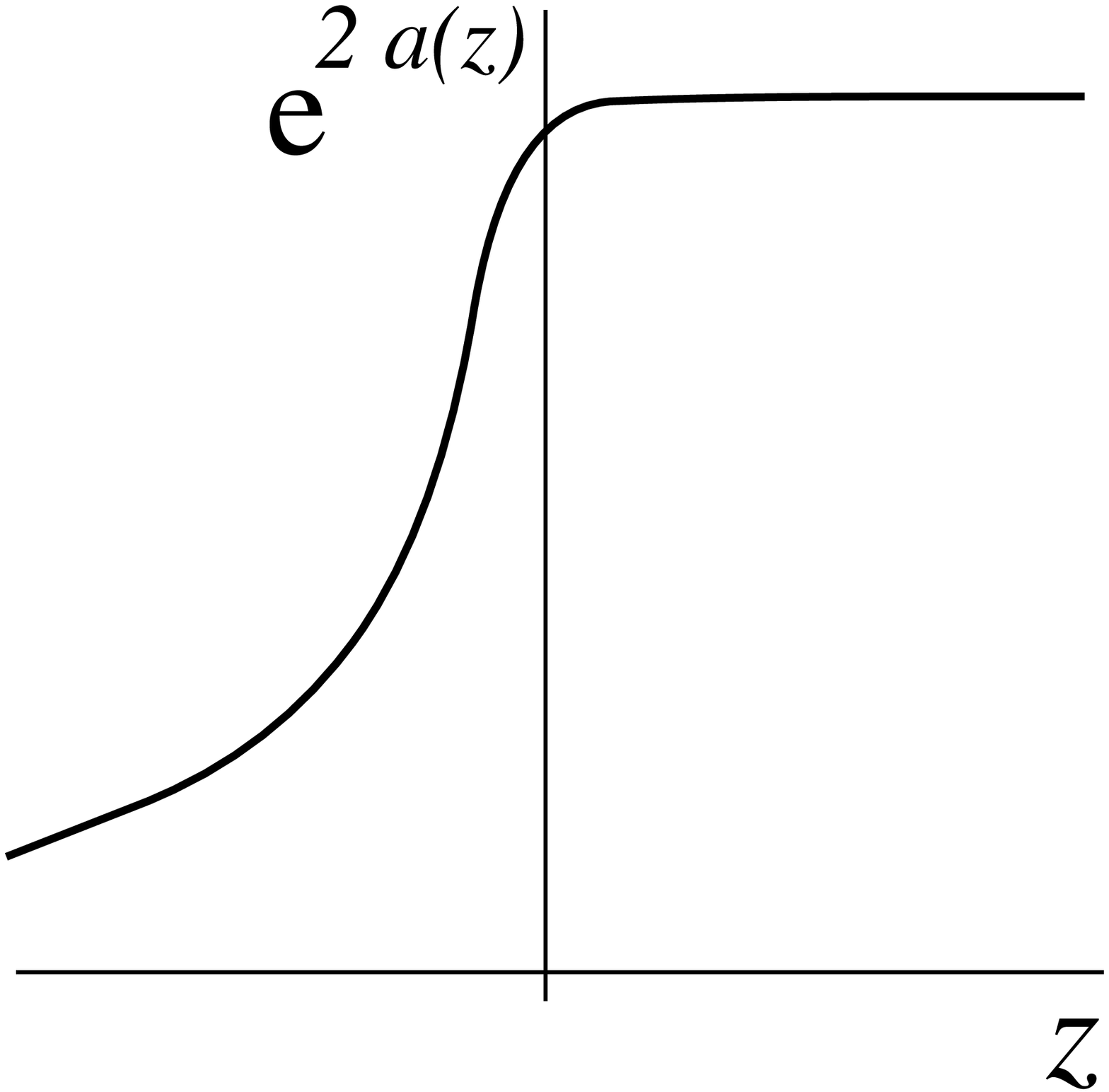,%
                   height=3.3cm,%
                   clip=}}
                   }
\caption{
The solutions for $\Phi(z)$ and
$\e^{2 a(z)}$ 
for the potential of Fig.~\protect\ref{Fig:ModInvariantV}.}  
\label{Fig:ThickWall}
\end{figure}

An illustrative example  of  an explicit  
solution \cite{CGR:NPB92} for the (finite size) domain 
wall  (with the 
$\mathrm{SL}(2,\mathbb{Z})$ symmetry due to one modulus 
field $T=T^{\text{mod}}$) is given  for the potential   with the   
following choice of  K\"ahler potential and superpotential \cite{CGR:NPB92}:
\begin{equation}
K=-3\ln \left[(T+T^{*})\, 
|\eta(T)|^4\right],\quad 
W=\Omega^3 J(T),
\label{Eq:kahlsuper}
\end{equation} 
where $J$ and $\eta$ are the absolute modular invariant 
and Dedekind function (the modular function 
with the modular weight $-1/2$), 
respectively \cite{Schoeneberg:Springer74}\@.\footnote{
For a review of the modular group $\mathrm{SL}(2,\mathbb{Z})$ 
and its modular functions see, \eg\ 
Ref.~\cite{Lehner:AMS64}. A large class of  
$\mathrm{SL}(2,\mathbb{Z})$  invariant potentials
are given in Ref.~\cite{CFILQ:NPB91}.} 
Here $\Omega$ corresponds  to the scale where 
non-perturbative effects, \eg\ gaugino 
condensation, which stabilize the dilaton, take place. 
In this case the potential has two  
supersymmetric isolated minima, one at $T=1$ 
($\mathbb{Z}_2$ Symmetric point of the fundamental domain, 
corresponding to the anti-de Sitter space--time) 
and the other one at $T=\e^{i\pi/6}$ ($\mathbb{Z}_3$ 
symmetric point of the fundamental domain, corresponding to the 
Minkowski space--time).    
The   domain wall is  then the  extreme Type I  vacuum domain  wall 
with the geodesic for $T(z) =\e^{i\Phi(z)}$ (\ie\ 
on the boundary of the fundamental domain). 
In  Figure~\ref{Fig:ModInvariantV} the potential along  $\Phi(z)$  
is given, while  the solution   for $\Phi(z)$  and the 
conformal  (metric) factor $\e^{2 a(z)}$ as a function of $z$   
are given in Figure~\ref{Fig:ThickWall}. 
Note that those correspond to 
explicit numerical solutions of the 
Bogomol'nyi  equations (\ref{boge}) (without dilaton field).  

Another example \cite{CD:PLB92}
is provided by choosing 
$W=\Omega^3\eta(T^{\text{mod}})^{-6}$. In this case the 
potential has the property \cite{CFILQ:NPB91} that  
the discrete  \emph{degenerate} minima with negative 
cosmological constant   break supersymmetry. 
The  type of walls in this case are the non-extreme 
(reflection symmetric)  vacuum domain walls between anti--de~Sitter 
vacua.  For further discussion of cosmological implications, 
including the possibility of inflation due to 
non-perturbative moduli potential, see Ref.~\cite{CD:PLB92}.

Another type of domain wall configuration could 
arise if the 
non-per\-tur\-ba\-tively induced potential for the dilaton field 
$S$ preserves the the weak--strong  
coupling duality symmetry, also referred to as 
$S$-duality  symmetry. 
 The strong-weak coupling duality conjecture assumes that 
the string vacua posses the $\mathrm{SL}(2,\mathbb{Z})$ 
symmetry associated with the dilaton  
$S$. Thus, in this case  the non-perturbatively 
induced potential for the  dilaton field respects  
the $\mathrm{SL}(2,\mathbb{Z})$ symmetry, and thus 
one can have  vacuum domain walls  associated with 
the dilaton field, whose features are 
analogous to those associated with the 
modulus field $T^{\text{mod}}$.   
The possibility of such walls (due to dilaton field)  
and their physical implications, including the 
possibility to account for inflation within domain walls 
\cite{Linde:PLB94,Vilenkin:PRL94},  
have been recently studied  in Refs.~\cite{HM:NPB94,BBSMS:PRD95}.

Another possibility may be that the non-perturbative 
effects in the superpotential are negligible, but 
the K\"ahler potential for the dilaton is modified \cite{BD:95}
so that it preserves the $SL(2,\mathbb{Z})$ symmetry. In this 
case the  domain wall cannot be formed 
due to the dilaton field. However, another 
field (with non-zero) superpotential   forms the 
wall, while the dilaton would modify  
the solution in such a way that it would 
describe  the domain wall with the self-dual 
dilaton coupling discussed in  Section~\ref{Sect:Self-Dual}.
 In this case 
the non-\ and ultra-extreme solutions  can reduce to the singularity-free 
vacuum domain wall solutions \cite{CS:PRD95}.

Further study of non-perturbative effects in 
string theory would shed light not only on aspects of supersymmetry 
breaking, but also on the  physical  
implications  of  domain walls in string theory.

           \addtocounter{equation}{-\value{equation}}
%

%
           \section{Conclusions}
\label{Sect:Conclusions}

In this  paper we have systematically 
reviewed domain wall 
configurations as  solutions of  general $N=1 $ supergravity theory. 
We have  given a thorough 
analysis of  
the
space--time structure of such domain 
walls and  emphasized the special 
r\^ole that supersymmetry 
is playing in determining the nature of such 
configurations.  Detailed results were given for both the 
vacuum 
domain walls, \ie\ configurations where on  
either side of the wall  all the matter fields  
assume  constant vacuum expectation  
values, as well as  for 
dilatonic domain walls, where the dilaton field, which  couples to
 the  scalar potential, varies  with a spatial separation from the wall. 

The vacuum domain walls can be classified according 
to
the energy density stored in the wall.  It turns out 
that  domain walls interpolating between 
supersymmetric minima of the matter potential correspond to the 
\emph{static}
planar  configurations, whose energy density 
(in the thin wall approximation) 
is $\sigma=\sigma_{\text{ext}}$, 
specified by  the value(s) of the 
cosmological constant(s) on each side of the wall. 
The  non-extreme walls  (with $\sigma>\sigma_{\text{ext}}$) 
correspond to expanding bubbles with two 
insides, while ultra-extreme  wall solutions  
with ($\sigma<\sigma_{\text{ext}}$) correspond to 
the false vacuum decay bubbles. 
The extreme solution  (with supersymmetric embedding) therefore 
provides a dividing line between the two 
classes of domain wall solutions.

Vacuum domain walls between vacua with 
non-positive  cosmological constants can belong 
to any of the  three classes. The 
domain walls with at least one of the 
cosmological constant positive correspond 
to the false vacuum decay walls, only.

Dilatonic domain walls can be classified 
analogously with extreme dilatonic domain 
walls  corresponding to static walls which 
interpolate  between  supersymmetric vacua 
with varying dilaton field and thus a 
space-time structure different from that 
of extreme vacuum domain walls.  The nature 
of the space-time crucially depends on on 
the value of the dilaton coupling to the matter potential.
The non-extreme and ultra-extreme dilatonic 
walls again correspond to expanding bubbles, 
however, now one always encounters one or more naked singularities.
The results due to non-perturbative corrections,  which induce the 
dilaton potential were also analysed.  

We also  reviewed  the properties of supergravity 
domain walls as they appear  within  effective $N=1$ 
supergravity from superstring theory with  or without 
inclusion of non-perturbative effects.   
Perturbative string vacua may allow for 
appearance of dilatonic domain walls. On the other 
hand, a non-perturbatively  induced potential for 
the moduli fields, which preserve discrete 
non-compact symmetry of the compactification 
space  ($T$-duality), as well as a
non-perturbatively  
induced potential for the  dilaton field, which 
preserves the   strong-weak coupling duality 
($S$-duality),   allow for vacuum domain walls 
due  to the modulus and/or dilaton field. Such 
domain walls may have  interesting  cosmological 
implications. However, before
a more detailed analysis of the physics of 
domain walls within $N=1$ string 
vacua can be  carried out, a  better understanding of the 
non-perturbative phenomena in string theory is needed.

Interestingly, the space-time structure of  
certain extreme (supersymmetric) domain walls has 
a space-time structure that is closely related to that 
of certain extreme  magnetically charged 
dilatonic black holes, some of them 
corresponding to the BPS-saturated states 
of $N=4$ (or $N=8$) superstring vacua.  Further  study of 
the connection between supergravity walls and 
other topological defects in supergravity 
(and superstring theory) is also  needed.

While this review provides a systematic analysis 
of the domain wall solutions within  $N=1$ supergravity theory, 
little has been said beyond  the  general statements,  about their implications 
for cosmology, in particular about mechanisms 
by which they can be formed in the early universe 
and their  implications for the early universe evolution.
Also the stability of such solutions as well 
as related dynamical questions have to be 
studied in more details. All of these 
questions await further investigations.

\subsubsection*{Acknowledgements}
We would  especially like to 
thank   S. Griffies for collaboration on 
a number of papers on vacuum domain walls in supergravity theory.  
We also benefited from useful discussions and 
collaborations with  R. Davis,  {\O}. Gr{\o}n, B. Jensen,
S.-J. Rey, and  D. Youm. The work of M.C.  is supported  by  
the Institute for Advanced Study funds and
J. Seward
Johnson foundation,  U.S. Department of Energy Grant No.\ 
DOE-EY-76-02-3071, the National Science
Foundation Career Advancement Award PHY95-12732 and 
the NATO collaborative research grant CGR 940870. 
The work of H.H.S is supported
by CERN funds.  
M.C. would also like to thank the Theory Divison at CERN, 
where a  part of the work was done, for the hospitality.

           \addtocounter{equation}{-\value{equation}}
%

%

\small

\bibliography{abbrevss,
              BlackHole,%
              Cosmology,%
              CosmicStrings,%
              Defect,%
              DomainWall,%
              EnergyExpr,%
              GRBooks,%
              Isometry,%
              Inflation,%
              QuantumGravity,%
              Phenomenology,%
              PBrane,%
              StringCosmology,%
              StringTheory,%
              SSB,%
              SuperGravity,%
              TwoPlusOne,%
              VacuumDecay,%
              VacuumLambda}

\end{document}